\documentclass[twocolumn,english,aps,prl,showpacs,superscriptaddress]{revtex4-2}
\usepackage[utf8]{inputenc}
\usepackage{geometry}
\usepackage{array}
\usepackage{booktabs}
\usepackage{amsmath}
\usepackage{amssymb}
\usepackage{graphicx}
\usepackage{esint}

\makeatletter

\providecommand{\tabularnewline}{\\}

\usepackage{babel}

\makeatother

\usepackage{babel}
\begin{document}
\title{High-fidelity and robust controlled-Z gates for superconducting transmon
qubits. %
\mbox{%
\mbox{%
II%
}%
}. System-parameter uncertainties and overall errors }
\author{Chia-Hsien Huang}
\affiliation{Department of Physics and Center for Theoretical Physics, National
Taiwan University, Taipei 10617, Taiwan}
\affiliation{Center for Quantum Science and Engineering, National Taiwan University,
Taipei 10617, Taiwan}
\author{Hsi-Sheng Goan}
\email{goan@phys.ntu.edu.tw}
\affiliation{Department of Physics and Center for Theoretical Physics, National
Taiwan University, Taipei 10617, Taiwan}
\affiliation{Center for Quantum Science and Engineering, National Taiwan University,
Taipei 10617, Taiwan}
\affiliation{Physics Division, National Center for Theoretical Sciences, Taipei 106319, Taiwan}

\date{\today}
\begin{abstract}
In the companion paper \citep{Huang2026}, we developed an optimal-control
framework to suppress intrinsic and decoherence-induced errors
in controlled-Z (CZ) gates between directly coupled, frequency-tunable
transmons. Here, we extend this framework to incorporate uncertainties
in coupling strength, anharmonicities, and zero-flux qubit frequencies,
enabling the design of high-fidelity gates robust against noise and
parameter uncertainties. We classify these uncertainties according
to whether they perturb the off-diagonal coupling strengths or diagonal
detunings of the Hamiltonian and identify distinct suppression mechanisms
for each class. Coupling errors are mitigated through dynamical decoupling
induced by strong detuning modulation. For a $30\text{-ns}$ gate,
this strategy reduces the coupling error to $6.21\times10^{-7}$ and
allows coupling-strength uncertainties up to $2\pi\times4\,\mathrm{MHz}$
while maintaining a total gate infidelity below $10^{-3}$. Detuning
errors are minimized near a symmetric-detuning operating point that
balances the two dominant leakage-mediated interactions. A $24\text{-ns}$
gate operated near this point tolerates individual uncertainties of
up to $2\pi\times5\,\mathrm{MHz}$ in each relevant diagonal parameter
under the same infidelity threshold. We further optimize gate performance
under realistic conditions where intrinsic errors, decoherence, and
parameter uncertainties coexist across six representative stages of
progressively improving hardware performance. The resulting stage-specific
controls achieve lower infidelity and greater robustness. At the most
advanced stage considered, the optimized gate reaches an intrinsic
error of $1.93\times10^{-8}$ and a total infidelity of $4.55\times10^{-6}$,
limited primarily by energy relaxation. Together with the companion
study, this framework provides a practical approach for balancing
speed, fidelity, and robustness while reducing the need for exhaustive
characterization and frequent recalibration in large-scale superconducting
quantum processors. 
\end{abstract}
\maketitle
\global\long\def\figurename{FIG.}%
\makeatletter 
\global\long\def\thefigure{\@arabic\c@figure}%
\makeatother

\section{Introduction}

\begin{figure}
\includegraphics[scale=0.42]{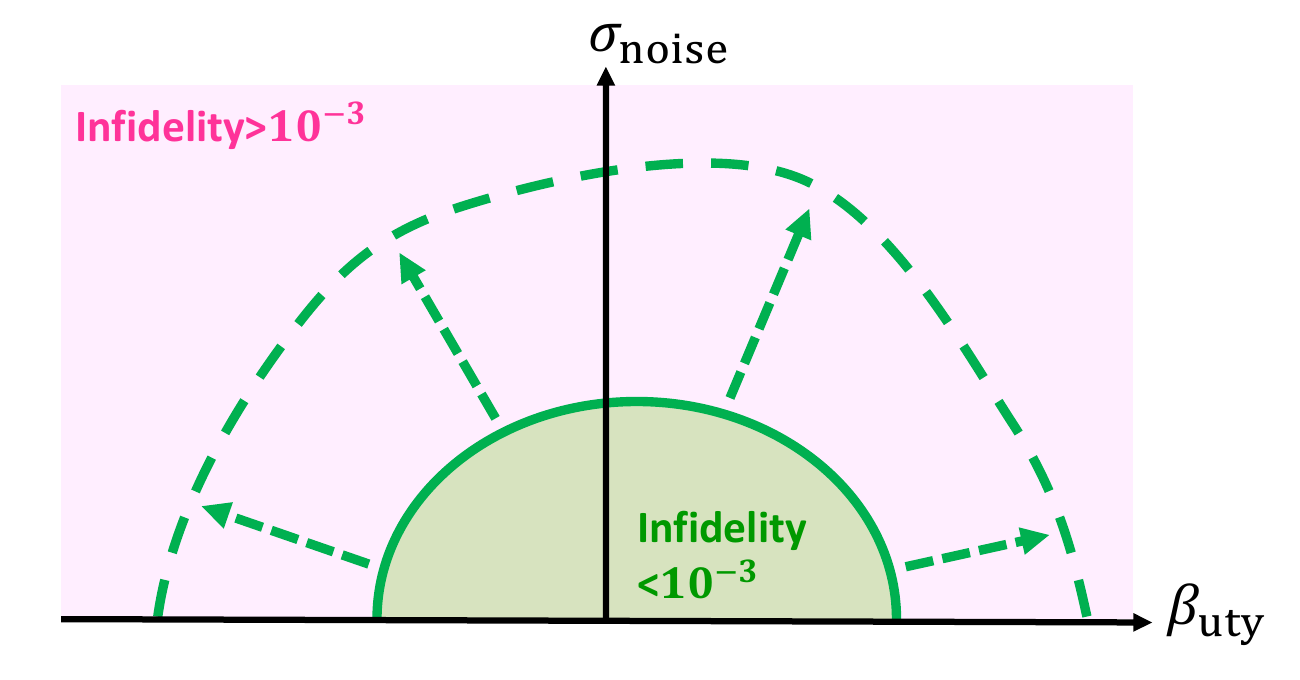}

\caption{Schematic illustration of the enlargement of the robustness window
against noise and system-parameter uncertainties. The vertical axis
represents the noise standard deviation, $\sigma_{\mathrm{noise}}$,
and the horizontal axis represents the system-parameter uncertainty,
$\beta_{\mathrm{uty}}$. Gate operations corresponding to parameter
combinations $\left(\beta_{\mathrm{uty}},\sigma_{\mathrm{noise}}\right)$
within the green region bounded by the solid green curve have infidelities
below the prescribed threshold of $10^{-3}$, whereas those outside
the solid green curve (pink region) exceed the threshold.
Robust control by engineering gate operations less sensitive to parameter variations and noise fluctuations can broaden the robustness window by expanding the green region from
the solid to the dashed green curve, thereby enabling the gate to
tolerate larger values of $\sigma_{\mathrm{noise}}$ and $\left|\beta_{\mathrm{uty}}\right|$
while maintaining the gate infidelity below the threshold of $10^{-3}$.}

\label{fig:robustness_window_scematical_illustration}
\end{figure}

High-fidelity and robust two-qubit gates are essential for fault-tolerant
quantum computation \citep{FowlerMariantoniMartinisEtAl2012,Acharya2025}.
High fidelity reduces the physical-qubit and circuit-depth overheads
associated with quantum error correction \citep{FowlerMariantoniMartinisEtAl2012,Acharya2025},
while robustness ensures that gate performance is maintained despite
imperfect knowledge, device-to-device variation, environmental fluctuations,
and temporal drift in hardware parameters \citep{Weidner2025,Proctor2020}.
The latter requirement, robustness, becomes increasingly important
as quantum processors scale. Although precise characterization and
frequent recalibration may be feasible for a small number of qubits,
applying the same procedure to every interacting qubit pair in a large
processor imposes a substantial operational burden. Moreover, system
parameters and noise environments can drift during the characterization
and calibration process itself, causing controls optimized at nominal
operating conditions to lose fidelity \citep{Kelly2016,Burnett2019,Proctor2020,Vepsaelaeinen2022,marciniak2026millisecondscalecalibrationbenchmarkingsuperconducting}.
Robust control addresses this challenge by engineering gate operations
that are intrinsically less sensitive to parameter uncertainties and
noise fluctuations \citep{Grace2012,HuangGoan2017,Weidner2025}. By
increasing the range of parameter perturbations and noise strengths
over which the gate error remains below a prescribed threshold, robust
control can substantially broaden the robustness window, as schematically
illustrated in Fig.\ \ref{fig:robustness_window_scematical_illustration},
thereby reducing the required characterization precision and calibration
frequency and facilitating the practical operation of large-scale
quantum hardware \citep{Carvalho2021,Le2023,Wright2026}. 

Superconducting transmon qubits constitute a leading platform for
quantum information processing \citep{KochYuGambettaEtAl2007,Place2021,Siddiqi2021},
with experimentally demonstrated two-qubit controlled-Z (CZ) gate
fidelities exceeding $99\%$, reaching as high as $99.93\%$ \citep{BarendsKellyMegrantEtAl2014,Barends2019,Rol2019,Negirneac2021,Foxen2020,Sung2021,Li2025a,Marxer2026,Li2024,Li2025}.
CZ gates on frequency-tunable transmons are commonly implemented by
tuning the computational state $\left|1_{B}1_{A}\right\rangle $ close
to an avoided crossing with a noncomputational state, such as $\left|0_{B}2_{A}\right\rangle $
or $\left|2_{B}0_{A}\right\rangle $ \citep{BarendsKellyMegrantEtAl2014,Krantz2019}.
Here, subscripts A and B denote qubit-A ($Q_{A}$) and qubit-B
($Q_{B}$). The resulting strong interaction enables rapid accumulation
of the required conditional phase, but it also makes the gate sensitive
to inaccuracies in the parameters determining the locations and strengths
of these avoided crossings \citep{Hwang2026}. Small deviations in
 coupling strength, anharmonicities, or qubit frequencies can alter
the accumulated conditional phase, modify population transfer, and
increase leakage. Consequently, a pulse that is highly accurate at
a nominal parameter point need not remain accurate when implemented
on a device whose parameters are known only within finite experimental
precision or whose parameters vary over time. 

This sensitivity to parameter deviations creates a fundamental challenge
for scalability. Conventional calibration procedures generally estimate
the relevant parameters of each coupled qubit pair and then optimize
the corresponding control pulses based on the measured nominal values.
However, as the number of qubits and couplings increases, exhaustive 
pairwise characterization becomes increasingly time-consuming. The
estimated parameters may also drift before the calibration of all gates
in the processor is completed, necessitating a continuous cycle of
measurement, optimization, and recalibration. A scalable control strategy
should therefore not require every parameter to be estimated with
extreme precision. Instead, it should produce gates whose fidelity
varies only weakly over experimentally relevant ranges of parameter
deviations. Such intrinsic insensitivity makes high-fidelity operation
less dependent on a precisely characterized nominal Hamiltonian, increases
the interval between recalibrations, and improves the portability
of optimized pulse designs across nominally similar devices.

In the companion paper \citep{Huang2026}, hereafter Part $\mathrm{I}$,
we developed an optimal-control framework for CZ gates in directly
coupled, frequency-tunable transmons that treats intrinsic and decoherence-induced
errors within a common formalism. Smooth single-shot flux pulses
were designed to suppress leakage, swap, and phase errors while accounting
for energy relaxation, fast dephasing associated with room-temperature
control electronics, and slow dephasing arising from $1/f$ flux noise.
By identifying noise-resilient operating regimes and optimizing the
corresponding pulse trajectories, Part $\mathrm{I}$ demonstrated
near-speed-limit gates with high fidelity and substantial tolerance
to dephasing noise. These results showed how pulse design can make
gate operations intrinsically less sensitive to environmental fluctuations,
rather than relying exclusively on improved noise characterization
and repeated calibration.

However, the analysis in Part $\mathrm{I}$ assumed that the system
parameters entering the nominal Hamiltonian were accurately known.
In practice, these parameters are inferred from measurements with
finite precision, vary across devices because of fabrication inhomogeneity,
and may drift over time. A complete robust-control framework must
therefore address both stochastic noise and uncertainty in the Hamiltonian
itself. Noise robustness protects the gate against variations in noise
strength, whereas parameter robustness protects it against imperfect
characterization and systematic deviations from nominal device values.
Their combination is necessary to reduce the calibration burden: suppressing
noise alone does not eliminate sensitivity to inaccurately estimated
Hamiltonian parameters, while tolerance to static parameter offsets
alone does not protect against stochastic noise.

Here, we extend the framework of Part $\mathrm{I}$ to uncertainties
in the direct coupling strength $g_{AB}$, the transmon anharmonicities
$\alpha_{A}$ and $\alpha_{B}$, and the zero-flux qubit frequencies
$\omega_{0,A}$ and $\omega_{0,B}$. These parameter uncertainties
influence the gate dynamics in physically distinct ways. Uncertainty
in $g_{AB}$ perturbs off-diagonal matrix elements of the Hamiltonian,
leading to coupling errors. By contrast, uncertainties in anharmonicities 
and zero-flux frequencies perturb diagonal matrix elements, shift
the relevant transition frequencies, and primarily produce single-qubit
phase errors and detuning errors. Single-qubit phase errors can
 be largely compensated for by calibrated virtual or physical single-qubit Z
gates. The residual effects are governed principally by the detuning
errors associated with the $\left|1_{B}1_{A}\right\rangle $$\leftrightarrow$$\left|0_{B}2_{A}\right\rangle $,
$\left|1_{B}1_{A}\right\rangle $$\leftrightarrow$$\left|2_{B}0_{A}\right\rangle $,
and $\left|0_{B}1_{A}\right\rangle $$\leftrightarrow$$\left|1_{B}0_{A}\right\rangle $
interactions. This decomposition reveals the mechanisms through which
different uncertainties degrade the CZ gate, thereby enabling the
design of control strategies tailored to each error class.

The distinction between coupling and detuning errors is central to
our approach. Near the strong $\left|1_{B}1_{A}\right\rangle $$\leftrightarrow$$\left|0_{B}2_{A}\right\rangle $
interaction, uncertainty in $g_{AB}$ acts approximately as a transverse
perturbation, whereas the controllable detuning provides a longitudinal
control field. Strong temporal modulation of the detuning can therefore
dynamically average out the coupling perturbation, thereby producing
a dynamical decoupling mechanism without requiring an additional control
channel. This mechanism makes the resulting operation
less sensitive to moderate deviations in the nominal coupling strength.
Using this strategy, we suppress the coupling error to $6.21\times10^{-7}$
for a $30\text{-ns}$ gate, compared with $3.54\times10^{-4}$ for
a conventional pulse without the designed modulation. The resulting
gate tolerates coupling-strength deviations of up to $\pm2\pi\times4\,\mathrm{MHz}$
while maintaining a total infidelity below $10^{-3}$. This broad
tolerance window directly relaxes the precision with which the coupling
of each qubit pair must be characterized.

Diagonal uncertainties require a different strategy because they enter
through the same Hamiltonian channels as the frequency controls, and
therefore their effects cannot be removed by the preceding detuning-modulation
mechanism. Instead, we identify a robust operating regime near the
symmetric-detuning point, $\delta_{A}\approx\delta_{B}$, where the
sensitivities of the $\left|1_{B}1_{A}\right\rangle $$\leftrightarrow$$\left|0_{B}2_{A}\right\rangle $
and $\left|1_{B}1_{A}\right\rangle $$\leftrightarrow$$\left|2_{B}0_{A}\right\rangle $
interactions are balanced. Here, $\delta_{A}$ and $\delta_{B}$ denote
the detunings associated with the respective transitions. Operating
near this point reduces the overall sensitivity of the gate dynamics
to detuning shifts and substantially suppresses the combined detuning
error arising from these two interactions. The resulting $24\text{-ns}$
gate tolerates individual diagonal-parameter uncertainties of up to
$2\pi\times5\,\mathrm{MHz}$ while maintaining a total infidelity
below $10^{-3}$. Thus, coupling and detuning errors are suppressed
through complementary mechanisms: dynamical decoupling for off-diagonal
perturbations and symmetry-assisted sensitivity reduction for diagonal
perturbations. Together, these mechanisms extend gate robustness beyond
a single nominal operating point to finite parameter ranges.

Robustness against parameter uncertainty generally competes with gate
speed. Strong modulation and operation near the symmetric-detuning
point may require longer gate durations than a pulse optimized solely
for nominal fidelity, increasing exposure to energy relaxation and
dephasing. The optimal solution must therefore minimize the overall
error, rather than any individual contribution, by balancing intrinsic
dynamics, decoherence, and parameter uncertainty. To make this trade-off
explicit, we consider six representative stages of progressively improving
hardware performance, characterized by reduced decoherence and parameter
uncertainty. Under present-day conditions, energy relaxation dominates,
favoring a gate near the speed limit of $12\,\mathrm{ns}$; the benefits
of a longer and more robust trajectory are insufficient to compensate for the
resulting increase in decoherence. However, as coherence improves, the uncertainty of the parameters becomes an increasingly significant contributor
to total infidelity, and robust gates with durations of approximately
$24\text{--}28\,\mathrm{ns}$ become preferable. This stage-specific
analysis provides practical guidance on when hardware improvements
should be accompanied by a transition from speed-oriented to robustness-oriented control.

At the most advanced stage considered, combining low decoherence with
small parameter uncertainties, fine-tuning within a five-level transmon
model, together with Gaussian waveform filtering, reduces the intrinsic
error to $1.93\times10^{-8}$. The resulting total infidelity is approximately 
$4.55\times10^{-6}$ and is limited primarily by energy relaxation.
More broadly, the analysis across the six stages shows that the optimal
gate is not universally the shortest gate or the one least sensitive
to a single parameter. Rather, the optimum depends on the hierarchy
of experimentally relevant errors. By decomposing the total infidelity
into modular contributions and adapting the cost function to the prevailing
hardware conditions, the framework provides a practical strategy for
selecting gate duration, operating regime, and pulse structure.

The principal contributions of this work are therefore threefold.
First, we formulate and quantify CZ-gate errors arising from uncertainties
in both the off-diagonal and diagonal Hamiltonian parameters. Second,
we identify two distinct physical error-suppression mechanisms---detuning-modulation-based
dynamical decoupling and the symmetric-detuning operating point---and
incorporate them into smooth, experimentally realizable flux pulses.
Third, we combine intrinsic, decoherence-induced, and parameter-uncertainty-induced
errors into a stage-specific framework that tracks the evolution of
the optimal control strategy as hardware performance improves. Together
with the noise-robust control developed in Part $\mathrm{I}$, the
resulting gates are less sensitive to both parameter
uncertainties and noise fluctuations. Their tolerance windows reduce
the precision with which individual coupled qubit pairs must be characterized,
extend the intervals between recalibrations, and directly address
a major obstacle to the implementation and maintenance of large-scale
superconducting quantum processors.

The remainder of this paper is organized as follows. Section\ \ref{subsec:uncertainty_error_analysis}
introduces the uncertain system Hamiltonian, defines the parameter-uncertainty-induced
error measures, and distinguishes between diagonal and off-diagonal
uncertainty channels. Secs.\ \ref{subsec:uty_robust_against_gAB_uncertainty}
and \ref{subsec:uty_robust_against_detuning_uncertainty} develop
the corresponding robustness strategies---dynamical suppression of
coupling errors and operation near the symmetric-detuning point to
reduce detuning errors, respectively. Section\ \ref{sec:overall_infidelity_improvement}
combines these mechanisms with the intrinsic-error and decoherence
models developed in Part $\mathrm{I}$ and presents optimization results
for six representative hardware stages. Section\ \ref{sec:Conclusion}
summarizes the principal conclusions and discusses extensions to tunable-coupling
architectures and additional control channels. Together with Part
$\mathrm{I}$, the present work establishes a unified framework for
fast, high-fidelity, and robust CZ gates that accommodate both stochastic
noise and parameter uncertainties, thereby reducing calibration overhead
and facilitating practical operation of large-scale fault-tolerant
superconducting quantum processors.

\section{System-parameter uncertainties}

\label{sec:robust_against_uncertainty}

In Sec.\ \ref{subsec:uncertainty_error_analysis}, we analyze the
contributions to the CZ-gate infidelity arising from uncertainties
in the system parameters $g_{AB}$, $\alpha_{A}$, $\alpha_{B}$,
$\omega_{0,A}$ and $\omega_{0,B}$ of the ideal Hamiltonian $\mathcal{H}_{I}$.
These contributions are classified into two dominant categories: coupling
and detuning errors. Coupling errors originate from uncertainties
in the off-diagonal Hamiltonian elements associated with $g_{AB}$.
By contrast, detuning errors result from uncertainties in the diagonal
elements associated with $\alpha_{A}$, $\alpha_{B}$, $\omega_{0,A}$
and $\omega_{0,B}$. In Secs.\ \ref{subsec:uty_robust_against_gAB_uncertainty}
and \ref{subsec:uty_robust_against_detuning_uncertainty}, we develop
control strategies to mitigate off-diagonal and diagonal uncertainties,
respectively, and demonstrate their robustness.

\subsection{Analysis of error contributions from system-parameter uncertainties}

\label{subsec:uncertainty_error_analysis}

\begin{figure}
\includegraphics[scale=0.52]{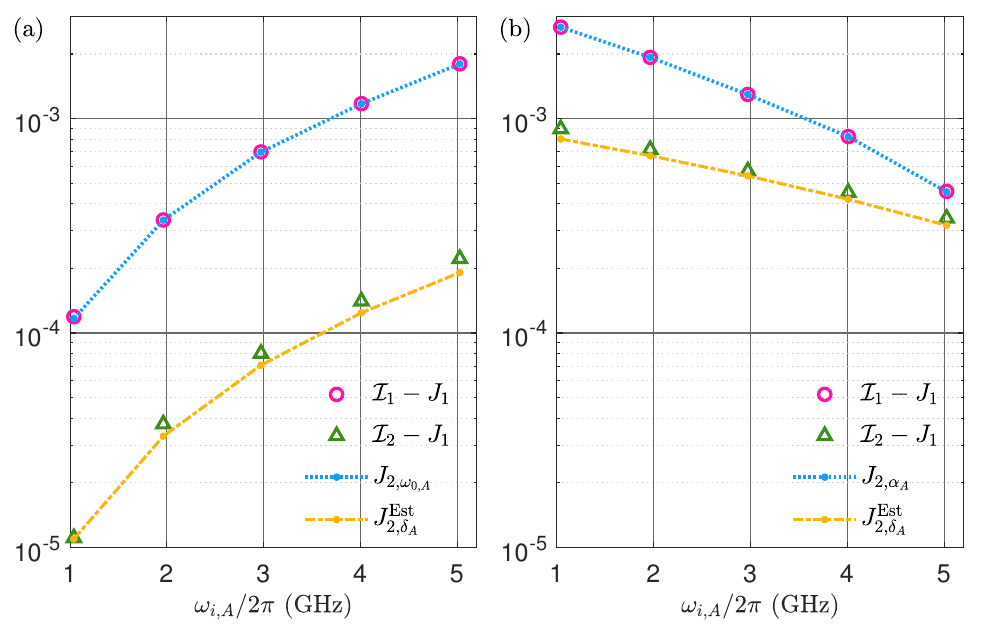}

\caption{Error contributions arising from the fixed uncertainties $\beta_{\omega_{0,A}}/2\pi=1\,\mathrm{MHz}$
(a) and $\beta_{\alpha_{A}}/2\pi=1\,\mathrm{MHz}$ (b) for optimized
CZ gates with gate time $t_{f}=12\,\mathrm{ns}$ and interaction frequency
$\omega_{i,A}/2\pi$ ranging from $\sim1$ to $\sim5\,\mathrm{GHz}$.
The ideal gate infidelity $J_{1}$ is evaluated using the ideal Hamiltonian
$\mathcal{H}_{I}$, whereas the realistic gate infidelities $\mathcal{I}_{1}$
and $\mathcal{I}_{2}$ are evaluated using $\mathcal{H}=\mathcal{H}_{I}+\mathcal{H}_{\omega_{0,A}}$
in (a) and $\mathcal{H}=\mathcal{H}_{I}+\mathcal{H}_{\alpha_{A}}$
in (b). The single-qubit-Z gate $U_{Z}$ applied after the two-qubit
interaction is calibrated using the ideal single-qubit phases for
$J_{1}$ and $\mathcal{I}_{1}$, and the realistic single-qubit phases
for $\mathcal{I}_{2}$. The infidelity difference $\mathcal{I}_{1}-J_{1}$
(pink circles) contains both the single-qubit phase and detuning errors,
whereas $\mathcal{I}_{2}-J_{1}$ (green triangles) is dominated by
the detuning error. The error contributions $J_{2,\omega_{0,A}}$
and $J_{2,\alpha_{A}}$, shown as blue dotted lines in (a) and (b),
respectively, are evaluated using Eq.\ (\ref{eq:J2_uty_general_format}).
The estimated detuning error $J_{2,\delta_{A}}^{\mathrm{Est}}$, shown
as yellow dash-dotted lines, is obtained from Eq.\ (\ref{eq:J2_detuning_00})
with $\delta_{A}=0$ and $t_{f}=12\,\mathrm{ns}$, using $\beta_{\delta_{A}}=\beta_{\omega_{0,A}}(\Delta_{A}-\alpha_{A})/(\omega_{0,A}-\alpha_{A})$
in (a) and $\beta_{\delta_{A}}=\beta_{\alpha_{A}}(\omega_{0,A}-\Delta_{A})/(\omega_{0,A}-\alpha_{A})+\beta_{\alpha_{A}}$
in (b).}

\label{fig:phase_error_and_coupling_err} 
\end{figure}

In Ref.\ \citep{Huang2026}, we employ the ideal Hamiltonian $\mathcal{H}_{I}$
in Eq.\ (1) therein, parameterized by experimentally characterized
system parameters $g_{AB}$, $\alpha_{A}$, $\alpha_{B}$, $\omega_{0,A}$
and $\omega_{0,B}$ \citep{BarendsKellyMegrantEtAl2014}, to simulate
the ideal CZ gate infidelity $J_{1}$. In the simulation, we treat
these system parameters as ideal and assume no associated uncertainties.
However, in realistic scenarios, the ideal system parameters $g_{AB}$,
$\alpha_{A}$, $\alpha_{B}$, $\omega_{0,A}$, and $\omega_{0,B}$
may each be subject to inherent uncertainties $\beta_{g_{AB}}$, $\beta_{\alpha_{A}}$,
$\beta_{\alpha_{B}}$, $\beta_{\omega_{0,A}}$, and $\beta_{\omega_{0,B}}$,
respectively, thereby introducing additional errors beyond $J_{1}$.
By substituting the realistic system parameters $g_{AB}+\beta_{g_{AB}}$,
$\alpha_{A}+\beta_{\alpha_{A}}$, $\alpha_{B}+\beta_{\alpha_{B}}$,
$\omega_{0,A}+\beta_{\omega_{0,A}}$, and $\omega_{0,B}+\beta_{\omega_{0,B}}$
for the ideal ones $g_{AB}$, $\alpha_{A}$, $\alpha_{B}$, $\omega_{0,A}$,
and $\omega_{0,B}$ in the ideal Hamiltonian $\mathcal{H}_{I}$, we
obtain the realistic Hamiltonian $\mathcal{H}=\mathcal{H}_{I}+\mathcal{H}_{g_{AB}}+\mathcal{H}_{\alpha_{A}}+\mathcal{H}_{\alpha_{B}}+\mathcal{H}_{\omega_{0,A}}+\mathcal{H}_{\omega_{0,B}}$.
Here 
\begin{align}
\mathcal{H}_{g_{AB}}/\hbar & =\beta_{g_{AB}}(-i)(a_{A}-a_{A}^{\dagger})(-i)(a_{B}-a_{B}^{\dagger}),\label{eq:uty_H_g}\\
\mathcal{H}_{\alpha_{A}}/\hbar & =\beta_{\alpha_{A}}\left[\dfrac{(\omega_{0,A}-\Delta_{A})}{(\omega_{0,A}-\alpha_{A})}a_{A}^{\dagger}a_{A}+\dfrac{1}{2}a_{A}^{\dagger}a_{A}(a_{A}^{\dagger}a_{A}-1)\right],\label{eq:uty_H_ahA}\\
\mathcal{H}_{\alpha_{B}}/\hbar & =\beta_{\alpha_{B}}\left[\dfrac{(\omega_{0,B}-\Delta_{B})}{(\omega_{0,B}-\alpha_{B})}a_{B}^{\dagger}a_{B}+\dfrac{1}{2}a_{B}^{\dagger}a_{B}(a_{B}^{\dagger}a_{B}-1)\right],\label{eq:uty_H_ahB}\\
\mathcal{H}_{\omega_{0,A}}/\hbar & =\beta_{\omega_{0,A}}\dfrac{(\Delta_{A}-\alpha_{A})}{(\omega_{0,A}-\alpha_{A})}a_{A}^{\dagger}a_{A},\label{eq:uty_H_w0A}\\
\mathcal{H}_{\omega_{0,B}}/\hbar & =\beta_{\omega_{0,B}}\dfrac{(\Delta_{B}-\alpha_{B})}{(\omega_{0,B}-\alpha_{B})}a_{B}^{\dagger}a_{B},\label{eq:uty_H_w0B}
\end{align}
denote the uncertainty Hamiltonians corresponding to the uncertainties
$\beta_{g_{AB}}$, $\beta_{\alpha_{A}}$, $\beta_{\alpha_{B}}$, $\beta_{\omega_{0,A}}$,
and $\beta_{\omega_{0,B}}$, respectively.

The additional error induced by uncertainty $\beta_{j}$, as described
by the uncertainty Hamiltonian $\mathcal{H}_{j}$, can be evaluated
as

\begin{align}
J_{2,j}= & \beta_{j}^{2}\int_{0}^{t_{f}}dt_{1}\int_{0}^{t_{1}}dt_{2}\nonumber \\
 & \left\{ \mathrm{Re}[\mathrm{Tr}\left(R_{j}(t_{1})R_{j}(t_{2})\right)_{4\times4}]/2-\right.\nonumber \\
 & \left.\mathrm{Tr}\left(R_{j,4\times4}(t_{1})\right)\cdot\mathrm{Tr}\left(R_{j,4\times4}(t_{2})\right)/8\right\} .\label{eq:J2_uty_general_format}
\end{align}
The expression $J_{2,j}$ is derived from $\left\langle J_{2,j}\right\rangle $,
defined in Eq.\ (29) of Ref.\ \citep{Huang2026}, which quantifies
the error contribution arising from stochastic noise $\beta_{j}(t)$.
Here, $\beta_{j}(t)$ is treated as a fixed parameter $\beta_{j}$
representing uncertainty, and the ensemble average $\left\langle \cdots\right\rangle $
is therefore not performed. In addition, the matrix $H_{j}(t)$ in
$R_{j}(t)=U_{I}^{\dagger}(t)H_{j}(t)U_{I}(t)$, as appearing in $J_{2,j}$,
is interpreted as the uncertainty-system coupling term defined via
the uncertainty Hamiltonian $\mathcal{H}_{j}(t)=\hbar\beta_{j}H_{j}(t)$.
We can substitute the uncertainty Hamiltonians $\mathcal{H}_{g_{AB}}$,
$\mathcal{H}_{\alpha_{A}}$, $\mathcal{H}_{\alpha_{B}}$, $\mathcal{H}_{\omega_{0,A}}$,
and $\mathcal{H}_{\omega_{0,B}}$ in Eqs.\ (\ref{eq:uty_H_g})-(\ref{eq:uty_H_w0B})
for $\mathcal{H}_{j}(t)$, respectively, to evaluate the corresponding
$J_{2,j}$, and we can also construct a composite uncertainty Hamiltonian
for $\mathcal{H}_{j}(t)$ by combining the elements of $\mathcal{H}_{g_{AB}}$,
$\mathcal{H}_{\alpha_{A}}$, $\mathcal{H}_{\alpha_{B}}$, $\mathcal{H}_{\omega_{0,A}}$,
and $\mathcal{H}_{\omega_{0,B}}$ to evaluate the associated $J_{2,j}$
for the composite uncertainty.

Next, we separate the uncertainties into two groups: the diagonal
uncertainties $\beta_{\alpha_{A}}$, $\beta_{\alpha_{B}}$, $\beta_{\omega_{0,A}}$,
and $\beta_{\omega_{0,B}}$, and the off-diagonal uncertainty $\beta_{g_{AB}}$,
and analyze their respective error contributions. The former enter
the diagonal elements of the ideal Hamiltonian $\mathcal{H}_{I}$,
whereas the latter enters its off-diagonal elements. In the analysis,
we employ the ideal Hamiltonian $\mathcal{H}_{I}$ in the dominant
basis states, together with its commuting decomposition components,
$\mathcal{H}_{I_{11}}$, $\mathcal{H}_{I_{12}}$, $\mathcal{H}_{I_{21}}$,
and $\mathcal{H}_{I_{22}}$, as defined in Eqs.\ (3)-(7) of Ref.\ \citep{Huang2026}.

The diagonal uncertainties $\beta_{\alpha_{A}}$, $\beta_{\alpha_{B}}$,
$\beta_{\omega_{0,A}}$, and $\beta_{\omega_{0,B}}$ give rise to
two distinct classes of errors: single-qubit phase errors and detuning
errors. Single-qubit phase errors originate from the composite uncertainties

\begin{align}
\beta_{\Delta_{A}} & =\beta_{\alpha_{A}}\dfrac{(\omega_{0,A}-\Delta_{A})}{(\omega_{0,A}-\alpha_{A})}+\beta_{\omega_{0,A}}\dfrac{(\Delta_{A}-\alpha_{A})}{(\omega_{0,A}-\alpha_{A})},\label{eq:uty_qbA_freq}\\
\beta_{\Delta_{B}} & =\beta_{\alpha_{B}}\dfrac{(\omega_{0,B}-\Delta_{B})}{(\omega_{0,B}-\alpha_{B})}+\beta_{\omega_{0,B}}\dfrac{(\Delta_{B}-\alpha_{B})}{(\omega_{0,B}-\alpha_{B})},\label{eq:uty_qbB_freq}
\end{align}
that accompany the parameters $\Delta_{A}$ and $\Delta_{B}$ appearing
in the Hamiltonians $\mathcal{H}_{I_{11}}$ and $\mathcal{H}_{I_{21}}$.
The uncertainty $\beta_{\Delta_{A}}$ is composed of the contributions
from the coefficients of $a_{A}^{\dagger}a_{A}$ in $\mathcal{H}_{\alpha_{A}}$
in Eq.\ (\ref{eq:uty_H_ahA}) and $\mathcal{H}_{\omega_{0,A}}$ in
Eq.\ (\ref{eq:uty_H_w0A}), and the uncertainty $\beta_{\Delta_{B}}$
is from the coefficients of $a_{B}^{\dagger}a_{B}$ in $\mathcal{H}_{\alpha_{B}}$
in Eq.\ (\ref{eq:uty_H_ahB}) and $\mathcal{H}_{\omega_{0,B}}$ in
Eq.\ (\ref{eq:uty_H_w0B}). Detuning errors are attributable to the
composite uncertainties

\begin{align}
\beta_{\delta_{\mathrm{01-10}}} & =\beta_{\Delta_{A}}-\beta_{\Delta_{B}},\label{eq:beta_01t10}\\
\beta_{\delta_{A}} & =\beta_{\Delta_{A}}-\beta_{\Delta_{B}}+\beta_{\alpha_{A}},\label{eq:beta_dtA}\\
\beta_{\delta_{B}} & =\beta_{\Delta_{B}}-\beta_{\Delta_{A}}+\beta_{\alpha_{B}},\label{eq:beta_dtB}
\end{align}
associated with the detunings $\delta_{\mathrm{01-10}}\equiv\Delta_{A}-\Delta_{B}$,
$\delta_{A}\equiv\Delta_{A}-\Delta_{B}+\alpha_{A}$, and $\delta_{B}\equiv\Delta_{B}-\Delta_{A}+\alpha_{B}$
appearing in the Hamiltonians $\mathcal{H}_{I_{12}}$ and $\mathcal{H}_{I_{22}}$.
In the CZ gate implementation described in Ref.\ \citep{Huang2026},
a single-qubit-Z gate, $U_{Z}$, is applied following the two-qubit
coupling process to compensate for the additional single-qubit phase
accumulation. Experimentally, $U_{Z}$ is calibrated using the actual
phases acquired during the coupling process, including contributions
from both the qubit frequencies $\Delta_{A/B}$ and their associated
uncertainties $\beta_{\Delta_{A/B}}$. As a result, the single-qubit
phase errors arising from $\beta_{\Delta_{A/B}}$ are largely canceled.
The residual errors associated with the diagonal uncertainties are
therefore dominated by the detuning errors.

To numerically analyze the single-qubit phase and detuning errors
induced by diagonal uncertainties, we simulate two types of realistic
CZ-gate infidelities, $\mathcal{I}_{1}$ and $\mathcal{I}_{2}$, using
the Hamiltonians $\mathcal{H}=\mathcal{H}_{I}+\mathcal{H}_{\omega_{0,A}}$
with $\beta_{\omega_{0,A}}=2\pi\times1\,\mathrm{MHz}$ {[}Fig.\ \ref{fig:phase_error_and_coupling_err}(a){]}
and $\mathcal{H}=\mathcal{H}_{I}+\mathcal{H}_{\alpha_{A}}$ with $\beta_{\alpha_{A}}=2\pi\times1\,\mathrm{MHz}$
{[}Fig.\ \ref{fig:phase_error_and_coupling_err}(b){]}. In the simulations,
CZ gates are constructed using the Qt-4 $J_{1}$ optimization described
in Sec.\ %
\mbox{%
II%
} of Ref.\ \citep{Huang2026} with gate time $t_{f}=12\,\mathrm{ns}$
and optimized $\omega_{i,A}/2\pi$ spanning approximately $1-5\,\mathrm{GHz}$.
The difference between $\mathcal{I}_{1}$ and $\mathcal{I}_{2}$ lies
in the calibration of $U_{Z}$. For $\mathcal{I}_{1}$, the calibration
accounts only for the ideal single-qubit phases determined by the
qubit frequencies $\Delta_{A/B}$. Consequently, the single-qubit
phase errors arising from the corresponding uncertainties $\beta_{\Delta_{A/B}}$
remain in $\mathcal{I}_{1}$. As a result, $\mathcal{I}_{1}$ contains
the intrinsic error $J_{1}$ and the contribution from diagonal uncertainties,
including the single-qubit phase and detuning errors. Accordingly,
the difference $\mathcal{I}_{1}-J_{1}$ is well described by the uncertainty-induced
errors $J_{2,\omega_{0,A}}$ and $J_{2,\alpha_{A}}$, associated with
$\beta_{\omega_{0,A}}$ and $\beta_{\alpha_{A}}$, respectively, as
shown in Figs.\ \ref{fig:phase_error_and_coupling_err}(a) and \ref{fig:phase_error_and_coupling_err}(b).
The quantities $J_{2,\omega_{0,A}}$ and $J_{2,\alpha_{A}}$ are obtained
from Eq.\ (\ref{eq:J2_uty_general_format}) by setting $\beta_{j}=\beta_{\omega_{0,A}}$
and $H_{j}=\mathcal{H}_{\omega_{0,A}}/(\hbar\beta_{\omega_{0,A}})$,
and $\beta_{j}=\beta_{\alpha_{A}}$ and $H_{j}=\mathcal{H}_{\alpha_{A}}/(\hbar\beta_{\alpha_{A}})$,
respectively. In contrast, for $\mathcal{I}_{2}$, following experimental
practice, $U_{Z}$ is calibrated using the actual single-qubit phases
associated with $\Delta_{A/B}$ and $\beta_{\Delta_{A/B}}$, thereby
eliminating the corresponding single-qubit phase errors. Consequently,
$\mathcal{I}_{2}$ is dominated by the intrinsic error $J_{1}$ and
the detuning error. As shown in Figs.\ \ref{fig:phase_error_and_coupling_err}(a)
and \ref{fig:phase_error_and_coupling_err}(b), the infidelity difference
$\mathcal{I}_{2}-J_{1}$ agrees well with the estimated detuning error

\begin{align}
J_{2,\delta_{A}}^{\mathrm{Est}}\approx & \dfrac{3}{16}\dfrac{\beta_{\delta_{A}}^{2}g_{AB}^{4}t_{f}^{2}}{\left(2g_{AB}^{2}+\delta_{A}^{2}/4\right)^{2}},\label{eq:J2_dA_est}
\end{align}
which is derived under the strong $\left|1_{B}1_{A}\right\rangle $$\leftrightarrow$$\left|0_{B}2_{A}\right\rangle $
coupling condition, as detailed in Appendix\ \ref{appendix:estimated_uncertainty_error}.
This verifies that the detuning error constitutes the dominant contribution
from diagonal uncertainties in a realistic CZ-gate implementation.

For the fixed $\beta_{\omega_{0,A}}$ case {[}Fig.\ \ref{fig:phase_error_and_coupling_err}(a){]},
both $\mathcal{I}_{1}-J_{1}$ and $\mathcal{I}_{2}-J_{1}$ increase
with $\omega_{i,A}$, whereas the opposite trend is observed for the
fixed $\beta_{\alpha_{A}}$ case {[}Fig.\ \ref{fig:phase_error_and_coupling_err}(b){]}.
In both cases, the single-qubit phase error in $\mathcal{I}_{1}-J_{1}$
is governed by the composite uncertainty $\beta_{\Delta_{A}}$ {[}Eq.\ (\ref{eq:uty_qbA_freq}){]},
while the detuning error in $\mathcal{I}_{1}-J_{1}$ and $\mathcal{I}_{2}-J_{1}$
are primarily determined by $\beta_{\delta_{A}}$ {[}Eq.\ (\ref{eq:beta_dtA}){]}.
For fixed $\beta_{\omega_{0,A}}$, $\beta_{\Delta_{A}}=\beta_{\delta_{A}}=\beta_{\omega_{0,A}}(\Delta_{A}-\alpha_{A})/(\omega_{0,A}-\alpha_{A})$
increases with $\omega_{i,A}$ through its dependence on $\Delta_{A}$,
leading to the observed rise in both error contributions. By contrast,
for fixed $\beta_{\alpha_{A}}$, both $\beta_{\Delta_{A}}=\beta_{\alpha_{A}}(\omega_{0,A}-\Delta_{A})/(\omega_{0,A}-\alpha_{A})$
and $\beta_{\delta_{A}}=\beta_{\alpha_{A}}+\beta_{\alpha_{A}}(\omega_{0,A}-\Delta_{A})/(\omega_{0,A}-\alpha_{A})$
decrease with $\omega_{i,A}$, resulting in the opposite trend. The
larger overall $\mathcal{I}_{2}-J_{1}$ in the fixed $\beta_{\alpha_{A}}$
case arises from the constant term $\beta_{\alpha_{A}}$ in $\beta_{\delta_{A}}$,
which uniformly offsets the $\beta_{\delta_{A}}$ dependence over
the range of $\omega_{i,A}$.

According to the estimated detuning error $J_{2,\delta_{A}}^{\mathrm{Est}}$
in Eq.\ (\ref{eq:J2_dA_est}), the maximum error, $\sim\frac{3}{64}\beta_{\delta_{A}}^{2}t_{f}^{2}$,
occurs under the resonant condition $\delta_{A}=0$ and decreases
with increasing $\left|\delta_{A}\right|$. However, a larger detuning
simultaneously weakens the effective two-qubit coupling strength.
Consequently, implementing a CZ gate requires a longer gate time $t_{f}$
to accumulate the target coupling phase $\phi_{AB}=\pi$, which in
turn increases $J_{2,\delta_{A}}^{\mathrm{Est}}$. This interplay
between reduced sensitivity to detuning uncertainty and diminished
coupling strength introduces a nontrivial trade-off. In Sec.\ \ref{subsec:uty_robust_against_detuning_uncertainty},
we examine this trade-off in detail and develop a control strategy
that suppresses detuning errors while enhancing robustness against
the diagonal uncertainties $\beta_{\alpha_{A}}$, $\beta_{\alpha_{B}}$,
$\beta_{\omega_{0,A}}$, and $\beta_{\omega_{0,B}}$.

The off-diagonal uncertainty $\beta_{g_{AB}}$ gives rise to a coupling
error, which can be estimated as

\begin{align}
J_{2,g_{AB}}^{\mathrm{Est}} & \approx\dfrac{2\beta_{g_{AB}}^{2}g_{AB}^{4}t_{f}^{2}}{\left(2g_{AB}^{2}+\delta_{A}^{2}/4\right)^{2}}\label{eq:J2_g_est}
\end{align}
in the strong $\left|1_{B}1_{A}\right\rangle $$\leftrightarrow$$\left|0_{B}2_{A}\right\rangle $
coupling regime, as derived in Appendix\ \ref{appendix:estimated_uncertainty_error}.
Similar to $J_{2,\delta_{A}}^{\mathrm{Est}}$ in Eq.\ (\ref{eq:J2_dA_est}),
$J_{2,g_{AB}}^{\mathrm{Est}}$ attains its maximum value, $\sim\frac{1}{2}\beta_{g_{AB}}^{2}t_{f}^{2}$,
at $\delta_{A}=0$ and decreases with increasing $\left|\delta_{A}\right|$.
However, its sensitivity to $\beta_{g_{AB}}$ at resonance is approximately
an order of magnitude larger than that of $J_{2,\delta_{A}}^{\mathrm{Est}}$,
namely $J_{2,g_{AB}}^{\mathrm{Est}}/\beta_{g_{AB}}^{2}\approx10\cdot J_{2,\delta_{A}}^{\mathrm{Est}}/\beta_{\delta_{A}}^{2}$.
In Sec.\ \ref{subsec:uty_robust_against_gAB_uncertainty}, we develop
an effective method to suppress the coupling error and enhance robustness
against the off-diagonal uncertainty $\beta_{g_{AB}}$.

\subsection{Robustness against off-diagonal uncertainties}

\label{subsec:uty_robust_against_gAB_uncertainty}

In this section, we focus on mitigating the coupling error to enhance
robustness against the off-diagonal uncertainty $\beta_{g_{AB}}$.
In Sec.\ \ref{subsubsec:coupling_error_suppression}, we analyze
the coupling error and develop an effective suppression method. Based
on this method, we construct robust CZ gates against $\beta_{g_{AB}}$
in Sec.\ \ref{subsubsec:robust_CZ_gates_against_beta_gAB}. Finally,
in Sec.\ \ref{subsubsec:robust_CZ_gates_against_beta_gAB_noises},
we extend the construction to two noise scenarios.

\subsubsection{Coupling error suppression}

\label{subsubsec:coupling_error_suppression}

\begin{figure}
\includegraphics[scale=0.52]{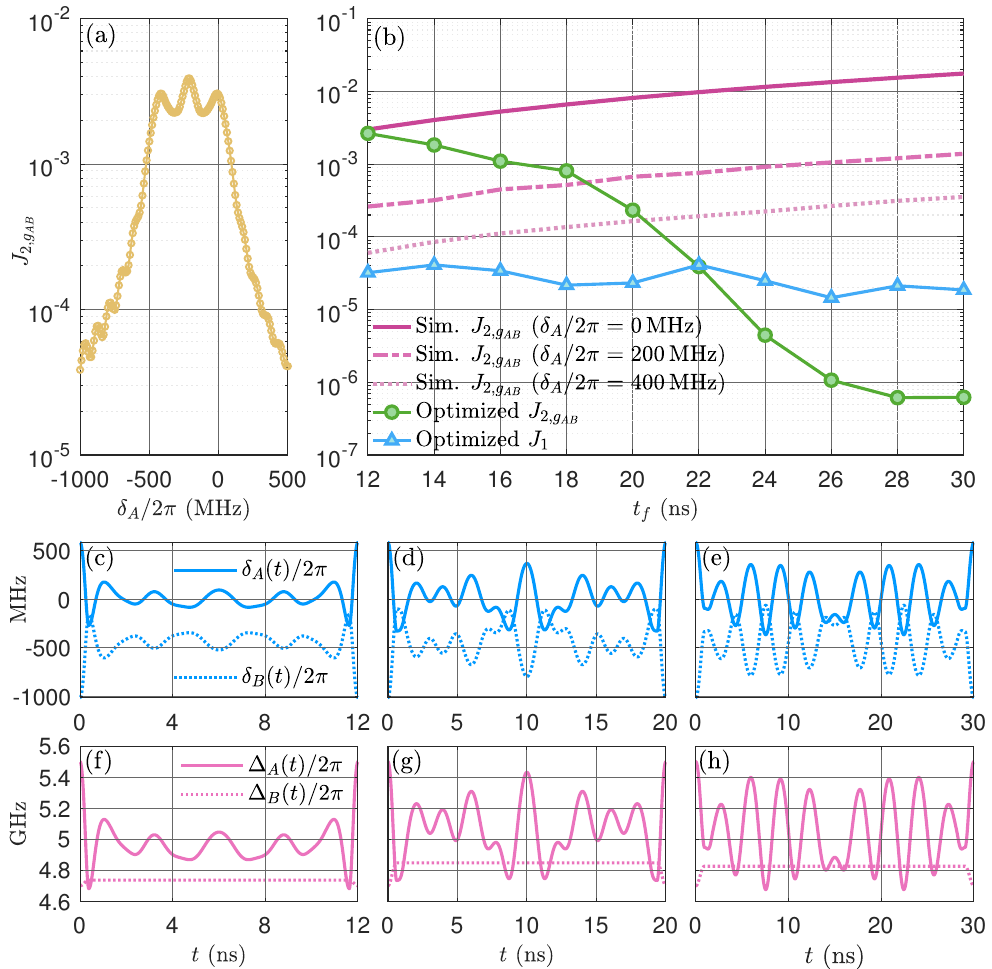}

\caption{Coupling error suppression. (a) Simulated $J_{2,g_{AB}}$ for $\beta_{g_{AB}}=2\pi\times1\,\mathrm{MHz}$
using the Pull-Back-Pulse $\Phi_{A/B}(t)=\Phi_{A/B}^{\mathrm{PBP}}(t)$
with a fixed gate time $t_{f}=12\,\mathrm{ns}$ and detuning $\delta_{A}/2\pi$
varied from $-1000$ to $500\,\mathrm{MHz}$. (b) Simulated $J_{2,g_{AB}}$
for $\beta_{g_{AB}}=2\pi\times1\,\mathrm{MHz}$ using $\Phi_{A/B}(t)=\Phi_{A/B}^{\mathrm{PBP}}(t)$
with fixed detunings $\delta_{A}/2\pi=0\,\mathrm{MHz}$ (purple solid),
$200\,\mathrm{MHz}$ (purple dash-dotted), and $400\,\mathrm{MHz}$
(purple dotted), while varying the gate time $t_{f}$ from $12$ to
$30\,\mathrm{ns}$. Also shown are the optimized $J_{2,g_{AB}}$ (green
circles, $\beta_{g_{AB}}=2\pi\times1\,\mathrm{MHz}$) and the optimized
$J_{1}$ (blue triangles) obtained from the optimization based on
the cost function $J_{1}+J_{2,g_{AB}}$. (c)--(e) Optimized detunings
$\delta_{A}(t)$ (blue solid) and $\delta_{B}(t)$ (blue dotted) for
gate times $t_{f}=12$, $20$, and $30\,\mathrm{ns}$, respectively.
(f)--(h) Corresponding optimized qubit frequencies $\Delta_{A}(t)$
(pink solid) and $\Delta_{B}(t)$ (pink dotted) for $t_{f}=12$, $20$,
and $30\,\mathrm{ns}$, respectively.}
\label{fig:uty_gAB_improve_pulse} 
\end{figure}

The estimated coupling error $J_{2,g_{AB}}^{\mathrm{Est}}$ in Eq.\ (\ref{eq:J2_g_est})
under the strong $\left|1_{B}1_{A}\right\rangle $$\leftrightarrow$$\left|0_{B}2_{A}\right\rangle $
coupling condition, where $\left|\delta_{A}\right|\ll\left|\delta_{B}\right|$
and $\left|\delta_{A}\right|\ll\left|\delta_{\mathrm{01-10}}\right|$,
indicates that the coupling error may be suppressed through appropriate
tuning of $\delta_{A}$. To investigate this possibility without imposing
the above constraints, we numerically evaluate the coupling error
$J_{2,g_{AB}}$ by sweeping $\delta_{A}/2\pi$ from $-1000\,\mathrm{MHz}$
to $500\,\mathrm{MHz}$ for the Pull-Back-Pulse $\Phi_{A/B}(t)=\Phi_{A/B}^{\mathrm{PBP}}(t)$
with a duration of $t_{f}=12\,\mathrm{ns}$. Here, $J_{2,g_{AB}}$
is evaluated using $J_{2,j}$ in Eq.\ (\ref{eq:J2_uty_general_format})
with $\beta_{j}=\beta_{g_{AB}}=2\pi\times1\,\mathrm{MHz}$ and $H_{j}=\mathcal{H}_{g_{AB}}/(\beta_{g_{AB}}\hbar)$.
The resulting $J_{2,g_{AB}}$, shown in Fig.\ \ref{fig:uty_gAB_improve_pulse}(a),
exhibits three pronounced peaks. The peak near $\delta_{A}=0$ originates
from the resonant $\left|1_{B}1_{A}\right\rangle $$\leftrightarrow$$\left|0_{B}2_{A}\right\rangle $
coupling and agrees well with the analytical estimate $J_{2,g_{AB}}^{\mathrm{Est}}$
in Eq.\ (\ref{eq:J2_g_est}). The two additional peaks, centered
at $\delta_{A}/2\pi\sim-200\,\mathrm{MHz}$ and $-400\,\mathrm{MHz}$,
originate from the resonances of the $\left|0_{B}1_{A}\right\rangle $$\leftrightarrow$$\left|1_{B}0_{A}\right\rangle $
coupling ($\delta_{\mathrm{01-10}}=0$) and the $\left|1_{B}1_{A}\right\rangle $$\leftrightarrow$$\left|2_{B}0_{A}\right\rangle $
coupling ($\delta_{B}=0$), respectively. In the region $-400\,\mathrm{MHz}$
$\lesssim\delta_{A}/2\pi\lesssim0\,\mathrm{MHz}$, where the system
is most sensitive to uncertainty $\beta_{g_{AB}}$, reducing $J_{2,g_{AB}}$
is therefore not advantageous.

In the regions $\delta_{A}/2\pi\gtrsim0\,\mathrm{MHz}$ and $\delta_{A}/2\pi\lesssim-400\,\mathrm{MHz}$,
$J_{2,g_{AB}}$ decreases monotonically with increasing $\left|\delta_{A}\right|$.
However, increasing $\left|\delta_{A}\right|$ weakens the coupling
strength and thereby increases the gate time $t_{f}$ required to
accumulate the target coupling phase $\phi_{AB}=\pi$ for a high-fidelity
CZ gate. The prolonged gate time, in turn, increases the coupling
error. To clarify this trade-off, we evaluate $J_{2,g_{AB}}$ by varying
$t_{f}$ for the Pull-Back-Pulse $\Phi_{A/B}(t)=\Phi_{A/B}^{\mathrm{PBP}}(t)$
with fixed $\delta_{A}/2\pi=0\,\mathrm{MHz}$, $200\,\mathrm{MHz}$,
and $400\,\mathrm{MHz}$, as shown in Fig.\ \ref{fig:uty_gAB_improve_pulse}(b).
For $\delta_{A}/2\pi=200\,\mathrm{MHz}$ and $400\,\mathrm{MHz}$,
the required gate times $t_{f}$ to achieve $\phi_{AB}=\pi$ increase
to $\sim56\,\mathrm{ns}$ and $\sim109\,\mathrm{ns}$, respectively,
as estimated using $t_{f}\sim\pi/[\sqrt{2g_{AB}^{2}+\frac{1}{4}\delta_{A}^{2}}-\frac{1}{2}\delta_{A}]$
in Sec.\ %
\mbox{%
II%
}\ A of Ref.\ \citep{Huang2026}. The corresponding coupling errors
$J_{2,g_{AB}}$ increase to $\sim4.5\times10^{-3}$ in both cases,
exceeding the peak value $J_{2,g_{AB}}\approx3.0\times10^{-3}$ at
$\delta_{A}/2\pi\approx0\,\mathrm{MHz}$ with $t_{f}=12\,\mathrm{ns}$.
This indicates that simply tuning $\delta_{A}$ provides limited improvement
in suppressing $J_{2,g_{AB}}$.

Next, we present an alternative strategy for mitigating the coupling
error. When the CZ gate operates near the strong $\left|1_{B}1_{A}\right\rangle $$\leftrightarrow$$\left|0_{B}2_{A}\right\rangle $
coupling condition ($\delta_{A}\approx0$), the dominant dynamics
of the system in the presence of a coupling-constant uncertainty $\beta_{g_{AB}}$
are described by the effective Hamiltonian $-\frac{1}{2}\delta_{A}(t)\sigma_{Z}+\sqrt{2}(g_{AB}+\beta_{g_{AB}})\sigma_{X}$,
expressed in the basis states $\left\{ \left|1_{B}1_{A}\right\rangle ,\left|0_{B}2_{A}\right\rangle \right\} $.
This Hamiltonian is derived from $\mathcal{H}_{I_{22}}$ in Eq.\ (7)
of Ref.\ \citep{Huang2026}, where $\sigma_{Z}$ and $\sigma_{X}$
denote the Pauli matrices. Notably, the controllable detuning $\delta_{A}(t)$
acts along the $\sigma_{Z}$ axis, whereas the coupling-constant uncertainty
$\beta_{g_{AB}}$ enters along the orthogonal $\sigma_{X}$ axis.
Consequently, modulating $\delta_{A}(t)$ about the resonance point
($\delta_{A}=0$) is expected to suppress the coupling error induced
by $\beta_{g_{AB}}$ through the mechanism of dynamical decoupling.
To validate this approach, we initialize the Pull-Back-Pulse $\Phi_{A/B}^{\mathrm{PBP}}(t)$
at zero detuning ($\delta_{A}=0$) by choosing the initial interaction
frequencies $\omega_{i,A/B}/2\pi=4.9/4.686\,\mathrm{GHz}$ and simultaneously
activate the Fine-Tune-Pulse $\Phi_{A}^{\mathrm{FTP}}(t)$ with $k_{\mathrm{max}}=10$
to induce oscillations of $\delta_{A}(t)$ about zero. Under these
conditions, CZ gates are optimized for symmetric transmons with $a_{E_{J}}=1.0$
using the rp-Qt-4 optimization procedure described in Sec.\ %
\mbox{%
II%
}\ B of Ref.\ \citep{Huang2026}, together with the cost function
$J_{1}+J_{2,g_{AB}}$. Figure\ \ref{fig:uty_gAB_improve_pulse}(b)
shows the optimized ideal gate infidelity $J_{1}$ and coupling error
$J_{2,g_{AB}}$ for $\beta_{g_{AB}}/2\pi=1\,\mathrm{MHz}$ as functions
of the gate duration $t_{f}$ over the range of $12\text{--}30\,\mathrm{ns}$.

The optimized coupling error $J_{2,g_{AB}}$ decreases with increasing
gate time $t_{f}$, whereas the optimized ideal gate infidelity $J_{1}$
remains within the range of $2\sim4\times10^{-5}$ over the same interval.
For $t_{f}=12\,\mathrm{ns}$, which is close to the CZ gate speed
limit of $t_{f}=\pi/(\sqrt{2}g_{AB})\cong11.6\,\mathrm{ns}$, the
optimized detuning $\delta_{A}(t)$ is constrained near the resonant
condition $\delta_{A}=0$ to maximize the coupling strength, as illustrated
in Fig.\ \ref{fig:uty_gAB_improve_pulse} (c). The corresponding
optimized qubit frequencies $\Delta_{A/B}(t)$ are shown in Fig.\ \ref{fig:uty_gAB_improve_pulse}(f).
Under these conditions, the optimized coupling error is $J_{2,g_{AB}}\approx2.65\times10^{-3}$,
only slightly smaller than the simulated value of $J_{2,g_{AB}}\approx3.0\times10^{-3}$
obtained without detuning modulation using the Pull-Back-Pulse alone,
$\Phi_{A/B}(t)=\Phi_{A/B}^{\mathrm{PBP}}(t)$ with $\delta_{A}/2\pi=0\,\mathrm{MHz}$
{[}Fig.\ \ref{fig:uty_gAB_improve_pulse}(b){]}. This result indicates
that the limited detuning modulation available near the speed limit
is insufficient to effectively suppress the coupling error. As $t_{f}$
increases, the required coupling strength decreases, allowing the
optimized detuning $\delta_{A}(t)$ to deviate farther from resonance
and oscillate more frequently. The resulting optimized detuning $\delta_{A}(t)$
for $t_{f}=20\,\mathrm{ns}$ and $30\,\mathrm{ns}$ are shown in Figs.\ \ref{fig:uty_gAB_improve_pulse}(d)
and \ref{fig:uty_gAB_improve_pulse}(e), respectively, with the corresponding
optimized qubit frequencies $\Delta_{A/B}(t)$ presented in Figs.\ \ref{fig:uty_gAB_improve_pulse}(g)
and \ref{fig:uty_gAB_improve_pulse}(h). The stronger detuning modulation
reduces the optimized coupling error at $t_{f}=30\,\mathrm{ns}$ to
$6.21\times10^{-7}$, more than two orders of magnitude smaller than
the simulated value of $J_{2,g_{AB}}\approx3.54\times10^{-4}$ obtained
using only the Pull-Back-Pulse, $\Phi_{A/B}(t)=\Phi_{A/B}^{\mathrm{PBP}}(t)$,
with $\delta_{A}/2\pi=400\,\mathrm{MHz}$ at the same gate duration.
These results demonstrate that strong modulation of $\delta_{A}(t)$
provides an effective strategy for mitigating the coupling error.

\subsubsection{Robust CZ gates against uncertainty $\beta_{g_{AB}}$}

\label{subsubsec:robust_CZ_gates_against_beta_gAB}

In this section, we introduce an additional procedure to identify
the most robust sample within an ensemble of optimized CZ gates with
suppressed $J_{2,g_{AB}}$, obtained via the $J_{1}+J_{2,g_{AB}}$
optimization described above. This additional selection step is not
required for constructing robust CZ gates against RTCE noise and flux
noise, as discussed in Sec.\ %
\mbox{%
III%
}\ C of Ref.\ \citep{Huang2026}. The error contributions from RTCE
noise and flux noise are effectively mitigated by operating the qubits
at their respective sweet spots, which minimize the noise--system
coupling terms $H_{j}(t)$, specifically $H_{\epsilon,A/B}(t)$ and
$H_{\phi,A/B}(t)$ in Eqs.\ (30) and (31) of Ref.\ \citep{Huang2026},
respectively. As a result, the leading second-order contributions
$\left\langle J_{2,\epsilon,A/B}\right\rangle $ and $\left\langle J_{2,\phi,A/B}\right\rangle $,
which are of order $\mathcal{O}(R_{j}^{2})$ with $R_{j}(t)=U_{I}^{\dagger}(t)H_{j}(t)U_{I}(t)$,
are strongly suppressed due to the reduced magnitudes of $H_{\epsilon,A/B}(t)$
and $H_{\phi,A/B}(t)$. Accordingly, higher-order contributions $\mathcal{O}(R_{j}^{n})$
with $n>2$ are expected to be negligible. Consequently, the most
robust sample is identified as the one minimizing $\left\langle J_{2,\epsilon,A/B}\right\rangle $
and/or $\left\langle J_{2,\phi,A/B}\right\rangle $.

\begin{figure}
\includegraphics[scale=0.52]{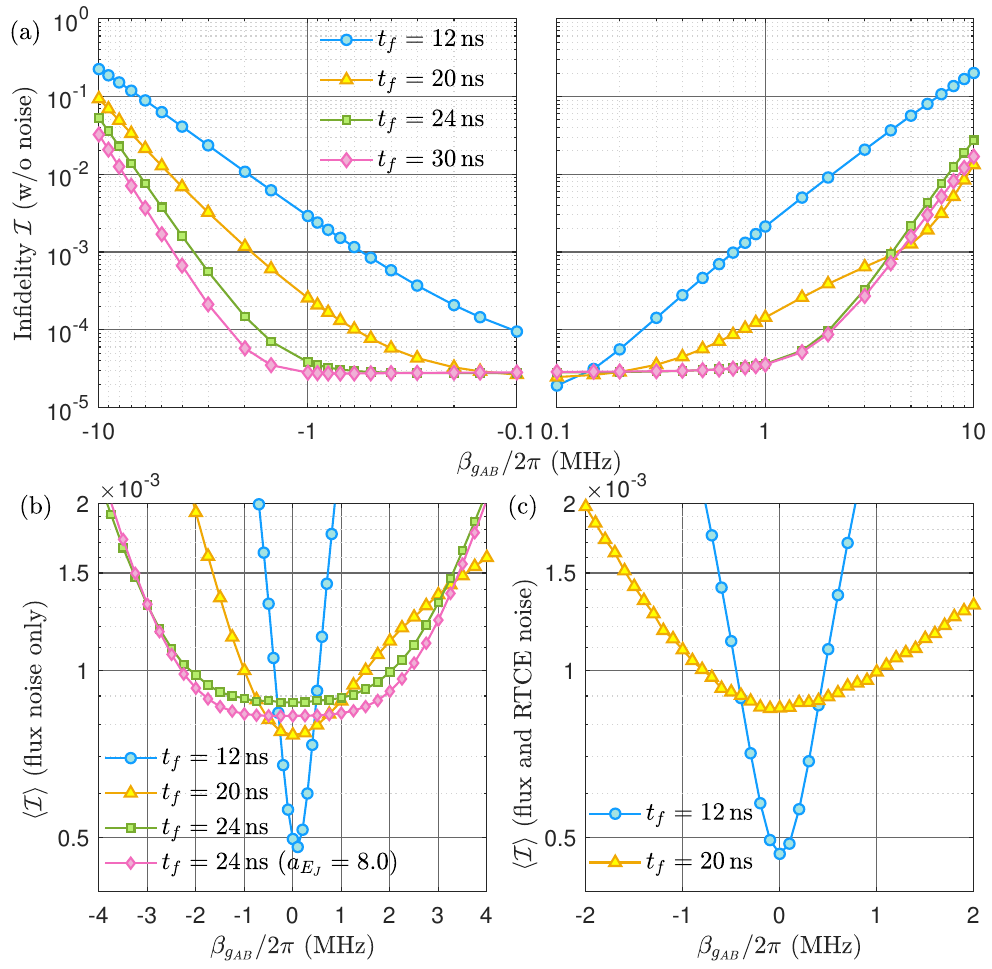}

\caption{Robustness against coupling-constant uncertainty $\beta_{g_{AB}}$.
(a) Gate infidelity $\mathcal{I}$, excluding noise contributions,
as a function of the coupling-constant uncertainty $\beta_{g_{AB}}/2\pi$,
over the ranges $-10$ to $-0.1\,\mathrm{MHz}$ (left) and $0.1$
to $10\,\mathrm{MHz}$ (right). Results are presented for optimized
CZ gates with gate times $t_{f}=12\,\mathrm{ns}$ (blue circles),
$20\,\mathrm{ns}$ (yellow triangles), $24\,\mathrm{ns}$ (green squares),
and $30\,\mathrm{ns}$ (pink diamonds), implemented on symmetric transmons
with $a_{E_{J}}=1.0$. (b) Ensemble-averaged gate infidelity $\left\langle \mathcal{I}\right\rangle $,
obtained from 1000 realizations of flux noise $\beta_{\phi,A/B}(t)$
with $T_{1}=30\,\mu\mathrm{s}$, as a function of $\beta_{g_{AB}}/2\pi$
over the interval $-4$ to $4\,\mathrm{MHz}$. Results are shown for
optimized CZ gates with $t_{f}=12\,\mathrm{ns}$ (blue circles), $20\,\mathrm{ns}$
(yellow triangles), and $24\,\mathrm{ns}$ (green squares) implemented
on symmetric transmons with $a_{E_{J}}=1.0$, together with the $t_{f}=24\,\mathrm{ns}$
gate (pink diamonds) implemented on asymmetric transmons with $a_{E_{J}}=8.0$.
(c) Ensemble-averaged gate infidelity $\left\langle \mathcal{I}\right\rangle $,
obtained from 1000 realizations of both flux noise $\beta_{\phi,A/B}(t)$
and RTCE noise $\beta_{\epsilon,A/B}(t)$ with $T_{1}=30\,\mu\mathrm{s}$,
as a function of the uncertainty $\beta_{g_{AB}}/2\pi$ over the interval
$-2$ to $2\,\mathrm{MHz}$. Results are presented for the optimized
CZ gate with $t_{f}=12\,\mathrm{ns}$ (blue circles) implemented on
asymmetric transmons with $a_{E_{J}}=1.31$, and for the optimized
CZ gate with $t_{f}=20\,\mathrm{ns}$ (yellow triangles) implemented
on asymmetric transmons with $a_{E_{J}}=1.5$.}

\label{fig:robustness_against_uty_gAB} 
\end{figure}

By contrast, for the coupling-constant uncertainty $\beta_{g_{AB}}$,
the uncertainty--system coupling term $H_{j}$ is the constant operator
$(-i)(a_{A}-a_{A}^{\dagger})(-i)(a_{B}-a_{B}^{\dagger})$. The second-order
contribution, $J_{2,g_{AB}}$, which scales as $\mathcal{O}(R_{g_{AB}}^{2})$
with $R_{g_{AB}}(t)=U_{I}^{\dagger}(t)(-i)(a_{A}-a_{A}^{\dagger})(-i)(a_{B}-a_{B}^{\dagger})U_{I}(t)$,
can be effectively suppressed by dynamical decoupling through strong
modulation of $\delta_{A}(t)$. Nevertheless, the suppression of $J_{2,g_{AB}}$
does not generally imply the suppression of contributions from other
orders, which may remain appreciable. To address this issue, we employ
the realistic Hamiltonian $\mathcal{H}=\mathcal{H}_{I}+\mathcal{H}_{g_{AB}}$
to evaluate the gate infidelity $\mathcal{I}$ at both positive and
negative values of $\beta_{g_{AB}}$ for each optimized CZ gate in
the ensemble with suppressed $J_{2,g_{AB}}$. Since the infidelity
$\mathcal{I}$ is computed using the full Hamiltonian, it incorporates
error contributions to all orders. The most robust gate is then identified
by jointly considering the infidelity at the two opposite values of
$\beta_{g_{AB}}$. Specifically, we evaluate $\mathcal{I}$ at $\beta_{g_{AB}}=\pm2\pi\times5\,\mathrm{MHz}$
to identify the most robust CZ gates for $t_{f}=12$, $20$, $24$,
and $30\,\mathrm{ns}$. Figure\ \ref{fig:robustness_against_uty_gAB}(a)
shows the resulting infidelity $\mathcal{I}$ as a function of $\beta_{g_{AB}}$
over the range rom $-2\pi\times10\,\mathrm{MHz}$ to $2\pi\times10\,\mathrm{MHz}$
for the selected gates.

As shown in Fig.\ \ref{fig:robustness_against_uty_gAB}(a), the robustness
window against the uncertainty $\beta_{g_{AB}}$ increases with increasing
$t_{f}$, consistent with the reduction of $J_{2,g_{AB}}$ with $t_{f}$
presented in Fig.\ \ref{fig:uty_gAB_improve_pulse}(b). For $t_{f}=12\,\mathrm{ns}$,
the optimized $J_{2,g_{AB}}\approx2.65\times10^{-3}$ at $\beta_{g_{AB}}/2\pi=1\,\mathrm{MHz}$
is still too large for higher-order error contributions to be observable
in the infidelity $\mathcal{I}$. Consequently, the infidelity $\mathcal{I}$
remains dominated by the second-order contribution $J_{2,g_{AB}}$,
which scales as $\beta_{g_{AB}}^{2}$ in the large $\left|\beta_{g_{AB}}\right|$
regime. By contrast, in the small $\left|\beta_{g_{AB}}\right|$ regime,
$\mathcal{I}$ exhibits a pronounced asymmetry: its value on the positive-$\beta_{g_{AB}}$
side is lower than that on the negative-$\beta_{g_{AB}}$ side. This
asymmetry originates from the first-order error contribution derived
in Ref.\ \citep{HuangGoan2017}. Such first-order terms are absent
in the case of stochastic noise $\beta_{j}(t)$ because the zero-mean
property of the noise, $\left\langle \beta_{j}(t)\right\rangle =0$,
eliminates them. Specifically, this first-order contribution is negative
for positive $\beta_{g_{AB}}$, where it partially cancels the $J_{2,g_{AB}}$
term and thereby suppresses $\mathcal{I}$. In contrast, it is positive
for negative $\beta_{g_{AB}}$, where it adds to contributions beyond
$J_{2,g_{AB}}$, resulting in an enhancement of $\mathcal{I}$. Accordingly,
under the error threshold of $10^{-3}$, the CZ gate is slightly more
robust to positive $\beta_{g_{AB}}$, tolerating values up to $2\pi\times0.7\,\mathrm{MHz}$,
than to negative $\beta_{g_{AB}}$, for which the robustness is limited
to $\left|\beta_{g_{AB}}\right|<2\pi\times0.6\,\mathrm{MHz}$.

For $t_{f}=20\,\mathrm{ns}$, the optimized $J_{2,g_{AB}}\approx2.35\times10^{-4}$
at $\beta_{g_{AB}}/2\pi=1\,\mathrm{MHz}$ is more than one order of
magnitude smaller than that for $t_{f}=12\,\mathrm{ns}$. As a result,
the emerging third-order contribution, which has opposite signs for
positive and negative $\beta_{g_{AB}}$, begins to influence the robustness
of the CZ gate. Under the error threshold of $10^{-3}$, the gate
tolerates positive $\beta_{g_{AB}}$ up to $2\pi\times4\,\mathrm{MHz}$,
whereas the magnitude of the allowable negative $\beta_{g_{AB}}$
is limited to below $2\pi\times2\,\mathrm{MHz}$. For $t_{f}\gtrsim24$
ns, the optimized $J_{2,g_{AB}}$ at $\beta_{g_{AB}}/2\pi=1\,\mathrm{MHz}$
is suppressed below the optimized $J_{1}\sim3\times10^{-5}$, as illustrated
in Fig.\ \ref{fig:uty_gAB_improve_pulse}(b). Consequently, the infidelity
$\mathcal{I}$ is dominated by $J_{1}$ and saturates at a constant
value for $\left|\beta_{g_{AB}}/2\pi\right|\lesssim1\,\mathrm{MHz}$.
For $\left|\beta_{g_{AB}}/2\pi\right|\gtrsim1\,\mathrm{MHz}$, $\mathcal{I}$
becomes nearly symmetric and is dominated by the symmetric fourth-order
contribution, with odd-order terms providing only a subleading asymmetric
contribution. For $t_{f}=30\,\mathrm{ns}$, the CZ gate tolerates
both positive and negative $\beta_{g_{AB}}$ with magnitudes up to
$2\pi\times4\,\mathrm{MHz}$ under the error threshold of $10^{-3}$,
representing nearly an order-of-magnitude improvement over the $t_{f}=12\,\mathrm{ns}$
case, where the tolerance is limited to approximately $2\pi\times0.5\,\mathrm{MHz}$,
as shown in Fig.\ \ref{fig:robustness_against_uty_gAB}(a).

\subsubsection{Robust CZ gates against uncertainty $\beta_{g_{AB}}$ in noisy environments}

\label{subsubsec:robust_CZ_gates_against_beta_gAB_noises}

In the preceding sections, we constructed robust CZ gates against
the off-diagonal uncertainty $\beta_{g_{AB}}$ by employing strong
modulation of the detuning $\delta_{A}(t)$ together with extended
gate times $t_{f}$. However, increasing $t_{f}$ inevitably leads
to the accumulation of decoherence-induced errors, $\left\langle J_{2,T_{1}}\right\rangle +\left\langle J_{2,\epsilon}\right\rangle +\left\langle J_{2,\phi}\right\rangle $,
introduced in Sec.\ %
\mbox{%
III%
}\ C of Ref.\ \citep{Huang2026}. The resulting increase in the overall
infidelity narrows the robustness window under a fixed error threshold.
For $t_{f}>30\,\mathrm{ns}$, the dominant decoherence contribution,
$\left\langle J_{2,T_{1}}\right\rangle =t_{f}/T_{1}$ with $T_{1}=30\,\mu\mathrm{s}$,
exceeds the error threshold of $10^{-3}$, thereby eliminating the
robustness window below this threshold. In the intermediate regime
$24\lesssim t_{f}\lesssim30\,\mathrm{ns}$, the robustness against
$\beta_{g_{AB}}$ remains nearly unchanged because the second-order
contribution $J_{2,g_{AB}}$ is strongly suppressed, leaving the fourth-order
error as the dominant contribution, as shown in Fig.\ \ref{fig:robustness_against_uty_gAB}(a).
Once decoherence is taken into account, however, the robustness deteriorates
with increasing $t_{f}$ owing to the continued accumulation of decoherence-induced
errors. Consequently, the optimal operating point in this regime is
expected to occur near $t_{f}\approx24\,\mathrm{ns}$. By contrast,
for $t_{f}<24\,\mathrm{ns}$, although the decoherence-induced error
continues to decrease, the coupling error $J_{2,g_{AB}}$ increases,
as shown in Fig.\ \ref{fig:uty_gAB_improve_pulse}(b). The robustness
is therefore determined by the trade-off between decoherence and coupling
errors. Accordingly, we focus on the regime $t_{f}\lesssim24\,\mathrm{ns}$
to identify the optimal control strategy for the two noise scenarios
considered.

In the first noise scenario, we consider energy relaxation and flux
noise, with the total gate infidelity given by $J_{1}+\left\langle J_{2,T_{1}}\right\rangle +\left\langle J_{2,\phi}\right\rangle +J_{2,g_{AB}}$.
Our strategy is to operate the CZ gate near the flux-noise sweet spot
to suppress the flux-noise contribution $\left\langle J_{2,\phi}\right\rangle $,
while simultaneously exploring longer gate times in the regime $t_{f}\lesssim24\,\mathrm{ns}$
to effectively reduce the coupling error $J_{2,g_{AB}}$. To this
end, we set the initial detuning to $\delta_{A}=0$ by choosing the
initial qubit frequencies $\omega_{i,A/B}/2\pi=5.2/4.986\,\mathrm{GHz}$,
which are close to the flux-noise sweet spots of symmetric transmons,
where $\Delta_{A/B}=\omega_{0,A/B}$. We then perform the rp-Qt-4
optimization using the cost function $J_{1}+J_{2,g_{AB}}$ for CZ
gates with $t_{f}=12$, $20$, and $24\,\mathrm{ns}$. The pulse settings
are $\Phi_{A}(t)=\Phi_{A}^{\mathrm{PBP}}(t)+\Phi_{A}^{\mathrm{FTP}}(t)$,
with $k_{\mathrm{max}}=10$ in $\Phi_{A}^{\mathrm{FTP}}(t)$, and
$\Phi_{B}(t)=\Phi_{B}^{\mathrm{PBP}}(t)$. Applying the selection
procedure introduced in Sec.\ \ref{subsubsec:robust_CZ_gates_against_beta_gAB},
we identify the most robust sample for each $t_{f}$ and present their
performance in Fig.\ \ref{fig:robustness_against_uty_gAB}(b). As
$t_{f}$ increases, the ensemble-averaged infidelity $\left\langle \mathcal{I}\right\rangle $
at $\beta_{g_{AB}}/2\pi=0\,\mathrm{MHz}$, which is dominated by $\left\langle J_{2,T_{1}}\right\rangle +\left\langle J_{2,\phi}\right\rangle $,
also increases. For $t_{f}=24\,\mathrm{ns}$, the accumulated infidelity
at $\beta_{g_{AB}}/2\pi=0\,\mathrm{MHz}$ reaches approximately $9\times10^{-4}$,
leaving an error budget of only $\sim10^{-4}$ for the coupling error
under the $10^{-3}$ threshold. Nevertheless, the optimized CZ gate
with $t_{f}=24\,\mathrm{ns}$ exhibits the greatest robustness due
to the strongest suppression of $J_{2,g_{AB}}$. Specifically, it
remains robust for both positive and negative $\beta_{g_{AB}}$ with
magnitudes up to $2\pi\times2\,\mathrm{MHz}$, representing more than
a fivefold improvement over the $t_{f}=12\,\mathrm{ns}$ gate, whose
robustness window is limited to less than $2\pi\times0.4\,\mathrm{MHz}$
under the same error threshold of $10^{-3}$.

Moreover, we construct robust CZ gates using asymmetric transmons
by varying the Josephson energy ratio $a_{E_{J}}$ and applying the
same optimization procedure at a working point near the second flux-noise
sweet spot, i.e., $\Delta_{A/B}=\Delta_{A/B,\mathrm{min}}$, with
$t_{f}=24\,\mathrm{ns}$. By contrast, the pulse settings here are
changed to $\Phi_{A}(t)=\Phi_{A}^{\mathrm{PBP}}(t)$ and $\Phi_{B}(t)=\Phi_{B}^{\mathrm{PBP}}(t)+\Phi_{B}^{\mathrm{FTP}}(t)$
with $k_{\mathrm{max}}=10$. As discussed in Sec.\ %
\mbox{%
III%
}\ C of Ref.\ \citep{Huang2026}, the flux-noise contribution $\left\langle J_{2,\phi}\right\rangle $
decreases with increasing $a_{E_{J}}$. However, the robustness against
the uncertainty $\beta_{g_{AB}}$ degrades rapidly for $a_{E_{J}}\gtrsim8.0$,
because the achievable amplitude of the detuning modulation $\delta_{A}(t)$
is limited by the reduced tunable qubit-frequency range of increasingly
asymmetric transmons. We identify the most robust solution for asymmetric
transmons at $a_{E_{J}}=8.0$, which exhibits robustness against $\beta_{g_{AB}}$
comparable to the symmetric-transmon case with $a_{E_{J}}=1.0$ at
$t_{f}=24\,\mathrm{ns}$, as shown in Fig.\ \ref{fig:robustness_against_uty_gAB}(b).
Notably, the optimized $\left\langle J_{2,\phi}\right\rangle \approx4.86\times10^{-6}$
for $a_{E_{J}}=8.0$ is reduced by more than an order of magnitude
compared with $\left\langle J_{2,\phi}\right\rangle \approx5.51\times10^{-5}$
for $a_{E_{J}}=1.0$. Consequently, the ensemble-averaged infidelity
$\left\langle \mathcal{I}\right\rangle $ at $\beta_{g_{AB}}/2\pi=0\,\mathrm{MHz}$
is correspondingly reduced, as shown in Fig.\ \ref{fig:robustness_against_uty_gAB}(b).
These results demonstrate that the optimized CZ gate with $a_{E_{J}}=8.0$
achieves robustness against both flux noise $\beta_{\phi_{A/B}}(t)$
and coupling-constant uncertainty $\beta_{g_{AB}}$.

In the second noise scenario, we incorporate the RTCE noises $\beta_{\epsilon,A/B}(t)$
into the first noise scenario, yielding a total gate infidelity of
$J_{1}+\left\langle J_{2,T_{1}}\right\rangle +\left\langle J_{2,\phi}\right\rangle +\left\langle J_{2,\epsilon}\right\rangle +J_{2,g_{AB}}$.
As discussed in Sec.\ %
\mbox{%
III%
}\ C of Ref.\ \citep{Huang2026}, the minimum infidelity, $J_{1}+\left\langle J_{2,T_{1}}\right\rangle +\left\langle J_{2,\phi}\right\rangle +\left\langle J_{2,\epsilon}\right\rangle \approx4.66\times10^{-4}$,
is achieved for the optimized CZ gate at $a_{E_{J}}=1.31$ and $t_{f}=12\,\mathrm{ns}$,
with the qubits operated near the second flux-noise sweet spots. Starting
from this operating point, our strategy is to increase the gate time
beyond $12\,\mathrm{ns}$ to enable stronger modulation of the detuning
$\delta_{A}(t)$, thereby suppressing the coupling error $J_{2,g_{AB}}$.
However, increasing $t_{f}$ also increases the accumulated decoherence-induced
error, $\left\langle J_{2,T_{1}}\right\rangle +\left\langle J_{2,\epsilon}\right\rangle +\left\langle J_{2,\phi}\right\rangle $.
In addition, the enhanced detuning modulation drives the flux pulse
$\Phi_{A/B}(t)/\Phi_{0}$ away from the second sweet spots, leading
to an increase in the flux-noise contribution $\left\langle J_{2,\phi}\right\rangle $.
This contribution can be mitigated by increasing $a_{E_{J}}$, which
lowers the peak values of $d\Delta_{A/B}/d(\Phi_{A/B}/\Phi_{0})$
and shifts the peaks away from the second sweet spots. However, larger
$a_{E_{J}}$ also enhances the RTCE-noise contribution $\left\langle J_{2,\epsilon}\right\rangle $
because of the reduced tunable qubit-frequency range, as illustrated
in Fig.\ 10 of Ref.\ \citep{Huang2026}. To balance these competing
error contributions, we vary both $a_{E_{J}}$ and $t_{f}$ and perform
the rp-Qt-4 optimization using the cost function $J_{1}+\left\langle J_{2,T_{1}}\right\rangle +\left\langle J_{2,\phi}\right\rangle +\left\langle J_{2,\epsilon}\right\rangle +J_{2,g_{AB}}$,
while maintaining operation near the second flux-noise sweet spots.
The most robust CZ gate is obtained for $a_{E_{J}}=1.5$ and $t_{f}=20\,\mathrm{ns}$,
tolerating both positive and negative $\beta_{g_{AB}}$ with magnitudes
up to approximately $2\pi\times0.8\,\mathrm{MHz}$ under the error
threshold of $10^{-3}$. This represents an approximately twofold
enhancement of the robustness window compared with the minimum-infidelity
case ($a_{E_{J}}=1.31$ and $t_{f}=12\,\mathrm{ns}$), which tolerates
uncertainties of only approximately $2\pi\times0.45\,\mathrm{MHz}$,
as shown in Fig.\ \ref{fig:robustness_against_uty_gAB}(c). For both
cases, $\Phi_{B}(t)=\Phi_{B}^{\mathrm{PBP}}(t)+\Phi_{B}^{\mathrm{FTP}}(t)$,
with $k_{\mathrm{max}}=10$ in $\Phi_{B}^{\mathrm{FTP}}(t)$, and
$\Phi_{A}(t)=\Phi_{A}^{\mathrm{PBP}}(t)$.

\subsection{Robustness against diagonal uncertainties}

\label{subsec:uty_robust_against_detuning_uncertainty}

In this section, we address the detuning errors to enhance the robustness
of the CZ gates against the diagonal uncertainties $\beta_{\alpha_{A}}$,
$\beta_{\alpha_{B}}$, $\beta_{\omega_{0,A}}$, and $\beta_{\omega_{0,B}}$.
Section\ \ref{subsubsec:detuning_error_suppression} analyzes the
detuning errors and develops an effective suppression method. Section\ \ref{subsubsec:robust_CZ_gates_against_dt_uty}
applies this method to construct robust CZ gates. Finally, Sec.\ \ref{subsubsec:robust_CZ_gates_against_dt_uty_noise}
incorporates noise-induced error contributions and realizes robust
CZ gates under two noise scenarios.

\subsubsection{Detuning error suppression}

\label{subsubsec:detuning_error_suppression}

The diagonal uncertainties $\beta_{\alpha_{A}}$, $\beta_{\alpha_{B}}$,
$\beta_{\omega_{0,A}}$, and $\beta_{\omega_{0,B}}$ enter the Hamiltonian
through the same channels as the control frequencies $\Delta_{A/B}(t)$,
whereas the off-diagonal uncertainty $\beta_{g_{AB}}$ is orthogonal
to these control fields. Consequently, the dynamical decoupling protocol
introduced in Sec.\ \ref{subsec:uty_robust_against_gAB_uncertainty},
which effectively suppresses the off-diagonal uncertainty $\beta_{g_{AB}}$,
is not applicable to diagonal uncertainties. An alternative strategy
is therefore required. As discussed in Sec.\ \ref{subsec:uncertainty_error_analysis},
diagonal uncertainties give rise to both single-qubit phase errors
and detuning errors. The former can be compensated by inserting the
experimentally calibrated single-qubit-Z gate $U_{Z}$. Accordingly,
the following analysis focuses on suppressing detuning errors. The
detuning uncertainties $\beta_{\delta_{01-10}}$, $\beta_{\delta_{A}}$,
and $\beta_{\delta_{B}}$ are linear combinations of the diagonal
uncertainties $\beta_{\alpha_{A}}$, $\beta_{\alpha_{B}}$, $\beta_{\omega_{0,A}}$,
and $\beta_{\omega_{0,B}}$, as defined in Eqs.\ (\ref{eq:beta_01t10}-\ref{eq:beta_dtB}),
and give rise to the corresponding detuning errors. Among these detuning
uncertainties, $\beta_{\delta_{01-10}}$ consists only of the frequency-dependent
contributions $\beta_{\Delta_{A}}$ and $\beta_{\Delta_{B}}$, which
vary with the qubit frequencies $\Delta_{A}$ and $\Delta_{B}$, respectively,
as shown in Eqs.\ (\ref{eq:uty_qbA_freq}) and (\ref{eq:uty_qbB_freq}).
In contrast, $\beta_{\delta_{A}}$ and $\beta_{\delta_{B}}$ contain
not only the frequency-dependent contributions $\beta_{\Delta_{A}}$
and $\beta_{\Delta_{B}}$ but also the constant contributions $\beta_{\alpha_{A}}$
and $\beta_{\alpha_{B}}$. Due to these constant contributions, $\beta_{\delta_{A}}$
and $\beta_{\delta_{B}}$ generally produce substantially larger detuning
errors than $\beta_{\delta_{01-10}}$. We therefore focus on suppressing
the dominant detuning errors arising from the constant contributions
to $\beta_{\delta_{A}}$ and $\beta_{\delta_{B}}$. Accordingly, in
the following analysis, $\beta_{\delta_{A}}$ and $\beta_{\delta_{B}}$
are treated as independent constants, thereby facilitating the development
of an effective suppression strategy.

\begin{figure}
\includegraphics[scale=0.52]{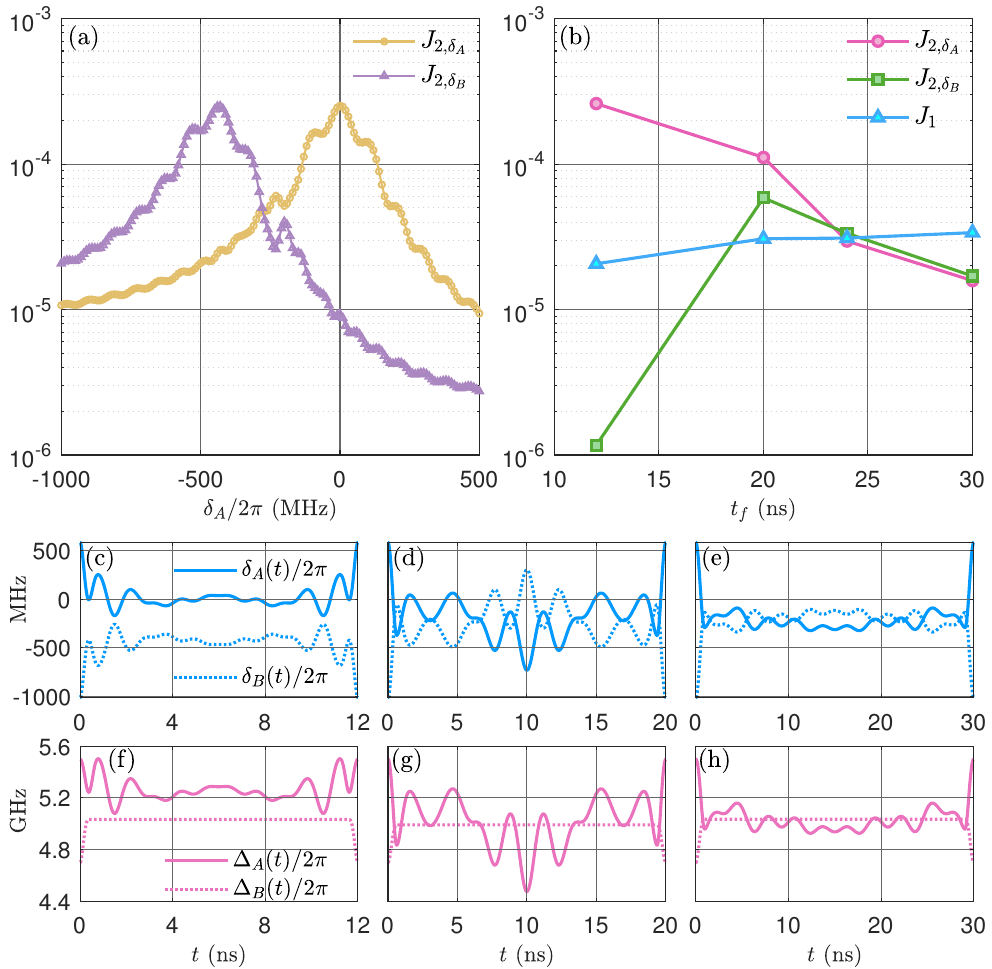}

\caption{Detuning error suppression. (a) Detuning errors $J_{2,\delta_{A}}$
with $\beta_{\delta_{A}}/2\pi=1\,\mathrm{MHz}$ (yellow-circle line)
and $J_{2,\delta_{B}}$ with $\beta_{\delta_{B}}/2\pi=1\,\mathrm{MHz}$
(purple-triangle line) for the Pull-Back-Pulse $\Phi_{A/B}(t)=\Phi_{A/B}^{\mathrm{PBP}}(t)$
with $t_{f}=12\,\mathrm{ns}$, plotted as functions of the detuning
$\delta_{A}$ varied from $-2\pi\times1000\,\mathrm{MHz}$ to $2\pi\times500\,\mathrm{MHz}$.
(b) Optimized detuning errors $J_{2,\delta_{A}}$ (pink circles) and
$J_{2,\delta_{B}}$ (green squares), together with the ideal gate
infidelity $J_{1}$ (blue triangles), for CZ gates with $t_{f}=12$,
$20$, $24$, and $30\,\mathrm{ns}$, obtained by minimizing the cost
function $J_{1}+J_{2,\delta_{A}}+J_{2,\delta_{B}}$. The initial detunings
and interaction frequencies are $\delta_{A/B}/2\pi=-214/-212\,\mathrm{MHz}$
and $\omega_{i,A/B}/2\pi=5.0/5.0\,\mathrm{GHz}$, respectively. (c)--(e)
Optimized detunings $\delta_{A}(t)$ (blue-solid line) and $\delta_{B}(t)$
(blue-dotted line) for $t_{f}=12$, $20$, and $30\,\mathrm{ns}$,
respectively. (f)--(h) Corresponding optimized frequencies $\Delta_{A}(t)$
(pink-solid line) and $\Delta_{B}(t)$ (pink-dotted line) for the
same gate durations as in panels (c)--(e).}

\label{fig:detuning_error_suppression} 
\end{figure}

The detuning errors $J_{2,\delta_{A}}$ and $J_{2,\delta_{B}}$, associated
with the detuning uncertainties $\beta_{\delta_{A}}$ and $\beta_{\delta_{B}}$,
respectively, are evaluated in the dominant basis states by substituting
$\beta_{j}=\beta_{\delta_{A/B}}$ and $H_{j}=\mathcal{H}_{\delta_{A/B}}/(\hbar\beta_{\delta_{A/B}})$
into Eq.\ (\ref{eq:J2_uty_general_format}) for $J_{2,j}$, where

\begin{align}
\mathcal{H}_{\delta_{A}}/\hbar & =\beta_{\delta_{A}}\left[\dfrac{1}{2}a_{A}^{\dagger}a_{A}(a_{A}^{\dagger}a_{A}-1)\right],\label{eq:uty_H_dtA}\\
\mathcal{H}_{\delta_{B}}/\hbar & =\beta_{\delta_{B}}\left[\dfrac{1}{2}a_{B}^{\dagger}a_{B}(a_{B}^{\dagger}a_{B}-1)\right].\label{eq:uty_H_dtB}
\end{align}
Figure\ \ref{fig:detuning_error_suppression}(a) presents the calculated
values of $J_{2,\delta_{A}}$ and $J_{2,\delta_{B}}$ for the Pull-Back-Pulse
$\Phi_{A/B}(t)=\Phi_{A/B}^{\mathrm{PBP}}(t)$, as $\delta_{A}/2\pi$
is varied from $-1000\,\mathrm{MHz}$ to $500\,\mathrm{MHz}$ with
$t_{f}=12\,\mathrm{ns}$ and $\beta_{\delta_{A}}/2\pi=\beta_{\delta_{B}}/2\pi=1\,\mathrm{MHz}$.
Pronounced peaks in $J_{2,\delta_{A}}$ and $J_{2,\delta_{B}}$ occur
at $\delta_{A}/2\pi=0$ and $-430\,\mathrm{MHz}$, respectively, corresponding
to the resonance conditions of the $\left|1_{B}1_{A}\right\rangle \leftrightarrow\left|0_{B}2_{A}\right\rangle $
and $\left|1_{B}1_{A}\right\rangle $$\leftrightarrow$$\left|2_{B}0_{A}\right\rangle $
transitions. Away from these resonance peaks ($\delta_{A}/2\pi>0\,\mathrm{MHz}$
and $\delta_{A}/2\pi<-430\,\mathrm{MHz}$), both $J_{2,\delta_{A}}$
and $J_{2,\delta_{B}}$ decrease monotonically with increasing $\left|\delta_{A}\right|$,
suggesting that larger values of $\left|\delta_{A}\right|$ are favorable
for suppressing the detuning errors. This apparent improvement, however,
is offset by the concomitant increase in the gate duration. As discussed
in Sec.\ %
\mbox{%
II%
}\ A of Ref.\ \citep{Huang2026}, the gate time required to accumulate
the target coupling phase $\phi_{AB}=\pi$ for implementing a CZ gate
scales as $t_{f}\sim(\pi/2)\cdot(\left|\delta_{A}\right|/g_{AB}^{2})$
in the large-$\left|\delta_{A}\right|$ regime. Consequently, the
error induced by $\beta_{\delta_{A}}$ accumulates over the prolonged
evolution, giving an estimated contribution of $(\pi^{2}/8)\cdot(\beta_{\delta_{A}}/g_{AB})^{2}\sim1.33\times10^{-3}$,
as obtained from Eqs.\ (\ref{eq:J2_detuning_00}-\ref{eq:J2_detuning_01_I}).
This accumulated error is nearly an order of magnitude larger than
the peak value of $J_{2,\delta_{A}}\approx2.52\times10^{-4}$ at $\delta_{A}=0$.
Taken together, these results indicate that operating outside the
resonance peaks offers little overall advantage for suppressing the
detuning errors.

In the regime between the resonance peaks, i.e., $-430<\delta_{A}/2\pi<0\,\mathrm{MHz}$,
$J_{2,\delta_{A}}$ increases whereas $J_{2,\delta_{B}}$ decreases
as $\delta_{A}$ decreases, leading to a local minimum of the total
detuning error $J_{2,\delta_{A}}+J_{2,\delta_{B}}$ at $\delta_{A}/2\pi\sim-200\,\mathrm{MHz}$,
as shown in Fig.\ \ref{fig:detuning_error_suppression}(a). We refer
to this operating point as the symmetric detuning point, characterized
by the approximate condition $\delta_{A}\approx\delta_{B}$. This
point is a promising candidate for a sweet spot because it not only
minimizes $J_{2,\delta_{A}}+J_{2,\delta_{B}}$ but also provides a
stronger effective coupling than the point at $\delta_{A}/2\pi\sim+200\,\mathrm{MHz}$
with the same $\left|\delta_{A}\right|$. At the symmetric detuning
point, the effective coupling receives comparable contributions from
both the $\left|1_{B}1_{A}\right\rangle \leftrightarrow\left|0_{B}2_{A}\right\rangle $
and $\left|1_{B}1_{A}\right\rangle \leftrightarrow\left|2_{B}0_{A}\right\rangle $
interactions, since $\left|\delta_{B}\right|\approx\left|\delta_{A}\right|$.
By contrast, at $\delta_{A}/2\pi\sim+200\,\mathrm{MHz}$, the latter
contribution is substantially weaker because $\left|\delta_{B}\right|\approx3\left|\delta_{A}\right|$.
The resulting enhancement of the effective coupling reduces the gate
time $t_{f}$ required to accumulate the target coupling phase, thereby
mitigating the accumulation of detuning errors and further supporting
the symmetric detuning point as a sweet spot for detuning-error suppression.

To verify the effectiveness of the symmetric detuning point, we construct
optimized CZ gates with $t_{f}=12$, $20$, $24$, and $30\,\mathrm{ns}$,
using the initial detuning $\delta_{A}/2\pi=-214\,\mathrm{MHz}$ near
the symmetric detuning point and the rp-Qt-4 optimization with the
cost function $J_{1}+J_{2,\delta_{A}}+J_{2,\delta_{B}}$. The pulse
settings are $\Phi_{A}(t)=\Phi_{A}^{\mathrm{PBP}}(t)+\Phi_{A}^{\mathrm{FTP}}(t)$,
with $k_{\mathrm{max}}=10$ in $\Phi_{A}^{\mathrm{FTP}}(t)$, and
$\Phi_{B}(t)=\Phi_{B}^{\mathrm{PBP}}(t)$. The initial interaction
frequencies are chosen as $\omega_{i,A/B}/2\pi=5.0/5.0\,\mathrm{GHz}$,
close to $\omega_{0,A/B}/2\pi=5.78/5.06\,\mathrm{GHz}$, to balance
the detuning-error contributions arising from the diagonal uncertainties
$\beta_{\alpha_{A}}$, $\beta_{\alpha_{B}}$, $\beta_{\omega_{0,A}}$,
and $\beta_{\omega_{0,B}}$. Assuming comparable strengths of the
diagonal uncertainties, the corresponding contributions depend on
the uncertainty-strength ratios $\beta_{\delta_{A/B}}/\beta_{\alpha_{A/B}}=1+(\omega_{0,A/B}-\Delta_{A/B})/(\omega_{0,A/B}-\alpha_{A/B})$
and $\beta_{\delta_{A/B}}/\beta_{\omega_{0,A/B}}=(\Delta_{A/B}-\alpha_{A/B})/(\omega_{0,A/B}-\alpha_{A/B})$,
as derived from Eqs.\ (\ref{eq:uty_qbA_freq}-\ref{eq:beta_dtB}).
As $\Delta_{A/B}$ approaches $\omega_{0,A/B}$, both ratios approach
unity, resulting in balanced detuning-error contributions from the
different diagonal uncertainties. By contrast, as $\Delta_{A/B}$
moves increasingly below $\omega_{0,A/B}$, the ratio $\beta_{\delta_{A/B}}/\beta_{\alpha_{A/B}}$
approaches $2$, whereas $\beta_{\delta_{A/B}}/\beta_{\omega_{0,A/B}}$
approaches zero, leading to an imbalance among the detuning-error
contributions. The combined effect of simultaneously considering all
diagonal uncertainties will be discussed later; however, such an imbalance
is expected to lead to a larger overall detuning error compared with
the balanced case. This consideration motivates our choice of the
initial interaction frequencies $\omega_{i,A/B}/2\pi=5.0/5.0\,\mathrm{GHz}$.

Figure\ \ref{fig:detuning_error_suppression}(b) presents the optimized
ideal gate infidelity $J_{1}$ and detuning errors $J_{2,\delta_{A}}$
and $J_{2,\delta_{B}}$ for CZ gates with $t_{f}=12$, $20$, $24$,
and $30\,\mathrm{ns}$. For $t_{f}=12\,\mathrm{ns}$, which is close
to the CZ gate speed limit, the optimized $\delta_{A}(t)$ departs
from the initial setting $\delta_{A}/2\pi=-214\,\mathrm{MHz}$ near
the symmetric detuning point and oscillates around $\delta_{A}/2\pi\sim0\,\mathrm{MHz}$
to maximize the $\left|1_{B}1_{A}\right\rangle \leftrightarrow\left|0_{B}2_{A}\right\rangle $
coupling, as shown in Fig.\ \ref{fig:detuning_error_suppression}(c).
Consequently, the optimized $J_{2,\delta_{A}}\approx2.61\times10^{-4}$
in Fig.\ \ref{fig:detuning_error_suppression}(b) is comparable to
the peak value of $J_{2,\delta_{A}}\approx2.52\times10^{-4}$ at $\delta_{A}/2\pi=0\,\mathrm{MHz}$
in Fig.\ \ref{fig:detuning_error_suppression}(a), obtained using
the Pull-Back-Pulse $\Phi_{A/B}(t)=\Phi_{A/B}^{\mathrm{PBP}}(t)$
with $t_{f}=12\,\mathrm{ns}$. Although the optimized $J_{2,\delta_{B}}\approx1.17\times10^{-6}$
is further reduced compared with $J_{2,\delta_{B}}\approx9.29\times10^{-6}$
at $\delta_{A}/2\pi=0\,\mathrm{MHz}$ in Fig.\ \ref{fig:detuning_error_suppression}(a),
the total optimized detuning error $J_{2,\delta_{A}}+J_{2,\delta_{B}}$
for $t_{f}=12\,\mathrm{ns}$ remains essentially unchanged relative
to that obtained with the Pull-Back-Pulse.

As $t_{f}$ increases from $12\,\mathrm{ns}$ to $20\,\mathrm{ns}$,
the optimized $J_{2,\delta_{A}}$ decreases to $1.11\times10^{-4}$,
whereas the optimized $J_{2,\delta_{B}}$ increases to $5.86\times10^{-5}$,
as shown in Fig.\ \ref{fig:detuning_error_suppression}(b). These
changes correspond to an increase in $\left|\delta_{A}(t)\right|$
and a decrease in $\left|\delta_{B}(t)\right|$, as shown in Figs.\ \ref{fig:detuning_error_suppression}(c)
and \ref{fig:detuning_error_suppression}(d). The optimized $\delta_{A}(t)$
and $\delta_{B}(t)$ for $t_{f}=20\,\mathrm{ns}$ can oscillate around
the initial setting near the symmetric detuning point, $\delta_{A}/2\pi\approx\delta_{B}/2\pi\sim-200\,\mathrm{MHz}$.
However, the coupling strength at the symmetric detuning point remains
insufficient for $t_{f}=20\,\mathrm{ns}$, causing large-amplitude
oscillations of $\delta_{A/B}(t)$ between the two resonance points,
$\delta_{A}/2\pi\sim0\,\mathrm{MHz}$ ($\delta_{B}/2\pi\sim-430\,\mathrm{MHz}$)
and $\delta_{A}/2\pi\sim-430\,\mathrm{MHz}$ ($\delta_{B}/2\pi\sim0\,\mathrm{MHz}$),
to enhance the coupling through both the $\left|1_{B}1_{A}\right\rangle \leftrightarrow\left|0_{B}2_{A}\right\rangle $
and $\left|1_{B}1_{A}\right\rangle $$\leftrightarrow$$\left|2_{B}0_{A}\right\rangle $
interactions, as shown in Fig.\ \ref{fig:detuning_error_suppression}(d).
As a result, the optimized $\delta_{A}(t)$ and $\delta_{B}(t)$ cannot
remain close to the symmetric detuning point throughout the entire
gate duration to minimize the detuning error, leading to only marginal
improvement in the optimized $J_{2,\delta_{A}}+J_{2,\delta_{B}}$
for $t_{f}=20\,\mathrm{ns}$.

For $t_{f}\geq24\,\mathrm{ns}$, the optimized $J_{2,\delta_{A}}$
and $J_{2,\delta_{B}}$ in Fig.\ \ref{fig:detuning_error_suppression}(b)
decrease with increasing $t_{f}$ and converge toward one another,
resembling the nearly identical $J_{2,\delta_{A}}$ and $J_{2,\delta_{B}}$
obtained for the Pull-Back-Pulse $\Phi_{A/B}(t)=\Phi_{A/B}^{\mathrm{PBP}}(t)$
at the symmetric detuning point $\delta_{A/B}/2\pi\sim-200\,\mathrm{MHz}$,
as shown in Fig.\ \ref{fig:detuning_error_suppression}(a). This
behavior indicates that, for $t_{f}\geq24\,\mathrm{ns}$, the coupling
strength at the symmetric detuning point is sufficient for the optimized
detunings $\delta_{A}(t)$ and $\delta_{B}(t)$ to undergo small-amplitude
oscillations around this point, as exemplified by the $t_{f}=30\,\mathrm{ns}$
case in Fig.\ \ref{fig:detuning_error_suppression}(e). Consequently,
the detuning errors in this regime are effectively suppressed. For
$t_{f}=30\,\mathrm{ns}$, the optimized $J_{2,\delta_{A}}+J_{2,\delta_{B}}$
is reduced to $3.28\times10^{-5}$, with contributions $J_{2,\delta_{A}}\approx1.58\times10^{-5}$
and $J_{2,\delta_{B}}\approx1.70\times10^{-5}$. This corresponds
to an approximately one-order-of-magnitude reduction compared with
the optimized $J_{2,\delta_{A}}+J_{2,\delta_{B}}\approx2.62\times10^{-4}$
for $t_{f}=12\,\mathrm{ns}$. These results demonstrate that the symmetric
detuning point $\delta_{A}\cong\delta_{B}$ serves as a sweet spot
for suppressing detuning errors when the gate time is sufficiently
long to maintain the optimized detuning $\delta_{A/B}(t)$ close to
this point. By contrast, the optimized $J_{1}$ slightly increases
from $2.07\times10^{-5}$ to $3.38\times10^{-5}$ as $t_{f}$ is extended
from $12\,\mathrm{ns}$ to $30\,\mathrm{ns}$, as shown in Fig.\ \ref{fig:detuning_error_suppression}(b).
This trend suggests that synchronizing the intrinsic errors to simultaneously
reach their minima at the symmetric detuning point becomes increasingly
challenging.

\subsubsection{Robust CZ gates against uncertainties $\beta_{\alpha_{A}}$, $\beta_{\alpha_{B}}$,
$\beta_{\omega_{0,A}}$, and $\beta_{\omega_{0,B}}$}

\label{subsubsec:robust_CZ_gates_against_dt_uty}

\begin{figure}
\includegraphics[scale=0.52]{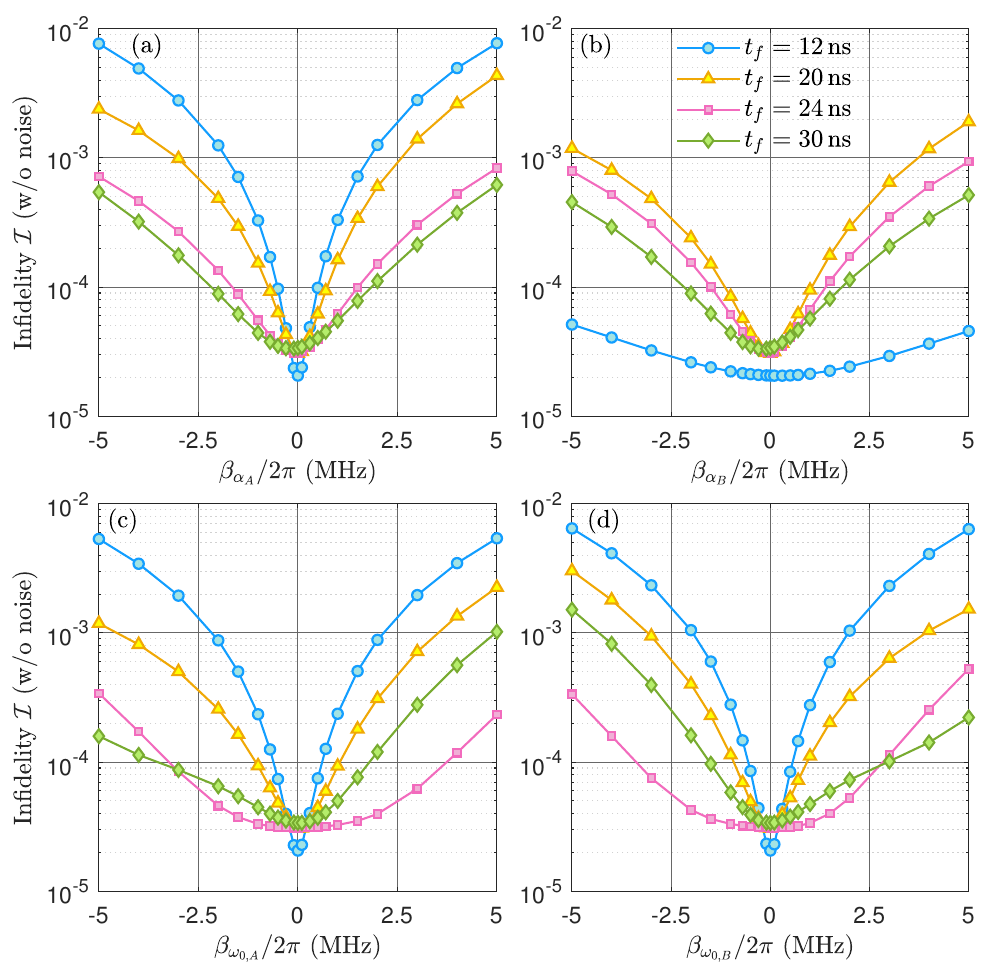}

\caption{Robustness of the optimized CZ gates against uncertainties in (a)
$\beta_{\alpha_{A}}$, (b) $\beta_{\alpha_{B}}$, (c) $\beta_{\omega_{0,A}}$,
and (d) $\beta_{\omega_{0,B}}$. For each gate time $t_{f}$, the
corresponding optimized CZ gate is evaluated under the four individual
uncertainty sources, with $t_{f}=12\,\mathrm{ns}$ (blue circles),
$20\,\mathrm{ns}$ (yellow triangles), $24\,\mathrm{ns}$ (pink squares),
and $30\,\mathrm{ns}$ (green diamonds). The infidelities $\mathcal{I}$
are obtained using the realistic Hamiltonians $\mathcal{H}_{I}+\mathcal{H}_{\alpha_{A}}$,
$\mathcal{H}_{I}+\mathcal{H}_{\alpha_{B}}$, $\mathcal{H}_{I}+\mathcal{H}_{\omega_{0,A}}$,
and $\mathcal{H}_{I}+\mathcal{H}_{\omega_{0,B}}$ for panels (a)--(d),
respectively.}

\label{fig:robustness_against_uty_dt} 
\end{figure}

In this section, we apply the additional selection procedure described
in Sec.\ \ref{subsubsec:robust_CZ_gates_against_beta_gAB} to identify
the most robust sample from an ensemble of optimized CZ gates with
suppressed detuning error $J_{2,\delta_{A}}+J_{2,\delta_{B}}$, obtained
through the $J_{1}+J_{2,\delta_{A}}+J_{2,\delta_{B}}$ optimization
in Sec.\ \ref{subsubsec:detuning_error_suppression}. Specifically,
the infidelities of each diagonal uncertainty evaluated at $\pm2\pi\times5\,\mathrm{MHz}$
are jointly considered. The selected samples with gate times $t_{f}=12$,
$20$, $24$, and $30\,\mathrm{ns}$ are then used to evaluate the
robustness against the uncertainties $\beta_{\alpha_{A}}$, $\beta_{\alpha_{B}}$,
$\beta_{\omega_{0,A}}$, and $\beta_{\omega_{0,B}}$ individually,
as shown in Figs.\ \ref{fig:robustness_against_uty_dt}(a)-\ref{fig:robustness_against_uty_dt}(d).
For each uncertainty source, the realistic infidelities $\mathcal{I}$
are calculated by simulating the corresponding Hamiltonian, $\mathcal{H}_{I}+\mathcal{H}_{\alpha_{A}}$,
$\mathcal{H}_{I}+\mathcal{H}_{\alpha_{B}}$, $\mathcal{H}_{I}+\mathcal{H}_{\omega_{0,A}}$,
or $\mathcal{H}_{I}+\mathcal{H}_{\omega_{0,B}}$, with the uncertainty
strength varied from $-2\pi\times5\,\mathrm{MHz}$ to $2\pi\times5\,\mathrm{MHz}$.

For the uncertainty $\beta_{\alpha_{A}}$, the detuning uncertainties
$\beta_{\mathrm{01-10}}=\beta_{\Delta_{A}}$, $\beta_{\delta_{A}}=\beta_{\Delta_{A}}+\beta_{\alpha_{A}}$,
and $\beta_{\delta_{B}}=-\beta_{\Delta_{A}}$ are given by Eqs.\ (\ref{eq:beta_01t10})-(\ref{eq:beta_dtB}).
The magnitude of $\beta_{\Delta_{A}}$ is estimated by substituting
$\Delta_{A}=2\pi\times5\,\mathrm{GHz}$ into $\beta_{\Delta_{A}}=\beta_{\alpha_{A}}(\omega_{0,A}-\Delta_{A})/(\omega_{0,A}-\alpha_{A})$
from Eq.\ (\ref{eq:uty_qbA_freq}), where $\Delta_{A}$ follows from
the initial condition $\omega_{i,A/B}=2\pi\times5\,\mathrm{GHz}$
used in the optimization procedure described in Sec.\ \ref{subsubsec:detuning_error_suppression}.
This yields $\left|\beta_{\Delta_{A}}\right|\sim0.13\left|\beta_{\alpha_{A}}\right|$.
Consequently, $\left|\beta_{\delta_{A}}\right|\sim\left|\beta_{\alpha_{A}}\right|$
is approximately one order of magnitude larger than $\left|\beta_{\delta_{B}}\right|=\left|\beta_{\mathrm{01-10}}\right|\sim0.13\left|\beta_{\alpha_{A}}\right|$.
This indicates that the optimized $J_{2,\delta_{A}}$ associated with
$\beta_{\delta_{A}}$ provides the dominant contribution to the total
detuning error. As a result, the enhanced robustness window against
the uncertainty $\beta_{\alpha_{A}}$ with increasing $t_{f}$ in
Fig.\ \ref{fig:robustness_against_uty_dt}(a) can be attributed to
the decrease in the optimized $J_{2,\delta_{A}}$ shown in Fig.\ \ref{fig:detuning_error_suppression}(b).

By contrast, for the uncertainty $\beta_{\alpha_{B}}$, the optimized
$J_{2,\delta_{B}}$ associated with $\beta_{\delta_{B}}$ dominates
the total detuning error and thus determines the robustness of the
CZ gate against $\beta_{\alpha_{B}}$. This is because $\beta_{\Delta_{B}}\approx0.011\beta_{\alpha_{B}}$,
estimated from $\beta_{\Delta_{B}}=\beta_{\alpha_{B}}(\omega_{0,B}-\Delta_{B})/(\omega_{0,B}-\alpha_{B})$
using Eq.\ (\ref{eq:uty_qbB_freq}) with $\Delta_{B}=2\pi\times5\,\mathrm{GHz}$,
which leads to the detuning uncertainty $\left|\beta_{\delta_{B}}\right|=\left|\beta_{\Delta_{B}}+\beta_{\alpha_{B}}\right|\sim\left|\beta_{\alpha_{B}}\right|$
being approximately two orders of magnitude larger than the other
two uncertainties $\left|\beta_{\delta_{A}}\right|=\left|-\beta_{\Delta_{B}}\right|$
and $\left|\beta_{\mathrm{01-10}}\right|=\left|-\beta_{\Delta_{B}}\right|$.
As shown in Fig.\ \ref{fig:robustness_against_uty_dt}(b), the most
robust CZ gate is achieved at $t_{f}=12\,\mathrm{ns}$, where the
gate remains robust over the uncertainty range $\left|\beta_{\alpha_{B}}\right|\leq2\pi\times5\,\mathrm{MHz}$
while maintaining the infidelity $\mathcal{I}$ below $10^{-4}$.
This behavior is consistent with the minimum optimized $J_{2,\delta_{B}}\approx1.17\times10^{-6}$
at $t_{f}=12\,\mathrm{ns}$, as shown in Fig.\ \ref{fig:detuning_error_suppression}(b).
As $t_{f}$ increases to $20\,\mathrm{ns}$, the robustness window
narrows and subsequently broadens with further increasing $t_{f}$,
in agreement with the behavior of the optimized $J_{2,\delta_{B}}$
in Fig.\ \ref{fig:detuning_error_suppression}(b).

For the uncertainty $\beta_{\omega_{0,A}}$, the associated detuning
uncertainties satisfy $\beta_{\delta_{A}}=\beta_{\Delta_{A}}$, $\beta_{\delta_{B}}=-\beta_{\Delta_{A}}$,
and $\beta_{\mathrm{01-10}}=\beta_{\Delta_{A}}$, all having the same
magnitude $\beta_{\Delta_{A}}=\beta_{\omega_{0,A}}(\Delta_{A}-\alpha_{A})/(\omega_{0,A}-\alpha_{A})\approx0.87\beta_{\omega_{0,A}}$,
where the estimate is obtained using $\Delta_{A}=2\pi\times5\,\mathrm{GHz}$.
Likewise, for the uncertainty $\beta_{\omega_{0,B}}$, the corresponding
detuning uncertainties satisfy $\beta_{\delta_{A}}=-\beta_{\Delta_{B}}$,
$\beta_{\delta_{B}}=+\beta_{\Delta_{B}}$, and $\beta_{\mathrm{01-10}}=-\beta_{\Delta_{B}}$,
with a common magnitude $\beta_{\Delta_{B}}=\beta_{\omega_{0,B}}(\Delta_{B}-\alpha_{B})/(\omega_{0,B}-\alpha_{B})\approx0.99\beta_{\omega_{0,B}}$
, where $\Delta_{B}=2\pi\times5\,\mathrm{GHz}$. These relations indicate
that the optimized cost function $J_{2,\delta_{A}}+J_{2,\delta_{B}}$,
which accounts only for the detuning uncertainties $\beta_{\delta_{A}}$
and $\beta_{\delta_{B}}$, cannot fully characterize the gate robustness
against the uncertainties $\beta_{\omega_{0,A}}$ and $\beta_{\omega_{0,B}}$.
This is primarily because the detuning error associated with $\beta_{\mathrm{01-10}}$,
whose magnitude is identical to those of $\beta_{\delta_{A}}$ and
$\beta_{\delta_{B}}$, is not included in the cost function. Furthermore,
the detuning uncertainties $\beta_{\delta_{A}}$ and $\beta_{\delta_{B}}$
are strongly correlated, satisfying $\beta_{\delta_{A}}=-\beta_{\delta_{B}}$,
whereas the simulations of $J_{2,\delta_{A}}+J_{2,\delta_{B}}$ in
Sec.\ \ref{subsubsec:detuning_error_suppression} treat them as independent.
Nevertheless, the robustness windows against $\beta_{\omega_{0,A}}$
and $\beta_{\omega_{0,B}}$ continue to broaden with increasing $t_{f}$
from $12\,\mathrm{ns}$ to $24\,\mathrm{ns}$, as shown in Figs.\ \ref{fig:robustness_against_uty_dt}(c)
and \ref{fig:robustness_against_uty_dt}(d), respectively. This behavior
is still captured by the decreasing trend of the optimized $J_{2,\delta_{A}}+J_{2,\delta_{B}}$
shown in Fig.\ \ref{fig:detuning_error_suppression}(b). For $t_{f}=24\,\mathrm{ns}$,
the optimized CZ gate maintains the infidelity $\mathcal{I}<6\times10^{-4}$
even when the uncertainty magnitude $\left|\beta_{\omega_{0,A/B}}\right|$
reaches $2\pi\times5\,\mathrm{MHz}$.

However, when the gate duration is increased to $t_{f}=30\,\mathrm{ns}$,
the robustness windows with respect to $\beta_{\omega_{0,A}}$ and
$\beta_{\omega_{0,B}}$ become significantly narrower. This behavior
can be attributed to the optimized detunings $\delta_{A}(t)/2\pi$
and $\delta_{B}(t)/2\pi$, which exhibit small oscillations around
the symmetric operating point $\delta_{A}/2\pi\approx\delta_{B}/2\pi\sim-200\,\mathrm{MHz}$,
as shown in Fig.\ \ref{fig:detuning_error_suppression}(e). At this
operating point, the $\left|0_{B}1_{A}\right\rangle \leftrightarrow\left|1_{B}0_{A}\right\rangle $
transition is nearly resonant, with $\delta_{01-10}/2\pi\sim0\,\mathrm{MHz}$,
where the interaction is most sensitive to the uncertainty $\beta_{\delta_{01-10}}$.
In addition, the induced uncertainties satisfy $\beta_{01-10}\approx0.87\beta_{\omega_{0,A}}$
and $\beta_{\mathrm{01-10}}\approx-0.99\beta_{\omega_{0,B}}$, with
magnitudes substantially larger than those induced by $\beta_{\alpha_{A}}$
and $\beta_{\alpha_{B}}$, for which $\beta_{01-10}\approx0.13\beta_{\alpha_{A}}$
and $\beta_{01-10}\approx0.011\beta_{\alpha_{B}}$. Consequently,
the contraction of the robustness windows is pronounced only for $\beta_{\omega_{0,A}}$
and $\beta_{\omega_{0,B}}$, whereas no comparable behavior is observed
for $\beta_{\alpha_{A}}$ and $\beta_{\alpha_{B}}$. Furthermore,
the apparent asymmetry of the infidelity $\mathcal{I}$ with respect
to positive and negative uncertainties in Figs.\ \ref{fig:robustness_against_uty_dt}(c)
and \ref{fig:robustness_against_uty_dt}(d) originates from the interplay
between the first- and third-order error contributions, as discussed
in Sec.\ \ref{subsubsec:robust_CZ_gates_against_beta_gAB}.

In summary, Figs.\ \ref{fig:robustness_against_uty_dt}(a)--\ref{fig:robustness_against_uty_dt}(d)
show that the most balanced robust operating condition is achieved
at $t_{f}=24\,\mathrm{ns}$, where the optimized CZ gate tolerates
diagonal uncertainties $\left|\beta_{\alpha_{A}}\right|$, $\left|\beta_{\alpha_{B}}\right|$,
$\left|\beta_{\omega_{0,A}}\right|$, and $\left|\beta_{\omega_{0,B}}\right|$
up to $2\pi\times5\,\mathrm{MHz}$ individually while keeping the
infidelity $\mathcal{I}$ below $10^{-3}$.

\subsubsection{Robust CZ gates against uncertainties $\beta_{\alpha_{A}}$, $\beta_{\alpha_{B}}$,
$\beta_{\omega_{0,A}}$, and $\beta_{\omega_{0,B}}$ in noisy environments}

\label{subsubsec:robust_CZ_gates_against_dt_uty_noise}

In this section, we investigate the realistic situation in which the
diagonal uncertainties $\beta_{\alpha_{A}}$, $\beta_{\alpha_{B}}$,
$\beta_{\omega_{0,A}}$, and $\beta_{\omega_{0,B}}$ are simultaneously
present under two noise scenarios to identify the optimal control
strategy for realizing robust CZ gates.

\paragraph{The first noise scenario}

For the first noise scenario, we consider energy relaxation and flux
noise. Energy relaxation contributes an error of $\left\langle J_{2,T_{1}}\right\rangle =t_{f}/T_{1}$,
where $T_{1}=30\,\mu\mathrm{s}$, thereby limiting the gate duration
to $t_{f}\lesssim30\,\mathrm{ns}$ to satisfy the $10^{-3}$ error
threshold. For flux noise, the sweet spot of symmetric transmons with
$a_{E_{J}}=1.0$ occurs at zero applied flux, $\Phi_{A/B}=0$, corresponding
to $\Delta_{A/B}=\omega_{0,A/B}$. This operating point coincides
with the initial interaction frequency setting adopted in the optimization
for suppressing detuning errors in Sec.\ \ref{subsubsec:detuning_error_suppression}.
There, $\omega_{i,A/B}/2\pi=5\,\mathrm{GHz}$ is chosen close to $\omega_{0,A/B}$
to balance the detuning error sensitivities to the diagonal uncertainties
$\beta_{\alpha_{A}}$, $\beta_{\alpha_{B}}$, $\beta_{\omega_{0,A}}$,
and $\beta_{\omega_{0,B}}$. Furthermore, Sec.\ \ref{subsubsec:robust_CZ_gates_against_dt_uty}
identifies the most balanced robust solution at $t_{f}=24\,\mathrm{ns}$,
which individually tolerates each of these uncertainties up to $2\pi\times5\,\mathrm{MHz}$
while maintaining the gate error below $10^{-3}$. Therefore, for
the first noise scenario, we focus on identifying the most robust
solution within the ensemble of optimized CZ gates with $t_{f}=24\,\mathrm{ns}$,
obtained via the rp-Qt-4 optimization using the cost function $J_{1}+J_{2,\delta_{A}}+J_{2,\delta_{B}}$
introduced in Sec.\ \ref{subsubsec:detuning_error_suppression}.

When the diagonal uncertainties $\beta_{\alpha_{A}}$, $\beta_{\alpha_{B}}$,
$\beta_{\omega_{0,A}}$, and $\beta_{\omega_{0,B}}$ are simultaneously
present, the allowable uncertainty tolerance for each is reduced to
approximately $2\pi\times1\,\mathrm{MHz}$ to satisfy the $10^{-3}$
error threshold, given the dominant decoherence contribution $\left\langle J_{2,T_{1}}\right\rangle =8.0\times10^{-4}$
at $t_{f}=24\,\mathrm{ns}$. We therefore evaluate the total detuning
error at an uncertainty strength of $2\pi\times1\,\mathrm{MHz}$ for
every optimized CZ gate in the ensemble. The total detuning error
is defined as the sum of the individual detuning errors associated
with $\beta_{\alpha_{A}}$, $\beta_{\alpha_{B}}$, $\beta_{\omega_{0,A}}$,
and $\beta_{\omega_{0,B}}$, each evaluated at the strength of $2\pi\times1\,\mathrm{MHz}$.
These individual contributions are extracted as $\mathcal{I}-J_{1}$,
where $\mathcal{I}$ is obtained from simulations using the respective
realistic Hamiltonians $\mathcal{H}_{I}+\mathcal{H}_{\alpha_{A}}$,
$\mathcal{H}_{I}+\mathcal{H}_{\alpha_{B}}$, $\mathcal{H}_{I}+\mathcal{H}_{\omega_{0,A}}$,
and $\mathcal{H}_{I}+\mathcal{H}_{\omega_{0,B}}$. Since the detuning
error for uncertainty strengths around $2\pi\times1\,\mathrm{MHz}$
is dominated by the second-order contribution, it scales approximately
quadratically with the uncertainty strength. Using this quadratic
scaling, we estimate, for each optimized CZ gate, the maximum uncertainty
strength that satisfies the $10^{-3}$ error threshold under the optimized
infidelity $J_{1}+\left\langle J_{2,T_{1}}\right\rangle +\left\langle J_{2,\phi}\right\rangle $.
The gate exhibiting the largest allowable uncertainty strength is
identified as the most robust solution.

\begin{figure}
\includegraphics[scale=0.52]{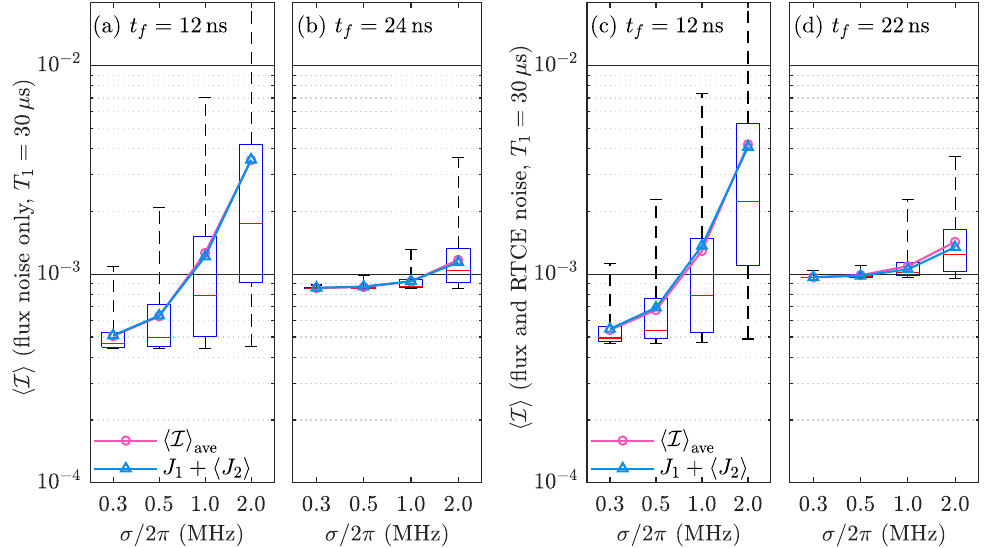}

\caption{Robustness against the simultaneous presence of diagonal uncertainties
$\beta_{\alpha_{A}}$, $\beta_{\alpha_{B}}$, $\beta_{\omega_{0,A}}$,
and $\beta_{\omega_{0,B}}$. Robust performance as a function of the
uncertainty level, $\sigma/2\pi=0.3$, $0.5$, $1.0$, and $2.0\,\mathrm{MHz}$,
for the optimal solutions with (a) $a_{E_{J}}=1.0$ and $t_{f}=12\,\mathrm{ns}$
and (b) $a_{E_{J}}=1.0$ with $t_{f}=24\,\mathrm{ns}$ under the first
noise scenario, and with (c) $a_{E_{J}}=1.31$ and $t_{f}=12\,\mathrm{ns}$
and (d) $a_{E_{J}}=2.5$ with $t_{f}=22\,\mathrm{ns}$ under the second
noise scenario. Each box plot represents an ensemble of 100 ensemble-averaged
infidelities, $\left\langle \mathcal{I}\right\rangle $, obtained
from normally distributed uncertainty sets $[\beta_{\alpha_{A}},\beta_{\alpha_{B}},\beta_{\omega_{0,A}},\beta_{\omega_{0,B}}]$
with standard deviation $\sigma/2\pi$. The whiskers indicate the
minimum and maximum values, the box spans the 25th to 75th percentiles,
and the red line denotes the median of the $\left\langle \mathcal{I}\right\rangle $
ensemble. The pink circles denote $\left\langle \mathcal{I}\right\rangle _{\mathrm{ave}}$,
the mean of the $\left\langle \mathcal{I}\right\rangle $ ensemble.
The blue triangles denote the optimized infidelity $J_{1}+\left\langle J_{2}\right\rangle $,
where $\left\langle J_{2}\right\rangle =\left\langle J_{2,T_{1}}\right\rangle +\left\langle J_{2,\phi}\right\rangle +J_{2,\delta_{A}+\delta_{B}}^{\Sigma}$
for panels (a) and (b), and $\left\langle J_{2}\right\rangle =\left\langle J_{2,T_{1}}\right\rangle +\text{\ensuremath{\left\langle J_{2,\epsilon}\right\rangle }}+\text{\ensuremath{\left\langle J_{2,\phi}\right\rangle }}+J_{2,\delta_{A}+\delta_{B}}^{\Sigma}$
for panels (c) and (d).}

\label{fig:infid_distribution_vs_uncertainty} 
\end{figure}

We then evaluate the ensemble-averaged infidelity $\left\langle \mathcal{I}\right\rangle $
of this solution using a relaxation time $T_{1}=30\,\mu\mathrm{s}$,
$1000$ realizations of the characterized flux noise $\beta_{\phi,A/B}(t)$,
and a representative uncertainty set $[\beta_{\alpha_{A}},\beta_{\alpha_{B}},\beta_{\omega_{0,A}},\beta_{\omega_{0,B}}]$.
Each uncertainty component in the set is independently sampled from
a normal distribution with standard deviation $\sigma$. For each
selected uncertainty level, $\sigma/2\pi=0.3$, $0.5$, $1.0$, and
$2.0\,\mathrm{MHz}$, we generate 100 independently sampled uncertainty
sets to characterize the corresponding distribution of $\left\langle \mathcal{I}\right\rangle $,
as shown in Fig.\ \ref{fig:infid_distribution_vs_uncertainty}(b).

In addition, we apply the same procedures to identify the most robust
solution from the ensemble of optimized $12\,\mathrm{ns}$ CZ gates.
For this solution, the gate duration is close to the CZ gate speed
limit, such that the decoherence error is nearly minimized and the
remaining error budget for detuning errors under the $10^{-3}$ threshold
is almost maximized. However, this increased margin is expected to
be rapidly consumed by detuning errors as the uncertainty strength
increases, due to the larger optimized detuning error for $t_{f}=12\,\mathrm{ns}$,
as discussed in Sec.\ \ref{subsubsec:detuning_error_suppression}.
The robust performance for $t_{f}=12\,\mathrm{ns}$ is shown in Fig.\ \ref{fig:infid_distribution_vs_uncertainty}(a).

To further characterize the gate performance, we evaluate the total
infidelity $J_{1}+\left\langle J_{2}\right\rangle $, where $\left\langle J_{2}\right\rangle =\left\langle J_{2,T_{1}}\right\rangle +\left\langle J_{2,\phi}\right\rangle +J_{2,\delta_{A}+\delta_{B}}^{\Sigma}$.
Here $J_{2,\delta_{A}+\delta_{B}}^{\Sigma}$ denotes the total detuning
error, defined as the sum of $J_{2,\delta_{A}+\delta_{B}}$ evaluated
individually for each uncertainty source $\beta_{\alpha_{A}}$, $\beta_{\alpha_{B}}$,
$\beta_{\omega_{0,A}}$, and $\beta_{\omega_{0,B}}$. For a given
uncertainty level, $J_{2,\delta_{A}+\delta_{B}}$ is evaluated by
substituting $\beta_{j}=\beta_{\alpha_{A}}$, $\beta_{\alpha_{B}}$,
$\beta_{\omega_{0,A}}$, and $\beta_{\omega_{0,B}}$, each with the
common strength $\sigma$, together with the corresponding $H_{j}=(\mathcal{H}_{\delta_{A}}+\mathcal{H}_{\delta_{B}})/(\hbar\beta_{j})$,
into $J_{2,j}$ defined in Eq.\ (\ref{eq:J2_uty_general_format}).
The corresponding forms of $\beta_{\delta_{A}}$ and $\beta_{\delta_{B}}$
in $\mathcal{H}_{\delta_{A}}+\mathcal{H}_{\delta_{B}}$ {[}Eqs.\ (\ref{eq:uty_H_dtA})-(\ref{eq:uty_H_dtB}){]}
are obtained by retaining only the selected uncertainty term $\beta_{j}$
in Eqs.\ (\ref{eq:beta_dtA})-(\ref{eq:beta_dtB}) and Eqs.\ (\ref{eq:uty_qbA_freq})-(\ref{eq:uty_qbB_freq}).
We employ $J_{2,\delta_{A}+\delta_{B}}$, rather than $J_{2,\delta_{A}}+J_{2,\delta_{B}}$
used in the optimization of Sec.\ \ref{subsubsec:detuning_error_suppression},
to provide a more accurate characterization of the total detuning
error. As shown in Figs.\ \ref{fig:infid_distribution_vs_uncertainty}(a)
and \ref{fig:infid_distribution_vs_uncertainty}(b), $J_{1}+\left\langle J_{2}\right\rangle $
accurately predicts the trend of $\left\langle \mathcal{I}\right\rangle _{\mathrm{ave}}$
as $\sigma$ increases. Here $\left\langle \mathcal{I}\right\rangle _{\mathrm{ave}}$
denotes the mean of the 100 ensemble-averaged infidelities $\left\langle \mathcal{I}\right\rangle $,
each obtained from an independently sampled uncertainty set $[\beta_{\alpha_{A}},\beta_{\alpha_{B}},\beta_{\omega_{0,A}},\beta_{\omega_{0,B}}]$
with standard deviation $\sigma$. This agreement indicates that,
in both cases, the dominant contribution to the total detuning error
arises from the detuning uncertainties $\beta_{\delta_{A}}$ and $\beta_{\delta_{B}}$.

As shown in Figs.\ \ref{fig:infid_distribution_vs_uncertainty}(a)
and \ref{fig:infid_distribution_vs_uncertainty}(b), for $\sigma/2\pi\lesssim0.5$
MHz, both the mean, $\left\langle \mathcal{I}\right\rangle _{\mathrm{ave}}$,
and the median of $\left\langle \mathcal{I}\right\rangle $ are higher
for $t_{f}=24\,\mathrm{ns}$ than for $t_{f}=12\,\mathrm{ns}$. In
this regime, the detuning error remains small, and the energy-relaxation
contribution $\left\langle J_{2,T_{1}}\right\rangle $ dominates $\left\langle \mathcal{I}\right\rangle $.
Because of its longer gate duration, the $24\,\mathrm{ns}$ gate exhibits
larger energy-relaxation and flux-noise contributions, with $\left\langle J_{2,T_{1}}\right\rangle =8.0\times10^{-4}$
and $\left\langle J_{2,\phi}\right\rangle \approx4.20\times10^{-5}$,
compared with $\left\langle J_{2,T_{1}}\right\rangle =4.0\times10^{-4}$
and $\left\langle J_{2,\phi}\right\rangle \approx1.80\times10^{-5}$
for the $12\,\mathrm{ns}$ gate. By contrast, the optimized ideal
gate infidelities are comparable, with $J_{1}\approx1.29\times10^{-5}$
and $J_{1}\approx2.10\times10^{-5}$ for $t_{f}=24\,\mathrm{ns}$
and $t_{f}=12\,\mathrm{ns}$, respectively. As $\sigma$ increases,
the total detuning error $J_{2,\delta_{A}+\delta_{B}}^{\Sigma}$ grows
and eventually becomes the dominant contribution to $\left\langle \mathcal{I}\right\rangle $.
At $\sigma/2\pi=1.0\,\mathrm{MHz}$, the optimized $J_{2,\delta_{A}+\delta_{B}}^{\Sigma}\approx7.12\times10^{-5}$
for $t_{f}=24\,\mathrm{ns}$ is more than an order of magnitude smaller
than the corresponding value of $7.76\times10^{-4}$ for $t_{f}=12\,\mathrm{ns}$.
Consequently, the optimized total infidelity $J_{1}+\left\langle J_{2}\right\rangle $
for the $24\,\mathrm{ns}$ gate increases much more slowly with increasing
$\sigma$ than that for the 12ns gate. Consistently, the corresponding
distribution of $\left\langle \mathcal{I}\right\rangle $ remains
substantially narrower for $t_{f}=24\,\mathrm{ns}$ than for $t_{f}=12\,\mathrm{ns}$,
as shown in Figs.\ \ref{fig:infid_distribution_vs_uncertainty}(a)
and \ref{fig:infid_distribution_vs_uncertainty}(b).

Under the first noise scenario, the selected $24\,\mathrm{ns}$ solution,
characterized by a reduced $J_{2,\delta_{A}+\delta_{B}}^{\Sigma}$,
effectively suppresses the spread of the $\left\langle \mathcal{I}\right\rangle $
distribution, allowing all simulated ensemble-averaged infidelities
$\left\langle \mathcal{I}\right\rangle $ to remain below the $10^{-3}$
error threshold for $\sigma/2\pi\lesssim0.5\,\mathrm{MHz}$, as illustrated
in Fig.\ \ref{fig:infid_distribution_vs_uncertainty}(b).

\paragraph{The second noise scenario}

In the second noise scenario, the RTCE noise is incorporated in addition
to the flux noise and energy relaxation considered in the first noise
scenario. The resulting increase in decoherence-induced errors reduces
the error budget available for detuning errors under a fixed target
infidelity threshold, thereby further narrowing the robustness window
against diagonal uncertainties. Consequently, suppressing decoherence-induced
errors becomes essential for maintaining robustness. In this regard,
Sec.\ %
\mbox{%
III%
} of Ref.\ \citep{Huang2026} reported an optimized $12\,\mathrm{ns}$
CZ gate that achieves the lowest infidelity, $J_{1}+\left\langle J_{2,T_{1}}\right\rangle +\text{\ensuremath{\left\langle J_{2,\epsilon}\right\rangle }}+\text{\ensuremath{\left\langle J_{2,\phi}\right\rangle }}\approx4.66\times10^{-4}$.
This performance was achieved by operating asymmetric transmons with
$a_{E_{J}}=1.31$ at a joint sweet spot for the flux and RTCE noise,
located at $\Phi_{A/B}/\Phi_{0}=0.5$, corresponding to $\Delta_{A/B}=\Delta_{A/B,\mathrm{min}}$.
At this operating point, decoherence-induced errors are substantially
suppressed, leaving a larger error budget for detuning errors and
thereby potentially expanding the robustness window. Motivated by
this observation, we initiate the search for control solutions robust
against diagonal uncertainties from this joint sweet spot.

At the noise sweet spot, where $\Delta_{A/B}=\Delta_{A/B,\mathrm{min}}$,
the corresponding detuning, $\delta_{A}=\Delta_{A,\mathrm{min}}-\Delta_{B,\mathrm{min}}+\alpha_{A}$,
becomes positive for $a_{E_{J}}\gtrsim1.2$. This behavior originates
from the increasing separation between $\Delta_{A,\mathrm{min}}$
and $\Delta_{B,\mathrm{min}}$ with increasing $a_{E_{J}}$, as shown
in Fig.\ 7(a) of Ref.\ \citep{Huang2026}. In contrast, suppression
of detuning errors is achieved near the symmetric detuning point,
$\delta_{A}\approx\delta_{B}\approx(\alpha_{A}+\alpha_{B})/2\sim-2\pi\times200\,\mathrm{MHz}$,
where the detuning is negative. Therefore, the noise sweet spot and
the detuning-error sweet spot cannot coincide. To identify an operating
point that balances these competing requirements while minimizing
performance degradation, we retain $\Delta_{A}\approx\Delta_{A,\mathrm{min}}$,
thereby preserving the low RTCE- and flux-noise contributions, $\text{\ensuremath{\left\langle J_{2,\epsilon,A}\right\rangle }}$
and $\text{\ensuremath{\left\langle J_{2,\phi,A}\right\rangle }}$,
achieved at the noise sweet spot of $Q_{A}$. By contrast, the noise-sweet-spot
condition of $Q_{B}$ is relaxed by increasing $\Delta_{B}$ toward
$\Delta_{A,\mathrm{min}}$. This adjustment shifts the operating point
closer to the detuning-error sweet spot at $\delta_{A}/2\pi\sim-200\,\mathrm{MHz}$,
thereby suppressing detuning errors.

As $\Delta_{B}$ departs from its noise sweet spot, $\Delta_{B,\mathrm{min}}$,
the flux-noise contribution $\text{\ensuremath{\left\langle J_{2,\phi,B}\right\rangle }}$
increases rapidly due to the nearby pronounced peak in the derivative
$d\Delta_{B}/d(\Phi_{B}/\Phi_{0})$, which is characteristic of weakly
asymmetric transmons with $a_{E_{J}}$ close to $1.0$. To mitigate
this increase in $\text{\ensuremath{\left\langle J_{2,\phi,B}\right\rangle }}$,
$a_{E_{J}}$ can be increased to reduce the magnitude of the derivative
peak, as illustrated in Figs.\ 10(a-2)-10(e-2) of Ref.\ \citep{Huang2026}.
However, increasing $a_{E_{J}}$ also narrows the tunable ranges of
$\Delta_{A/B}$ {[}see Figs.\ 10(a-1)-10(e-1) of Ref.\ \citep{Huang2026}{]}
and enhances the RTCE-noise contribution $\text{\ensuremath{\left\langle J_{2,\epsilon}\right\rangle }}$
through the corresponding increase in $\Delta_{A/B,\mathrm{min}}$.
Conversely, the reduced tunable ranges of $\Delta_{A/B}$ bring $\Delta_{A/B,\mathrm{min}}$
closer to $\omega_{0,A/B}$, which provides a favorable condition
for balancing the sensitivities of detuning errors to diagonal uncertainties,
as discussed in Sec.\ \ref{subsubsec:detuning_error_suppression}.
These observations demonstrate an intrinsic trade-off among the flux-noise
contribution $\text{\ensuremath{\left\langle J_{2,\phi}\right\rangle }}$,
RTCE-noise contribution $\text{\ensuremath{\left\langle J_{2,\epsilon}\right\rangle }}$,
and detuning errors through tuning $a_{E_{J}}$.

A second trade-off arises from the gate duration $t_{f}$. As discussed
in Sec.\ \ref{subsubsec:detuning_error_suppression}, a longer gate
duration allows the optimized detuning $\delta_{A/B}(t)$ to remain
closer to the symmetric-detuning point, $\delta_{A/B}/2\pi\sim-200\,\mathrm{MHz}$,
leading to more effective suppression of detuning errors. However,
under a fixed target error threshold, increasing $t_{f}$ also increases
the accumulated decoherence errors, $\left\langle J_{2,T_{1}}\right\rangle +\left\langle J_{2,\epsilon}\right\rangle +\left\langle J_{2,\phi}\right\rangle $,
leaving a smaller error budget for detuning errors and thus reducing
the robustness window against diagonal uncertainties. Consequently,
identifying an optimal control strategy also requires accounting for
the trade-off between decoherence and detuning errors through tuning
$t_{f}$.

To balance the above trade-offs, we tune $a_{E_{J}}$ and $t_{f}$
to perform the rp-Qt-4 optimization using the cost function $J_{1}+\text{\ensuremath{\left\langle J_{2,\epsilon}\right\rangle }}+\text{\ensuremath{\left\langle J_{2,\phi}\right\rangle }+\ensuremath{J_{2,\delta_{A}}}+\ensuremath{J_{2,\delta_{B}}}}$.
The initial interaction frequencies are chosen as $\omega_{i,A}=\omega_{i,B}=\Delta_{A,\mathrm{min}}$,
yielding an initial detuning of $\delta_{A}/2\pi=-214\,\mathrm{MHz}$,
close to the symmetric detuning point. We set $\Phi_{A}(t)=\Phi_{A}^{\mathrm{PBP}}(t)$
to remain near the noise sweet spot of $Q_{A}$, while $\Phi_{B}(t)=\Phi_{B}^{\mathrm{PBP}}(t)+\Phi_{B}^{\mathrm{FTP}}(t)$,
with $k_{\mathrm{max}}=10$ in $\Phi_{B}^{\mathrm{FTP}}(t)$, steers
the operating point toward the optimal balance between the noise and
detuning-error sweet spots. The same selection procedure used for
the first noise scenario is then applied to identify the most robust
solution.

For the second noise scenario, the most robust solution is identified
at $a_{E_{J}}=2.5$ and $t_{f}=22\,\mathrm{ns}$, which provides the
optimal balance between the total detuning error $J_{2,\delta_{A}+\delta_{B}}^{\Sigma}$
and the infidelity $J_{1}+\left\langle J_{2,T_{1}}\right\rangle +\left\langle J_{2,\epsilon}\right\rangle +\left\langle J_{2,\phi}\right\rangle $,
thereby maximizing the robustness window against diagonal uncertainties.
Departing from this optimal balance in either direction degrades the
robustness. Increasing $t_{f}$ further allows the optimized detuning
$\delta_{A/B}(t)$ to remain closer to the symmetric detuning point,
thereby suppressing the total detuning error below its optimized value
of $J_{2,\delta_{A}+\delta_{B}}^{\Sigma}\approx9.77\times10^{-5}$
at $\sigma/2\pi=1.0\,\mathrm{MHz}$. However, this improvement is
offset by the enhanced accumulation of decoherence errors, since the
optimized infidelity $J_{1}+\left\langle J_{2,T_{1}}\right\rangle +\left\langle J_{2,\epsilon}\right\rangle +\left\langle J_{2,\phi}\right\rangle \approx9.58\times10^{-4}$
is already close to the threshold of $10^{-3}$. Consequently, further
increasing $t_{f}$ rapidly drives the infidelity beyond the threshold
and eliminates the robustness window. Conversely, reducing $t_{f}$
or shifting $\Delta_{A/B}$ toward the noise sweet spot decreases
decoherence errors at the expense of increased detuning errors, likewise
narrowing the robustness window. At this optimal operating point,
the individual contributions are $J_{1}\approx3.15\times10^{-5}$,
$\left\langle J_{2,T_{1}}\right\rangle \approx7.33\times10^{-4}$,
$\left\langle J_{2,\epsilon}\right\rangle \approx1.19\times10^{-4}$,
and $\left\langle J_{2,\phi}\right\rangle \approx7.44\times10^{-5}$.

Figures\ \ref{fig:infid_distribution_vs_uncertainty}(c) and \ref{fig:infid_distribution_vs_uncertainty}(d)
demonstrate the robust performance of the solution with the lowest
optimized infidelity, obtained at $a_{E_{J}}=1.31$ and $t_{f}=12\,\mathrm{ns}$,
and the most robust solution, obtained at $a_{E_{J}}=2.5$ and $t_{f}=22\,\mathrm{ns}$,
respectively. For the $12\,\mathrm{ns}$ gate, the optimized infidelity,
$J_{1}+\left\langle J_{2,T_{1}}\right\rangle +\left\langle J_{2,\epsilon}\right\rangle +\left\langle J_{2,\phi}\right\rangle \approx4.66\times10^{-4}$,
is nearly a factor of two lower than that of the $22\,\mathrm{ns}$
gate, $9.58\times10^{-4}$. However, the corresponding total detuning
error, $J_{2,\delta_{A}+\delta_{B}}^{\Sigma}\approx9.03\times10^{-4}$
at $\sigma/2\pi=1.0\,\mathrm{MHz}$, is approximately one order of
magnitude larger than the optimized value of $J_{2,\delta_{A}+\delta_{B}}^{\Sigma}\approx9.77\times10^{-5}$
for the $22\,\mathrm{ns}$ gate. Consequently, in the small-$\sigma$
regime, the infidelity $J_{1}+\left\langle J_{2,T_{1}}\right\rangle +\text{\ensuremath{\left\langle J_{2,\epsilon}\right\rangle }}+\text{\ensuremath{\left\langle J_{2,\phi}\right\rangle }}$
dominates, resulting in a lower $\left\langle \mathcal{I}\right\rangle _{\mathrm{ave}}$
for the $12\,\mathrm{ns}$ gate than for the $22\,\mathrm{ns}$ gate.
As $\sigma$ increases, however, the total detuning error $J_{2,\delta_{A}+\delta_{B}}^{\Sigma}$
becomes dominant, leading to a rapid increase in $\left\langle \mathcal{I}\right\rangle _{\mathrm{ave}}$
and a pronounced broadening of the $\left\langle \mathcal{I}\right\rangle $
distribution for the $12\,\mathrm{ns}$ gate, whereas the $22\,\mathrm{ns}$
gate exhibits a much slower increase in $\left\langle \mathcal{I}\right\rangle _{\mathrm{ave}}$
together with a substantially narrower $\left\langle \mathcal{I}\right\rangle $
distribution.

In summary, for the second noise scenario, the most robust solution
with $a_{E_{J}}=2.5$ and $t_{f}=22\,\mathrm{ns}$ effectively mitigates
the growth of $\left\langle \mathcal{I}\right\rangle _{\mathrm{ave}}$
while significantly suppressing the spread of the $\left\langle \mathcal{I}\right\rangle $
distribution. Specifically, at $\sigma/2\pi=0.5\,\mathrm{MHz}$, $\left\langle \mathcal{I}\right\rangle _{\mathrm{ave}}\approx9.90\times10^{-4}$,
the 75th percentile of the $\left\langle \mathcal{I}\right\rangle $
samples remains below $10^{-3}$ , and all samples are below $1.1\times10^{-3}$.

\section{Overall errors}

\label{sec:overall_infidelity_improvement}

\subsection{Preliminaries}

In Sec.\ %
\mbox{%
II%
} of Ref.\ \citep{Huang2026}, we analyze the intrinsic error mechanisms
of the CZ gate and reduce the ideal gate infidelity $J_{1}$. In Sec.\ %
\mbox{%
III%
} of Ref.\ \citep{Huang2026}, we characterize the dominant noises
in the system and mitigate their contributions to the CZ gate infidelity.
Specifically, the energy relaxation error $\left\langle J_{2,T_{1}}\right\rangle $
is suppressed by shortening the CZ gate time close to its speed limit,
while the errors $\left\langle J_{2,\epsilon}\right\rangle $ and
$\left\langle J_{2,\phi}\right\rangle $, arising from RTCE noise
$\beta_{\epsilon}(t)$ and flux noise $\beta_{\phi}(t)$, respectively,
are minimized by operating the CZ gates at their corresponding sweet
spots. In Sec.\ \ref{subsec:uty_robust_against_gAB_uncertainty},
we further enhance robustness against the off-diagonal uncertainty
$\beta_{g_{AB}}$ by employing strong detuning $\delta_{A}(t)$ modulations
to suppress the associated error contribution $J_{2,g_{AB}}$. In
Sec.\ \ref{subsec:uty_robust_against_detuning_uncertainty}, robustness
against the diagonal uncertainties $\beta_{\alpha_{A}}$, $\beta_{\alpha_{B}}$,
$\beta_{\omega_{0,A}}$, and $\beta_{\omega_{0,B}}$ is improved by
operating the CZ gate close to the symmetric detuning point, thereby
minimizing the corresponding dominant error contribution $J_{2,\delta_{A}}+J_{2,\delta_{B}}$.

In this section, we simultaneously account for all contributions to
the gate infidelity arising from intrinsic errors, noises, and uncertainties.
Under this comprehensive error model, we determine optimal control
solutions that enable high-fidelity CZ gate operations at different
stages, subject to the noise conditions characterized by $T_{1}$,
$\sigma_{\epsilon}$, and $\sigma_{\phi,\mathrm{eff}}$ , as well
as the uncertainty levels $\sigma_{g_{AB}}$ and $\sigma_{\delta}$,
summarized in Table\ \ref{Table:noise_uty_strength_at_stages}. The
quantities $\sigma_{g_{AB}}$ and $\sigma_{\delta}$ denote the standard
deviations of the normally distributed uncertainties $\beta_{g_{AB}}$
and $\beta_{\delta}$, respectively. The parameter $\beta_{\delta}$
collectively represents the diagonal uncertainties $\beta_{\alpha_{A}}$,
$\beta_{\alpha_{B}}$, $\beta_{\omega_{0,A}}$, and $\beta_{\omega_{0,B}}$,
which are assumed to share a common standard deviation $\sigma_{\delta}$.

To reliably identify the optimal control solution under a comprehensive
error budget, we incorporate more accurate error modeling into the
optimization procedure. The cost functions $J_{2,\delta_{A}}+J_{2,\delta_{B}}$
and $J_{2,g_{AB}}$, introduced in Secs.\ \ref{subsec:uty_robust_against_detuning_uncertainty}
and \ref{subsec:uty_robust_against_gAB_uncertainty}, respectively,
are effective in enhancing robustness against the diagonal uncertainties
$\beta_{\alpha_{A}}$, $\beta_{\alpha_{B}}$, $\beta_{\omega_{0,A}}$,
and $\beta_{\omega_{0,B}}$, as well as the off-diagonal uncertainty
$\beta_{g_{AB}}$. However, when both types of uncertainties are simultaneously
included in the optimization, these cost functions do not provide
sufficient resolution to distinguish the true optimal solution. We
therefore replace $J_{2,\delta_{A}}+J_{2,\delta_{B}}$ and $J_{2,g_{AB}}$
with more accurate ensemble-averaged cost functions $\left\langle J_{2,\delta}\right\rangle $
and $\left\langle J_{2,g_{AB}}\right\rangle $ in the optimization
procedure. The uncertainties $\beta_{g_{AB}}$, $\beta_{\alpha_{A}}$,
$\beta_{\alpha_{B}}$, $\beta_{\omega_{0,A}}$, and $\beta_{\omega_{0,B}}$
are assumed to be statistically independent and normally distributed
with zero mean. Under this assumption, the corresponding ensemble-averaged
errors can be evaluated independently and subsequently combined to
characterize the overall infidelity.

The cost functions $J_{2,\delta_{A}}$ and $J_{2,\delta_{B}}$ represent
the detuning error arising from the constant detuning uncertainties
$\beta_{\delta_{A}}$ and $\beta_{\delta_{B}}$ in the $\left\{ \left|1_{B}1_{A}\right\rangle ,\left|0_{B}2_{A}\right\rangle ,\left|2_{B}0_{A}\right\rangle \right\} $
subspace. To improve the accuracy of the detuning-error evaluation,
we additionally include the contribution from the detuning uncertainty
$\beta_{\delta_{01-10}}$ in the $\left\{ \left|0_{B}1_{A}\right\rangle ,\left|1_{B}0_{A}\right\rangle \right\} $
subspace. Furthermore, we employ the detailed $\Delta_{A/B}$-dependent
detuning uncertainties $\beta_{\delta_{A}}$, $\beta_{\delta_{B}}$,
and $\beta_{\delta_{01-10}}$, evaluated using Eqs.\ (\ref{eq:beta_01t10})-(\ref{eq:beta_dtB}),
to calculate the detuning error associated with each diagonal uncertainty
$\beta_{\delta}\in\left\{ \beta_{\alpha_{A}},\beta_{\alpha_{B}},\beta_{\omega_{0,A}},\beta_{\omega_{0,B}}\right\} $.
The resulting error contributions are independently ensemble averaged
and subsequently combined to obtain the ensemble-averaged cost function
$\left\langle J_{2,\delta}\right\rangle $.

Relative to the cost function $J_{2,g_{AB}}$, the realistic infidelity
simulation yields a lower coupling error when $J_{2,g_{AB}}$ is reduced
to sufficiently small values. This behavior suggests that the uncertainty
$\beta_{g_{AB}}$ also generates additional local phases within the
computational subspace, resulting in a small error contribution to
$J_{2,g_{AB}}$. This residual contribution cannot be mitigated by
dynamical decoupling through strong $\delta_{A}(t)$ modulation but
is compensated by the characterized single-qubit-Z gate $U_{Z}$ in
the realistic infidelity simulation. However, within the $J_{2,j}$
framework, this residual error cannot be explicitly removed from $J_{2,g_{AB}}$
and may therefore lead to an inaccurate error estimate during optimization
in the presence of multiple error sources. To address this issue,
we employ the realistic infidelity simulation with $U_{Z}$ compensation
to obtain a more accurate coupling error, which is then ensemble averaged
to construct the ensemble-averaged cost function $\left\langle J_{2,g_{AB}}\right\rangle $.
The full derivations of $\left\langle J_{2,\delta}\right\rangle $
and $\left\langle J_{2,g_{AB}}\right\rangle $ are provided in Appendix\ \ref{appendix:precise_uty_cost_function}.

To compare with the optimal solutions at each stage, we choose as
the reference solution the one that achieves the lowest infidelity
$J_{1}+\left\langle J_{2,T_{1}}\right\rangle +\left\langle J_{2,\epsilon}\right\rangle +\left\langle J_{2,\phi}\right\rangle $
while maintaining a near-speed-limit gate time of $t_{f}=12\,\mathrm{ns}$,
obtained for asymmetric transmons with $a_{E_{J}}=1.31$, as reported
in Sec.\ %
\mbox{%
III%
}\ C of Ref.\ \citep{Huang2026}. The performance of the reference
solution at each stage is discussed in Sec.\ \ref{subsec:against_all_reference_control},
while the stage-specific optimal solutions are developed in Sec.\ \ref{subsec:against_all_optimized_control}.

Figure\ \ref{fig:overall_infidelity_performance}(a) shows the distributions
of the ensemble-averaged infidelity $\left\langle \mathcal{I}\right\rangle $
for the reference and optimal solutions at each stage under the specified
noise parameters $T_{1}$, $\sigma_{\epsilon}$, and $\sigma_{\phi,\mathrm{eff}}$,
together with the uncertainty levels $\sigma_{g_{AB}}$ and $\sigma_{\delta}$,
listed in Table\ \ref{Table:noise_uty_strength_at_stages}. Each
value of $\left\langle \mathcal{I}\right\rangle $ is obtained by
averaging $1000$ infidelity values $\mathcal{I}$, each simulated
using a noise realization $\left\{ \beta_{\epsilon}(t),\beta_{\phi}(t)\right\} $
generated according to $\left\{ \sigma_{\epsilon},\sigma_{\phi,\mathrm{eff}}\right\} $,
while fixing $T_{1}$ and a specific uncertainty realization $\left\{ \beta_{g_{AB}},\beta_{\alpha_{A}},\beta_{\alpha_{B}},\beta_{\omega_{0,A}},\beta_{\omega_{0,B}}\right\} $.
For each stage, $100$ uncertainty realizations are generated according
to the corresponding standard deviations $\left\{ \sigma_{g_{AB}},\sigma_{\delta},\sigma_{\delta},\sigma_{\delta},\sigma_{\delta}\right\} $,
each producing one value of $\left\langle \mathcal{I}\right\rangle $,
thereby forming a distribution of $\left\langle \mathcal{I}\right\rangle $
consisting of $100$ samples.

Figures\ \ref{fig:overall_infidelity_performance}(b) and \ref{fig:overall_infidelity_performance}(c)
present the individual error contributions at each stage for the reference
and optimal solutions, respectively. For both solutions, the total
infidelity $\left\langle \mathcal{I}\right\rangle _{\mathrm{ave}}$,
obtained from full Hamiltonian simulations, is in excellent agreement
with the total infidelity obtained by summing the decomposed cost
functions, $J_{1}+\left\langle J_{2}\right\rangle $, where $\left\langle J_{2}\right\rangle =\left\langle J_{2,T_{1}}\right\rangle +\left\langle J_{2,\epsilon}\right\rangle +\left\langle J_{2,\phi}\right\rangle +\left\langle J_{2,g_{AB}}\right\rangle +\left\langle J_{2,\delta}\right\rangle $.
Here, $\left\langle \mathcal{I}\right\rangle _{\mathrm{ave}}$ is
the average of $100$ $\left\langle \mathcal{I}\right\rangle $ samples
used to construct the $\left\langle \mathcal{I}\right\rangle $ distribution
shown in Fig.\ \ref{fig:overall_infidelity_performance}(a). This
agreement validates the approximations underlying the proposed error
models.

\subsection{Reference solution}

\label{subsec:against_all_reference_control}

At Stage 1, we adopt the noise parameters experimentally characterized
in Sec.\ %
\mbox{%
III%
} of Ref.\ \citep{Huang2026}, namely $\sigma_{\epsilon}=\sigma_{\epsilon}^{0}$,
$\sigma_{\phi,\mathrm{eff}}=\sigma_{\phi,\mathrm{eff}}^{0}$, and
$T_{1}=30\,\mu\mathrm{s}$, together with currently achievable experimental
uncertainty levels $\sigma_{g_{AB}}/2\pi=0.1\,\mathrm{MHz}$ and $\sigma_{\delta}/2\pi=0.1\,\mathrm{MHz}$.
At this stage, for the reference solution, the total infidelity $\left\langle \mathcal{I}\right\rangle _{\mathrm{ave}}$
is dominated by the energy-relaxation contribution $\left\langle J_{2,T_{1}}\right\rangle =4\times10^{-4}$,
which exceeds all other error contributions by more than an order
of magnitude, including the uncertainty-induced error contributions
$\left\langle J_{2,g_{AB}}\right\rangle $ and $\left\langle J_{2,\delta}\right\rangle $,
as shown in Fig.\ \ref{fig:overall_infidelity_performance}(b). This
leads to a narrow distribution of $\left\langle \mathcal{I}\right\rangle $,
as illustrated by the box plot in Fig.\ \ref{fig:overall_infidelity_performance}(a).

From Stage 1 to Stage 2 and from Stage 2 to Stage 3, the noise standard
deviations $\sigma_{\epsilon}$ and $\sigma_{\phi,\mathrm{eff}}$
are reduced by one order of magnitude, and the energy relaxation time
$T_{1}$ is increased by a factor of $10$, as summarized in Table\ \ref{Table:noise_uty_strength_at_stages}.
Consequently, for the reference solution, between these consecutive
stages, the noise-induced error contributions $\left\langle J_{2,\epsilon}\right\rangle $
and $\left\langle J_{2,\phi}\right\rangle $, proportional to $\sigma_{\epsilon}^{2}$
and $\sigma_{\phi,\mathrm{eff}}^{2}$, respectively, decrease by two
orders of magnitude, while the energy-relaxation contribution $\left\langle J_{2,T_{1}}\right\rangle =t_{f}/T_{1}$
decreases by one order of magnitude, as shown in Fig.\ \ref{fig:overall_infidelity_performance}(b).
From Stage 1 through Stage 3, the uncertainty standard deviations
$\sigma_{g_{AB}}$ and $\sigma_{\delta}$ are held fixed at $2\pi\times0.1\,\mathrm{MHz}$,
and the corresponding error contributions remain constant at $\left\langle J_{2,g_{AB}}\right\rangle \approx2.57\times10^{-5}$
and $\left\langle J_{2,\delta}\right\rangle \approx9.16\times10^{-6}$.
The intrinsic error $J_{1}$, which is independent of noise and uncertainty,
remains at $2.46\times10^{-5}$ across all stages. At Stages 2 and
3, the reduction in the decoherence-induced error contributions $\left\langle J_{2,T_{1}}\right\rangle $,
$\left\langle J_{2,\epsilon}\right\rangle $, and $\left\langle J_{2,\phi}\right\rangle $
causes the uncertainty-induced error contributions $\left\langle J_{2,g_{AB}}\right\rangle $
and $\left\langle J_{2,\delta}\right\rangle $ to account for an increasing
fraction of the total infidelity $\left\langle \mathcal{I}\right\rangle _{\mathrm{ave}}$,
leading to broader distributions of $\left\langle \mathcal{I}\right\rangle $
observed in the box plots of Fig.\ \ref{fig:overall_infidelity_performance}(a).

After Stage 3, the noise parameters are held fixed at their Stage-3
values, namely $\sigma_{\epsilon}=0.01\cdot\sigma_{\epsilon}^{0}$,
$\sigma_{\phi,\mathrm{eff}}=0.01\cdot\sigma_{\phi,\mathrm{eff}}^{0}$,
and $T_{1}=3\,\mathrm{ms}$. Consequently, the corresponding error
contributions remain constant at $\left\langle J_{2,\epsilon}\right\rangle \approx1.02\times10^{-9}$,
$\left\langle J_{2,\phi}\right\rangle \approx3.11\times10^{-9}$,
and $\left\langle J_{2,T_{1}}\right\rangle =4\times10^{-6}$. At Stage
4, the off-diagonal uncertainty standard deviation $\sigma_{g_{AB}}$
is maintained at $2\pi\times0.1\,\mathrm{MHz}$, as in Stages 1--3,
while the diagonal uncertainty standard deviation $\sigma_{\delta}$
is improved to $2\pi\times0.01\,\mathrm{MHz}$. As a result, the detuning
error $\left\langle J_{2,\delta}\right\rangle \propto\sigma_{\delta}^{2}$
is reduced by two orders of magnitude to $9.16\times10^{-8}$ compared
with Stages 1--3. Consequently, the total infidelity $\left\langle \mathcal{I}\right\rangle _{\mathrm{ave}}$
is primarily determined by the coupling error $\left\langle J_{2,g_{AB}}\right\rangle $
and intrinsic error $J_{1}$, with $\left\langle J_{2,g_{AB}}\right\rangle $
contributing to the broad distribution of $\left\langle \mathcal{I}\right\rangle $
in Fig.\ \ref{fig:overall_infidelity_performance}(a). Conversely,
at Stage 5, $\sigma_{\delta}=2\pi\times0.1\,\mathrm{MHz}$ while $\sigma_{g_{AB}}=2\pi\times0.01\,\mathrm{MHz}$.
In this case, the coupling error $\left\langle J_{2,g_{AB}}\right\rangle \propto\sigma_{g_{AB}}^{2}$
is reduced to $2.57\times10^{-7}$, and $\left\langle J_{2,\delta}\right\rangle $
becomes the dominant contribution, resulting in a broad distribution
of $\left\langle \mathcal{I}\right\rangle $. At Stage 6, both $\sigma_{g_{AB}}$
and $\sigma_{\delta}$ are improved to $2\pi\times0.01\,\mathrm{MHz}$.
With both $\left\langle J_{2,g_{AB}}\right\rangle $ and $\left\langle J_{2,\delta}\right\rangle $
substantially reduced, $\left\langle \mathcal{I}\right\rangle _{\mathrm{ave}}$
becomes determined solely by the intrinsic error $J_{1}$, and the
distribution of $\left\langle \mathcal{I}\right\rangle $ in the box
plot becomes tightly concentrated.

\subsection{Stage-specific optimal solutions}

\label{subsec:against_all_optimized_control}

\subsubsection{Stage 1 and preparation for Stages 2-6}

At Stage 1, the short $T_{1}=30\,\mu\mathrm{s}$ causes the energy-relaxation
contribution $\left\langle J_{2,T_{1}}\right\rangle $ to dominate
the total infidelity $\left\langle \mathcal{I}\right\rangle _{\mathrm{ave}}$.
As a result, the most effective strategy for Stage 1 is to reduce
the gate time $t_{f}$ to minimize $\left\langle J_{2,T_{1}}\right\rangle =t_{f}/T_{1}$.
The reference solution reduces the gate time to $t_{f}=12\,\mathrm{ns}$,
close to the CZ gate speed limit, and achieves the minimum infidelity
$J_{1}+\left\langle J_{2,T_{1}}\right\rangle +\left\langle J_{2,\epsilon}\right\rangle +\left\langle J_{2,\phi}\right\rangle $.
In this case, $\left\langle J_{2,T_{1}}\right\rangle $ dominates
$\left\langle \mathcal{I}\right\rangle _{\mathrm{ave}}$ and exceeds
all other contributions by more than an order of magnitude, including
the uncertainty-induced contributions $\left\langle J_{2,g_{AB}}\right\rangle $
and $\left\langle J_{2,\delta}\right\rangle $, as discussed in Sec.\ \ref{subsec:against_all_reference_control}.
Attempts to enhance robustness against uncertainties by further suppressing
$\left\langle J_{2,g_{AB}}\right\rangle $ and $\left\langle J_{2,\delta}\right\rangle $
through a longer gate time $t_{f}$ instead lead to a larger $\left\langle J_{2,T_{1}}\right\rangle $.
This results in a higher $\left\langle \mathcal{I}\right\rangle _{\mathrm{ave}}$
despite the narrower $\left\langle \mathcal{I}\right\rangle $ distribution.
Consequently, robust control against uncertainties provides no benefit
at Stage 1. The reference solution is therefore the optimal solution
for Stage 1.

By contrast, beyond Stage 1, the impact of noise is substantially
reduced, as reflected by the noise parameters listed in Table\ \ref{Table:noise_uty_strength_at_stages}.
Consequently, the noise contributions $\left\langle J_{2,T_{1}}\right\rangle $,
$\left\langle J_{2,\epsilon}\right\rangle $, and $\left\langle J_{2,\phi}\right\rangle $
decrease, whereas the uncertainty contributions $\left\langle J_{2,g_{AB}}\right\rangle $
and $\left\langle J_{2,\delta}\right\rangle $ become increasingly
important and eventually dominate the total infidelity $\left\langle \mathcal{I}_{\mathrm{ave}}\right\rangle $,
resulting in broader distributions of $\left\langle \mathcal{I}\right\rangle $.
The reduced noise contributions free up error budget, allowing moderate
increases in $\left\langle J_{2,T_{1}}\right\rangle $, $\left\langle J_{2,\epsilon}\right\rangle $,
and $\left\langle J_{2,\phi}\right\rangle $ as the gate time $t_{f}$
is extended to suppress $\left\langle J_{2,g_{AB}}\right\rangle $
and $\left\langle J_{2,\delta}\right\rangle $, thereby yielding optimal
solutions with enhanced robustness against uncertainties.

Moreover, the intrinsic error $J_{1}$ becomes another dominant contribution
as the impact of noise is suppressed. As shown in Fig.\ 4 of Ref.\ \citep{Huang2026},
the optimized $J_{1}$ is primarily limited by the jump-2-quanta errors,
including the $\left|0_{B}0_{A}\right\rangle \leftrightarrow\left|1_{B}1_{A}\right\rangle $
swap error, as well as the Qt-3 and Qt-4 leakage errors. By reducing
the interaction frequencies $\omega_{i,A/B}$ in $\Delta_{A/B}$ through
tuning the interaction flux $\Phi_{i,A/B}$ in the Pull-Back-Pulse
$\Phi_{A/B}^{\mathrm{PBP}}(t)$, the oscillation frequencies of the
jump-2-quanta interactions are likewise reduced. This reduction facilitates
the synchronization of the associated errors at their minima. Consequently,
we search for optimal control solutions for Stages 2--6 using symmetric
transmons with $a_{E_{J}}=1.0$, which provide large qubit-frequency
tunability. Nevertheless, as $\omega_{i,A/B}$ becomes too low, the
synchronization performance degrades because additional leakage channels
become accessible, thereby increasing $J_{1}$. In addition, the flux-noise
contribution $\left\langle J_{2,\phi}\right\rangle $ is enhanced
due to the increase in the derivatives $d\Delta_{A/B}/d(\Phi_{A/B}/\Phi_{0})$
with decreasing $\omega_{i,A/B}$, as shown in Fig.\ 10 of Ref.\ \citep{Huang2026}.
These factors are therefore taken into account in the optimization
for Stages 2--6.

\subsubsection{Stages 2 and 3}

For Stages 2 and 3, the increasingly relaxed noise conditions with
fixed uncertainty levels $\sigma_{g_{AB}}=\sigma_{\delta}=2\pi\times0.1\,\mathrm{MHz}$
enable us to employ the rp-Qt-4 optimization with the cost function
$J_{1}+\left\langle J_{2,g_{AB}}\right\rangle +\left\langle J_{2,\delta}\right\rangle $
to identify the optimal control solutions. To address the increased
complexity of the optimization, we activate Fine-Tune-Pulse $\Phi_{A/B}^{\mathrm{FTP}}(t)$
with the control parameter number $k_{\mathrm{max}}=30$ for both
qubits. The initial $\delta_{A}$ is first tuned in the optimization
to determine the optimal value, $\delta_{A}=-2\pi\times400\,\mathrm{MHz}$,
for which $\left\langle J_{2,g_{AB}}\right\rangle +\left\langle J_{2,\delta}\right\rangle $
is minimized. This choice induces pronounced oscillations of $\delta_{B}(t)$
around $\delta_{B}=0$, which effectively suppresses the dominant
contribution to the coupling error $\left\langle J_{2,g_{AB}}\right\rangle $
arising from the $\left\{ \left|1_{B}1_{A}\right\rangle ,\left|2_{B}0_{A}\right\rangle \right\} $
subspace. In addition, this setting allows $\delta_{A}(t)$ and $\delta_{B}(t)$
to spend more time near the symmetric detuning point $\delta_{A}\approx\delta_{B}\approx-2\pi\times200\,\mathrm{MHz}$,
thereby mitigating the detuning error $\left\langle J_{2,\delta}\right\rangle $.
These features can be observed in the optimized detunings $\delta_{A/B}(t)$
for Stage 3, as shown in Fig.\ \ref{fig:overall_infidelity_performance}(e).

Subsequently, we tune the initial interaction flux $\Phi_{i,A/B}$
of the Pull-Back-Pulse $\Phi_{A/B}^{\mathrm{PBP}}(t)$ in the optimization
to determine the optimal initial interaction frequency $\omega_{i,A/B}$
that minimizes $J_{1}$. For Stage 3, the optimal initial $\omega_{i,A}/2\pi$
is found to be $2\,\mathrm{GHz}$, yielding an optimized $J_{1}\approx6.46\times10^{-6}$,
as shown in Fig.\ \ref{fig:overall_infidelity_performance}(c). The
corresponding optimized qubit frequencies $\Delta_{A/B}(t)$ are presented
in Fig.\ \ref{fig:overall_infidelity_performance}(d). For Stage
2, the flux-noise parameter $\sigma_{\phi,\mathrm{eff}}=0.1\cdot\sigma_{\phi,\mathrm{eff}}^{0}$
is an order of magnitude larger than $\sigma_{\phi,\mathrm{eff}}=0.01\cdot\sigma_{\phi,\mathrm{eff}}^{0}$
for Stage 3. Consequently, the flux-noise contribution $\left\langle J_{2,\phi}\right\rangle $
becomes more significant as $\omega_{i,A/B}$ is reduced, limiting
the optimal initial interaction frequency for Stage 2 to $\omega_{i,A}/2\pi=4.5\,\mathrm{GHz}$.
Under this constraint, $J_{1}$ can only be reduced to approximately
$1.59\times10^{-5}$. Finally, using these optimal settings in the
optimization, we obtain an ensemble of optimized CZ gates with suppressed
$J_{1}+\left\langle J_{2,g_{AB}}\right\rangle +\left\langle J_{2,\delta}\right\rangle $.
An optimal solution is then selected from this ensemble by taking
into account both the total infidelity $J_{1}+\left\langle J_{2}\right\rangle $,
which determines $\left\langle \mathcal{I}\right\rangle _{\mathrm{ave}}$,
and the uncertainty contributions $\left\langle J_{2,g_{AB}}\right\rangle +\left\langle J_{2,\delta}\right\rangle $,
which determine the distribution of $\left\langle \mathcal{I}\right\rangle $.

At Stage 2, the optimal solution with $t_{f}=27\,\mathrm{ns}$ yields
$\left\langle \mathcal{I}\right\rangle <2\times10^{-4}$ for all samples,
with a significantly narrowed $\left\langle \mathcal{I}\right\rangle $
distribution due to the suppressed uncertainty contributions $\left\langle J_{2,g_{AB}}\right\rangle +\left\langle J_{2,\delta}\right\rangle \approx7.32\times10^{-6}$.
This contrasts with the reference solution, where $\left\langle J_{2,g_{AB}}\right\rangle +\left\langle J_{2,\delta}\right\rangle \approx3.49\times10^{-5}$,
resulting in a broader $\left\langle \mathcal{I}\right\rangle $ distribution,
as shown in Fig.\ \ref{fig:overall_infidelity_performance}(a). For
the optimal solution, the total infidelity $\left\langle \mathcal{I}\right\rangle _{\mathrm{ave}}\approx1.22\times10^{-4}$
is slightly higher than that of the reference solution, $\left\langle \mathcal{I}\right\rangle _{\mathrm{ave}}\approx9.64\times10^{-5}$,
primarily due to the increased energy-relaxation contribution $\left\langle J_{2,T_{1}}\right\rangle =9\times10^{-5}$
associated with the longer gate time $t_{f}$, as shown in Figs.\ \ref{fig:overall_infidelity_performance}(b)
and \ref{fig:overall_infidelity_performance}(c).

At Stage 3, the optimal solution with $t_{f}=28\,\mathrm{ns}$ reduces
the total infidelity to $\left\langle \mathcal{I}\right\rangle _{\mathrm{ave}}\approx2.22\times10^{-5}$,
while all $\left\langle \mathcal{I}\right\rangle $ samples remain
below $6\times10^{-5}$. This represents a substantial improvement
over the reference solution, which yields $\left\langle \mathcal{I}\right\rangle _{\mathrm{ave}}\approx6.11\times10^{-5}$
and a maximum $\left\langle \mathcal{I}\right\rangle $ exceeding
$2\times10^{-4}$, as shown in Figs.\ \ref{fig:overall_infidelity_performance}(a),
\ref{fig:overall_infidelity_performance}(b), and \ref{fig:overall_infidelity_performance}(c).
The relaxed noise parameters at Stage 3 ($\sigma_{\epsilon}=0.01\sigma_{\epsilon}^{0}$,
$\sigma_{\phi,\mathrm{eff}}=0.01\sigma_{\phi,\mathrm{eff}}^{0}$,
and $T_{1}=3\,\mathrm{ms}$), compared with those at Stage 2 ($\sigma_{\epsilon}=0.1\sigma_{\epsilon}^{0}$,
$\sigma_{\phi,\mathrm{eff}}=0.1\sigma_{\phi,\mathrm{eff}}^{0}$, and
$T_{1}=300\,\mu\mathrm{s}$), reduce the corresponding noise contributions
$\left\langle J_{2,\epsilon}\right\rangle $, $\left\langle J_{2,\phi}\right\rangle $,
and $\left\langle J_{2,T_{1}}\right\rangle $. This provides additional
error budget to accommodate the resulting increases in these contributions
as $J_{1}$ and $\left\langle J_{2,g_{AB}}\right\rangle +\left\langle J_{2,\delta}\right\rangle $
are further suppressed. Consequently, at Stage 3 the optimized $J_{1}$
is reduced to $6.46\times10^{-6}$, compared with $1.59\times10^{-5}$
at Stage 2; both values are lower than $J_{1}\approx2.46\times10^{-5}$
for the reference solution. The optimized $\left\langle J_{2,g_{AB}}\right\rangle $
at Stage 3 is reduced to $9.97\times10^{-8}$, more than one order
of magnitude below the Stage 2 value of $1.23\times10^{-6}$, and
more than two orders of magnitude below the value of $2.57\times10^{-5}$
for the reference solution. By contrast, the optimized $\left\langle J_{2,\delta}\right\rangle $
exhibits only a modest improvement from Stage 2 to Stage 3, decreasing
from $6.09\times10^{-6}$ to $5.46\times10^{-6}$. This saturation
indicates that $\left\langle J_{2,\delta}\right\rangle $ has been
suppressed close to the limit achievable under the present control
settings while minimizing the total uncertainty contribution $\left\langle J_{2,g_{AB}}\right\rangle +\left\langle J_{2,\delta}\right\rangle $.
Nevertheless, the optimized $\left\langle J_{2,\delta}\right\rangle \approx5.46\times10^{-6}$
is approximately $40\%$ lower than the value of $9.16\times10^{-6}$
for the reference solution. Collectively, these improvements account
for the enhanced $\left\langle \mathcal{I}\right\rangle $ performance
at Stage 3, where the $\left\langle \mathcal{I}\right\rangle $ distribution
is dominated by $\left\langle J_{2,\delta}\right\rangle $.

\subsubsection{Stages 4 and 5}

At Stage 4, the diagonal uncertainty standard deviation $\sigma_{\delta}$
is relaxed to $2\pi\times0.01\,\mathrm{MHz}$, while the remaining
parameters, $\sigma_{g_{AB}}$, $T_{1}$, $\sigma_{\epsilon}$, and
$\sigma_{\phi,\mathrm{eff}}$, are kept identical to those in Stage
3. Under these conditions, the optimization can focus on suppressing
the dominant error contributions---the intrinsic error $J_{1}$ and
the coupling error $\left\langle J_{2,g_{AB}}\right\rangle $---using
the control strategy introduced in Sec.\ \ref{subsec:uty_robust_against_gAB_uncertainty}.
Specifically, we implement the rp-Qt-4 optimization with the cost
function $J_{1}+\left\langle J_{2,g_{AB}}\right\rangle $. Owing to
the reduced complexity of the optimization, the number of control
parameters in each Fine-Tune-Pulse $\Phi_{A/B}^{\mathrm{FTP}}(t)$
is reduced to $20$. Instead of using the initial value $\delta_{A}/2\pi=-400\,\mathrm{MHz}$
adopted in Stage 3, we set the initial $\delta_{A}/2\pi$ to $0\,\mathrm{MHz}$
at Stage 4 because this setting results in a larger fraction of optimization
runs converging to solutions with simultaneously lower values of $J_{1}$,
$\left\langle J_{2,g_{AB}}\right\rangle $, and $\left\langle J_{2,\delta}\right\rangle $.
Moreover, compared with the optimal initial $\omega_{i,A}/2\pi=2\,\mathrm{GHz}$
obtained in Stage 3, a higher initial $\omega_{i,A}$ is required
at Stage 4. This is because, for the same initial $\omega_{i,A}$,
the initial $\omega_{i,B}$ is reduced by an additional $2\pi\times400\,\mathrm{MHz}$
relative to Stage 3, according to $\omega_{i,B}=\omega_{i,A}+\alpha_{A}-\delta_{A}$,
with $\delta_{A}=0\,\mathrm{MHz}$ in Stage 4 and $\delta_{A}=-2\pi\times400\,\mathrm{MHz}$
in Stage 3. A lower initial $\omega_{i,B}$ increases the number of
accessible leakage channels and exacerbates the synchronization burden,
thereby increasing the leakage error. To mitigate this effect, we
increase the initial frequency $\omega_{i,A}$ and identify $\omega_{i,A}/2\pi=3\,\mathrm{GHz}$
as the optimal initial value for Stage 4. Under the above settings,
we perform the optimization and analyze the total infidelity $J_{1}+\left\langle J_{2}\right\rangle $
together with the uncertainty contributions $\left\langle J_{2,g_{AB}}\right\rangle +\left\langle J_{2,\delta}\right\rangle $
to identify the optimal Stage-4 solution with $t_{f}=24\,\mathrm{ns}$.
The optimal solution reduces the average infidelity to $\left\langle \mathcal{I}\right\rangle _{\mathrm{ave}}\approx1.63\times10^{-5}$,
which is dominated by the optimized $J_{1}\approx7.63\times10^{-6}$
and $\left\langle J_{2,T_{1}}\right\rangle =8\times10^{-6}$. This
represents a substantial improvement over the reference solution,
which yields $\left\langle \mathcal{I}\right\rangle _{\mathrm{ave}}\approx5.17\times10^{-5}$.
Furthermore, the optimized uncertainty contributions, $\left\langle J_{2,g_{AB}}\right\rangle \approx2.84\times10^{-8}$
and $\left\langle J_{2,\delta}\right\rangle \approx2.98\times10^{-7}$,
are suppressed to more than an order of magnitude below $\left\langle \mathcal{I}\right\rangle _{\mathrm{ave}}\approx1.63\times10^{-5}$.
As a result, the $\left\langle \mathcal{I}\right\rangle $ distribution
is significantly narrowed, with all values of $\left\langle \mathcal{I}\right\rangle $
below $2\times10^{-5}$, which is even lower than the minimum value
of $\left\langle \mathcal{I}\right\rangle $ for the reference solution,
as shown in Fig.\ \ref{fig:overall_infidelity_performance}(a).

At Stage 5, the off-diagonal uncertainty standard deviation $\sigma_{g_{AB}}$
is relaxed to $2\pi\times0.01\,\mathrm{MHz}$, while the other parameters
$\sigma_{\delta}$, $T_{1}$, $\sigma_{\epsilon}$, and $\sigma_{\phi,\mathrm{eff}}$
remain the same as in Stage 3. Consequently, the detuning error $\left\langle J_{2,\delta}\right\rangle $
and the intrinsic error $J_{1}$ dominate the total infidelity. To
suppress these errors, we apply the rp-Qt-4 optimization using the
cost function $J_{1}+\left\langle J_{2,\delta}\right\rangle $. The
initial detuning is chosen as $\delta_{A}/2\pi=-214\,\mathrm{MHz}$,
corresponding to the symmetric detuning point, which effectively reduces
the detuning error in accordance with the control strategy described
in Sec.\ \ref{subsec:uty_robust_against_detuning_uncertainty}. At
Stage 5, the optimal initial $\omega_{i,A}$ coincides with that of
Stage 3, taking the value $\omega_{i,A}/2\pi=2\,\mathrm{GHz}$. This
agreement can be attributed to the comparable minimum values of the
qubit frequency $\Delta_{B}(t)$ achieved in the two stages. Stage
5 uses a larger initial detuning, $\delta_{A}/2\pi=-214\,\mathrm{MHz}$,
than Stage 3, where $\delta_{A}/2\pi=-400\,\mathrm{MHz}$. Consequently,
the initial $\omega_{i,B}=\omega_{i,A}+\alpha_{A}-\delta_{A}$ is
lower in Stage 5. However, Stage 5 employs weaker $\delta_{A}(t)$
modulation to keep the system near the symmetric detuning point and
thus reduce the detuning error, whereas Stage 3 uses stronger $\delta_{A}(t)$
modulation to suppress the coupling error. Thus, despite the different
initial $\delta_{A}$ settings, the two stages still yield comparable
minimum values of $\Delta_{B}(t)$. In addition, we employ the same
Fine-Tune-Pulse $\Phi_{A/B}^{\mathrm{FTP}}(t)$ setting used in Stage
4, comprising 20 control parameters per pulse. Under the above settings,
we perform the optimization and then identify the optimal Stage-5
solution with $t_{f}=24\,\mathrm{ns}$ by jointly considering the
total infidelity $J_{1}+\left\langle J_{2}\right\rangle $ and the
uncertainty contributions $\left\langle J_{2,g_{AB}}\right\rangle +\left\langle J_{2,\delta}\right\rangle $.
This optimal solution yields a substantial improvement, with $J_{1}\approx3.92\times10^{-6}$
and $\left\langle J_{2,\delta}\right\rangle \approx7.73\times10^{-7}$,
compared with the reference values $J_{1}\approx2.46\times10^{-5}$
and $\left\langle J_{2,\delta}\right\rangle \approx9.16\times10^{-6}$.
As a result, the total infidelity is reduced to $\left\langle \mathcal{I}\right\rangle _{\mathrm{ave}}\approx1.40\times10^{-5}$
from $\left\langle \mathcal{I}\right\rangle _{\mathrm{ave}}\approx3.81\times10^{-5}$
for the reference solution, despite an increased energy-relaxation
contribution $\left\langle J_{2,T_{1}}\right\rangle =8\times10^{-6}$
associated with the longer gate time $t_{f}$. The combination of
the reduced $\left\langle \mathcal{I}\right\rangle _{\mathrm{ave}}$
and suppressed uncertainty contributions, $\left\langle J_{2,g_{AB}}\right\rangle +\left\langle J_{2,\delta}\right\rangle \approx1.62\times10^{-6}$,
shifts the entire $\left\langle \mathcal{I}\right\rangle $ distribution
downward relative to that of the reference solution, for which $\left\langle J_{2,g_{AB}}\right\rangle +\left\langle J_{2,\delta}\right\rangle \approx9.42\times10^{-6}$,
as illustrated by the box plots in Fig.\ \ref{fig:overall_infidelity_performance}(a).

\subsubsection{Stage 6 and summary}

At Stage 6, the uncertainty standard deviations $\sigma_{g_{AB}}$
and $\sigma_{\delta}$ are both relaxed to $2\pi\times0.01\,\mathrm{MHz}$,
while the noise parameters $T_{1}$, $\sigma_{\epsilon}$, and $\sigma_{\phi,\mathrm{eff}}$
are kept identical to those in Stage 3. Under these conditions, the
effects of noise and uncertainty are substantially mitigated, and
the intrinsic error $J_{1}$ becomes the dominant contribution to
the total infidelity $\left\langle \mathcal{I}\right\rangle _{\mathrm{ave}}$.
Consequently, the Stage-6 optimization primarily focuses on suppressing
$J_{1}$. However, to prevent excessive dispersion in the distribution
of $\left\langle \mathcal{I}\right\rangle $ as $J_{1}$ is reduced,
the uncertainty contributions $\left\langle J_{2,g_{AB}}\right\rangle +\left\langle J_{2,\delta}\right\rangle $
are also included in the optimization. We adopt the same optimization
settings and procedures as in Stage 3. Specifically, we perform the
rp-Qt-4 optimization with initial values of $\delta_{A}=-2\pi\times400\,\mathrm{MHz}$
and $\omega_{i,A}/2\pi=2\,\mathrm{GHz}$, employing the cost function
$J_{1}+\left\langle J_{2,g_{AB}}\right\rangle +\left\langle J_{2,\delta}\right\rangle $.
The optimal Stage-6 solution with $t_{f}=12\,\mathrm{ns}$ is then
identified.

Next, we perform an additional fine-tuning optimization to further
suppress $J_{1}$ for the optimal solution. The optimized control
parameters of this solution are used as the initial values, and the
cost function consists solely of $J_{1}$. Here, the simulation is
extended to five energy levels per qubit, beyond the subspace spanned
by the Qt-0 to Qt-4 states used in the rp-Qt-4 optimization, to evaluate
$J_{1}$ more accurately by incorporating additional leakage channels
and thereby facilitating further suppression of leakage errors. Moreover,
as $J_{1}$ is further reduced, the fidelity degradation arising from
the filtering of high-frequency pulse components becomes non-negligible
for the short gate time $t_{f}=12\,\mathrm{ns}$ and should be taken
into account. To capture this effect, filtered control pulses are
generated using a transfer function corresponding to a Gaussian filter
$F(\omega)$ with cutoff frequency $\omega_{c}/2\pi=25.5\,\mathrm{GHz}$
when evaluating $J_{1}$ during the fine-tuning optimization, following
the procedure introduced in Sec.\ %
\mbox{%
II%
} E of Ref.\ \citep{Huang2026}. As a result, the optimized intrinsic
error is reduced to $J_{1}\approx1.93\times10^{-8}$, with the dominant
contributions arising from Qt-3 and Qt-4 leakage errors, which are
$1.11\times10^{-8}$ and $5.67\times10^{-9}$, respectively. The fine-tuning
optimization has a negligible impact on the other error contributions
$\left\langle J_{2,\epsilon}\right\rangle $, $\left\langle J_{2,\phi}\right\rangle $,
$\left\langle J_{2,g_{AB}}\right\rangle $, and $\left\langle J_{2,\delta}\right\rangle $,
because the control pulses are only slightly modified.

\begin{figure}
\includegraphics[scale=0.52]{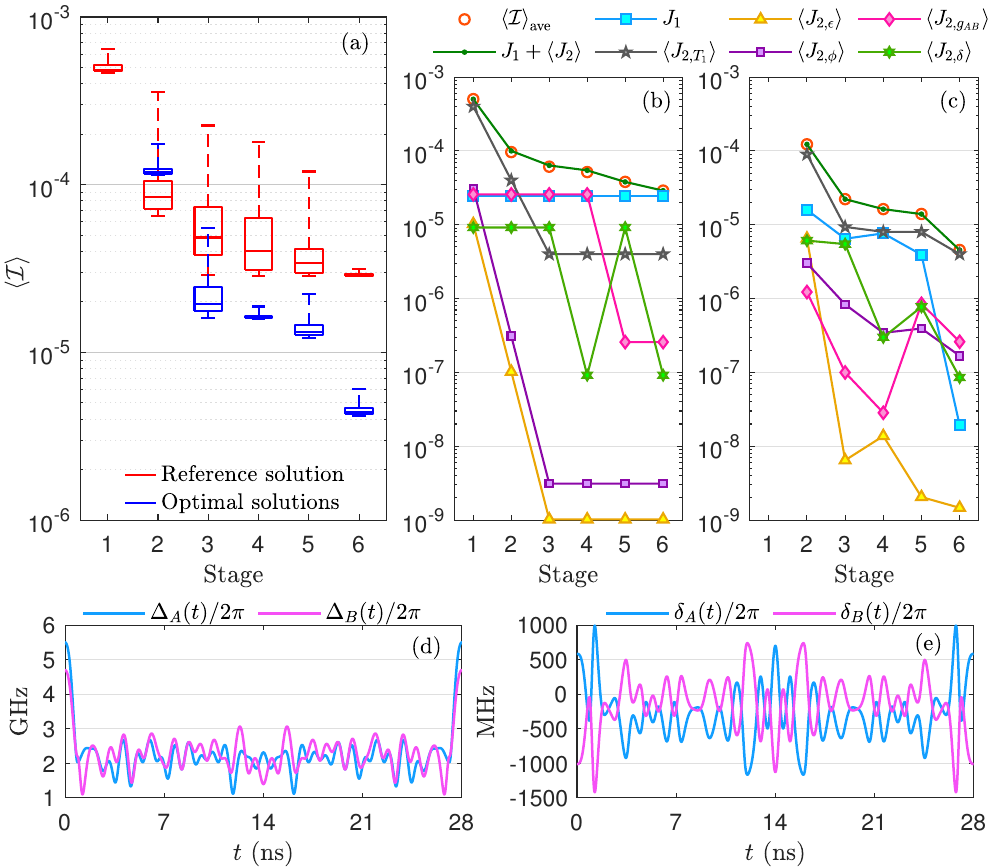}

\caption{CZ-gate performance from Stage 1 to Stage 6 for the reference solution
and the stage-specific optimal solutions. (a) Distributions of the
infidelity $\left\langle \mathcal{I}\right\rangle $ across the stages,
shown as box plots for the reference solution (red) and the optimal
solutions (blue). In each box plot, the whiskers indicate the minimum
and maximum values, the box spans the 25th to 75th percentiles, and
the central line denotes the median. (b,c) Individual error contributions
for the reference solution (b) and the optimal solutions (c). The
total infidelity $\left\langle \mathcal{I}\right\rangle _{\mathrm{ave}}$,
obtained from full Hamiltonian simulations, is indicated by orange
circles, while the total infidelity $J_{1}+\left\langle J_{2}\right\rangle $,
obtained by summing the individual error contributions, is represented
by green dotted lines, where $\left\langle J_{2}\right\rangle =\left\langle J_{2,T_{1}}\right\rangle +\left\langle J_{2,\epsilon}\right\rangle +\left\langle J_{2,\phi}\right\rangle +\left\langle J_{2,g_{AB}}\right\rangle +\left\langle J_{2,\delta}\right\rangle $.
The intrinsic error $J_{1}$, energy-relaxation contribution $\left\langle J_{2,T_{1}}\right\rangle $,
RTCE-noise contribution $\left\langle J_{2,\epsilon}\right\rangle $,
flux-noise contribution $\left\langle J_{2,\phi}\right\rangle $,
coupling error $\left\langle J_{2,g_{AB}}\right\rangle $, and detuning
error $\left\langle J_{2,\delta}\right\rangle $ are shown by blue-square,
gray-pentagram, yellow-triangle, purple-square, pink-diamond, and
green-hexagram lines, respectively. (d,e) Optimized qubit frequencies
$\Delta_{A}(t)$ (blue) and $\Delta_{B}(t)$ (pink), and the corresponding
optimized detunings $\delta_{A}(t)$ (blue) and $\delta_{B}(t)$ (pink),
respectively, for the Stage-3 optimal solution.}

\label{fig:overall_infidelity_performance} 
\end{figure}

Finally, for the optimal Stage-6 solution after the additional fine-tuning
procedure, the total infidelity is reduced to $\left\langle \mathcal{I}\right\rangle _{\mathrm{ave}}\approx4.55\times10^{-6}$,
approaching the energy-relaxation-limited value $\left\langle J_{2,T_{1}}\right\rangle =4\times10^{-6}$.
All other contributions, $J_{1}$, $\left\langle J_{2,\epsilon}\right\rangle $,
$\left\langle J_{2,\phi}\right\rangle $, $\left\langle J_{2,g_{AB}}\right\rangle $,
and $\left\langle J_{2,\delta}\right\rangle $, are more than one
order of magnitude smaller than $\left\langle J_{2,T_{1}}\right\rangle $,
as shown in Fig.\ \ref{fig:overall_infidelity_performance}(c). By
contrast, the reference value $\left\langle \mathcal{I}\right\rangle _{\mathrm{ave}}\approx2.90\times10^{-5}$
is dominated by $J_{1}$, as shown in Fig.\ \ref{fig:overall_infidelity_performance}(b).
In addition, the reduced total infidelity $\left\langle \mathcal{I}\right\rangle _{\mathrm{ave}}\approx4.55\times10^{-6}$,
together with the suppressed uncertainty contributions $\left\langle J_{2,g_{AB}}\right\rangle +\left\langle J_{2,\delta}\right\rangle \approx3.45\times10^{-7}$,
ensures that all $\left\langle \mathcal{I}\right\rangle $ samples
remain below $\sim6\times10^{-6}$, which is significantly smaller
than the minimum $\left\langle \mathcal{I}\right\rangle \approx2.85\times10^{-5}$
of the reference solution, as shown in Fig.\ \ref{fig:overall_infidelity_performance}(a).

In summary, we have successfully reduced the CZ-gate infidelity and
enhanced its robustness across the stages defined in Table\ \ref{Table:noise_uty_strength_at_stages}
using the optimal solutions presented here, as demonstrated in Fig.\ \ref{fig:overall_infidelity_performance}(a).
This improvement stems from a detailed understanding of the underlying
error mechanisms, precise error characterization, and effective error-suppression
strategies, building on our previous analyses of intrinsic errors
in Sec.\ %
\mbox{%
II%
} of Ref.\ \citep{Huang2026}, noise effects in Sec.\ %
\mbox{%
III%
} of Ref.\ \citep{Huang2026}, and uncertainty contributions in Sec.\ \ref{sec:robust_against_uncertainty}.
On this basis, we design stage-specific control strategies that target
the dominant error sources and obtain optimal solutions for each stage.
The proposed control framework is flexible and comprehensive, extending
beyond the stages listed in Table\ \ref{Table:noise_uty_strength_at_stages}.
It can be readily adapted to identify optimal solutions for arbitrary
realistic noise and uncertainty conditions.

\begin{table}
\caption{Noise parameters $T_{1}$, $\sigma_{\epsilon}$, and $\sigma_{\phi,\mathrm{eff}}$,
and uncertainty levels $\sigma_{g_{AB}}$ and $\sigma_{\delta}$ for
Stages 1--6.}

\%begin{tabular}{c>{\centering}p{0.8cm}>{\centering}p{0.8cm}>{\centering}p{0.8cm}>{\centering}p{0.8cm}>{\centering}p{0.8cm}>{\centering}p{0.8cm}}
\begin{tabular}{c>{\centering\arraybackslash}p{0.8cm}>{\centering\arraybackslash}p{0.8cm}>{\centering\arraybackslash}p{0.8cm}>{\centering\arraybackslash}p{0.8cm}>{\centering\arraybackslash}p{0.8cm}>{\centering\arraybackslash}p{0.8cm}}
\toprule 
{\footnotesize{}Stage}  & {\footnotesize{}$1$}  & {\footnotesize{}$2$}  & {\footnotesize{}$3$}  & {\footnotesize{}$4$}  & {\footnotesize{}$5$}  & {\footnotesize{}$6$}\tabularnewline
\midrule 
{\footnotesize{}$T_{1}$ ($\mu s$)}  & {\footnotesize{}$30$}  & {\footnotesize{}$300$}  & {\footnotesize{}$3000$}  & {\footnotesize{}$3000$}  & {\footnotesize{}$3000$}  & {\footnotesize{}$3000$}\tabularnewline
{\footnotesize{}$\sigma_{\epsilon}/\sigma_{\epsilon}^{0}$}  & {\footnotesize{}$1$}  & {\footnotesize{}$0.1$}  & {\footnotesize{}$0.01$}  & {\footnotesize{}$0.01$}  & {\footnotesize{}$0.01$}  & {\footnotesize{}$0.01$}\tabularnewline
{\footnotesize{}$\sigma_{\phi,\mathrm{eff}}/\sigma_{\phi,\mathrm{eff}}^{0}$}  & {\footnotesize{}$1$}  & {\footnotesize{}$0.1$}  & {\footnotesize{}$0.01$}  & {\footnotesize{}$0.01$}  & {\footnotesize{}$0.01$}  & {\footnotesize{}$0.01$}\tabularnewline
{\footnotesize{}$\sigma_{g_{AB}}/2\pi$ (MHz)}  & {\footnotesize{}$0.1$}  & {\footnotesize{}$0.1$}  & {\footnotesize{}$0.1$}  & {\footnotesize{}$0.1$}  & {\footnotesize{}$0.01$}  & {\footnotesize{}$0.01$}\tabularnewline
{\footnotesize{}$\sigma_{\delta}/2\pi$ (MHz)}  & {\footnotesize{}$0.1$}  & {\footnotesize{}$0.1$}  & {\footnotesize{}$0.1$}  & {\footnotesize{}$0.01$}  & {\footnotesize{}$0.1$}  & {\footnotesize{}$0.01$}\tabularnewline
\bottomrule
\end{tabular}

\label{Table:noise_uty_strength_at_stages} 
\end{table}

\section{Conclusion}

\label{sec:Conclusion}

We have extended the optimal-control framework developed in the companion
paper (Part $\mathrm{I}$) \citep{Huang2026} to account for uncertainties
in the system parameters of directly coupled, frequency-tunable transmons.
By incorporating deviations from the characterized values of the coupling
strength, anharmonicities, and zero-flux qubit frequencies, the present
work provides a unified treatment of intrinsic errors, decoherence,
and parameter uncertainties in CZ-gate design. This extension is important
for large-scale quantum hardware, where device-to-device inhomogeneity,
finite measurement precision, environmental fluctuations, and temporal
drift make exhaustive parameter characterization and frequent gate
recalibration increasingly challenging.

Robust control addresses this challenge by engineering gate operations
that are intrinsically less sensitive to parameter uncertainties and
noise fluctuations. Rather than optimizing a pulse only at a precisely
characterized nominal point, our approach constructs control trajectories
that preserve high fidelity over experimentally relevant ranges of
parameter deviations and noise strengths. This intrinsic robustness
relaxes the need for precise characterization and frequent recalibration,
allowing high-fidelity operation to be maintained despite deviations
from nominal operating conditions. By combining the noise-robust control
developed in Part $\mathrm{I}$ with the parameter-robust control
developed here, we obtain a CZ gate that is robust to both noise and
parameter uncertainties, thereby substantially reducing calibration
overhead and facilitating the practical operation of large-scale quantum
hardware.

Our analysis reveals that parameter uncertainties naturally decompose
into off-diagonal and diagonal uncertainties, which predominantly
give rise to coupling and detuning errors, respectively. Each error
class requires a distinct control strategy. For off-diagonal uncertainty,
specifically uncertainty in the coupling strength $g_{AB}$, strong
modulation of the controllable detuning produces a dynamical-decoupling
effect that suppresses sensitivity to the transverse coupling perturbation.
At a gate duration of $30\,\mathrm{ns}$, this strategy reduces the
coupling error to $6.21\times10^{-7}$, compared with $3.54\times10^{-4}$
for a conventional pulse without the designed modulation. The optimized
gate maintains a total infidelity below $10^{-3}$ for coupling-strength
deviations as large as $\pm2\pi\times4\,\mathrm{MHz}$. Consequently,
high-fidelity operation does not depend on determining and stabilizing
the coupling strength at a single narrowly defined value.

Diagonal uncertainties, such as uncertainties in anharmonicities ($\alpha_{A/B}$)
and zero-flux qubit frequencies ($\omega_{0,A/B}$), cannot be suppressed
by the same mechanism because they enter through the same Hamiltonian
channels as the controls themselves. Instead, their effects are minimized
near the symmetric-detuning point, $\delta_{A}\approx\delta_{B}$,
where the error contributions from the $\left|1_{B}1_{A}\right\rangle $$\leftrightarrow$$\left|0_{B}2_{A}\right\rangle $
and $\left|1_{B}1_{A}\right\rangle $$\leftrightarrow$$\left|2_{B}0_{A}\right\rangle $
interactions are balanced. This operating regime supports individual
diagonal-parameter deviations of up to $\pm2\pi\times5\,\mathrm{MHz}$
for a $24\text{-ns}$ gate while keeping the total infidelity below
$10^{-3}$. The symmetric-detuning point therefore provides a complementary
mechanism for reducing sensitivity to diagonal uncertainties. 

By incorporating intrinsic dynamics, energy relaxation, dephasing,
and parameter uncertainty, our framework identifies the optimal control
strategy for each of the six representative stages of progressively
improving hardware performance, based on the dominant error contributions.
Under present-day conditions, energy relaxation dominates, favoring
a near-speed-limit $12\text{-ns}$ gate, while extending the gate
duration to improve robustness provides little net benefit. As energy-relaxation
times increase and dephasing noise strengths are reduced, however,
parameter uncertainties become comparatively more important, and robust
solutions with gate durations of $24\text{--}28\,\mathrm{ns}$ yield
lower overall errors. In the most advanced stage considered, where
decoherence and parameter uncertainty are both reduced, fine-tuning
with a five-level transmon model and Gaussian waveform filtering reduces
the intrinsic error to $1.93\times10^{-8}$ and yields a total infidelity
of $4.55\times10^{-6}$, with energy relaxation remaining the principal
limiting factor. These stage-specific optimal solutions balance intrinsic
error, exposure to energy relaxation, sensitivity to dephasing noise,
and sensitivity to parameter uncertainties, thereby achieving a favorable
trade-off between infidelity and robustness for each operating stage.
These results demonstrate that our control framework can be readily
adapted to different operating conditions and provides practical and
effective design strategies for realizing robust, high-fidelity CZ
gates in superconducting quantum processors.

Together, Parts $\mathrm{I}$ and $\mathrm{II}$ provide a modular
framework for designing fast, high-fidelity, and robust CZ gates under
realistic experimental conditions. Part $\mathrm{I}$ addresses intrinsic
and decoherence-induced errors and establishes robustness against
noise fluctuations, whereas the present study (Part $\mathrm{II}$)
develops complementary strategies to suppress errors arising from
uncertain system parameters and enhance robustness against such uncertainties.
The resulting gate operations remain accurate over finite ranges of
both noise and parameter variations, reducing reliance on exhaustive
characterization and frequent recalibration. The control strategies
considered here involve only time-dependent, tunable magnetic flux
controls on qubits. However, the framework can be readily extended
to architectures with tunable coupling, where treating $g_{AB}(t)$
as an additional control may further reduce gate duration and leakage,
as well as to schemes incorporating microwave-driven transverse controls
for the active suppression of diagonal errors. The framework, through
the error decomposition and stage-specific optimization developed
here, provides experimentally actionable guidance for adapting two-qubit
control as superconducting hardware advances toward large-scale fault-tolerant
quantum computing.
\begin{acknowledgments}
H.-S. Goan acknowledges support from the NSTC, Taiwan, under Grants
No. NSTC 113-2112-M-002-022-MY3, No. NSTC 113-2119M-002-021, No. NSTC
114-2119-M-002-018, No. NSTC 114-2119-M-002-017-MY3, and from the
National Taiwan University under Grants No. NTU-CC-115L8937, No. NTUCC115L893704
and No. NTU-CC-115L8512. H.-S. Goan is also grateful for the support
of the ``Center for Advanced Computing and Imaging in Biomedicine
(NTU-115L900702)'' through the Featured Areas Research Center Program
within the framework of the Higher Education Sprout Project by the
Ministry of Education (MOE), Taiwan, the support of Taiwan Semiconductor
Research Institute (TSRI) through the Joint Developed Project (JDP)
and the support from the Physics Division, National Center for Theoretical
Sciences, Taiwan. 
\end{acknowledgments}

\appendix

\section{Estimated detuning and coupling errors}

\label{appendix:estimated_uncertainty_error}

In this appendix, we derive the estimated detuning and coupling errors,
$J_{2,\delta_{A}}^{\mathrm{Est}}$ and $J_{2,g_{AB}}^{\mathrm{Est}}$,
as shown in Eqs.\ (\ref{eq:J2_dA_est}) and (\ref{eq:J2_g_est})
of the main text, respectively. As discussed in the main text, these
errors primarily originate from the uncertainties in the system parameters
associated with the ideal Hamiltonians $\mathcal{H}_{I_{12}}$ and
$\mathcal{H}_{I_{22}}$, defined in the subspaces $\left\{ \left|0_{B}1_{A}\right\rangle ,\left|1_{B}0_{A}\right\rangle \right\} $
and $\left\{ \left|1_{B}1_{A}\right\rangle ,\left|0_{B}2_{A}\right\rangle ,\left|2_{B}0_{A}\right\rangle \right\} $,
respectively, as given in Eqs.\ (5) and (7) of Ref.\ \citep{Huang2026}.
Under the strong $\left|1_{B}1_{A}\right\rangle $$\leftrightarrow$$\left|0_{B}2_{A}\right\rangle $
coupling condition, where $\left|\delta_{A}\right|\ll\left|\delta_{01-10}\right|$
and $\left|\delta_{A}\right|\ll\left|\delta_{B}\right|$, the sensitivity
to system-parameter uncertainties in the subspace $\left\{ \left|1_{B}1_{A}\right\rangle ,\left|0_{B}2_{A}\right\rangle \right\} $
is significantly greater than that in the remaining subspaces. We
therefore consider the ideal Hamiltonian $\mathcal{H}_{I_{22}}$ projected
onto this effective two-level subspace,

\begin{equation}
\mathcal{H}_{I_{22}}^{2\times2}/\hbar=\delta_{A}\left(\frac{1}{2}I-\frac{1}{2}\sigma_{Z}\right)+\sqrt{2}g_{AB}\sigma_{X},\label{eq:ideal_H_2x2}
\end{equation}
where $\sigma_{Z}$ and $\sigma_{X}$ denote the Pauli matrices, and
$I$ is the identity matrix. The corresponding ideal propagator is
\begin{equation}
U_{I}(t)=\cos\theta\cdot I-i\sin\theta\cdot\left(\dfrac{\sqrt{2}g_{AB}\sigma_{X}-\dfrac{1}{2}\delta_{A}\sigma_{Z}}{\sqrt{2g_{AB}^{2}+\delta_{A}^{2}/4}}\right)\label{eq:ideal_U_2x2}
\end{equation}
with $\theta=\sqrt{2g_{AB}^{2}+\delta_{A}^{2}/4}\cdot t$.

Next, we substitute the parameters $\delta_{A}$ and $g_{AB}$ in
the ideal Hamiltonian $\mathcal{H}_{I_{22}}^{2\times2}$ of Eq.\ (\ref{eq:ideal_H_2x2})
with their perturbed values $\delta_{A}+\beta_{\delta_{A}}$ and $g_{AB}+\beta_{g_{AB}}$,
respectively, where $\beta_{\delta_{A}}$ and $\beta_{g_{AB}}$ denote
the corresponding parameter uncertainties. The resulting realistic
Hamiltonian is written as $\mathcal{H}=\mathcal{H}_{I_{22}}^{2\times2}+\mathcal{H}_{N,\delta_{A}}+\mathcal{H}_{N,g_{AB}}$,
where $\mathcal{H}_{N,\delta_{A}}$ and $\mathcal{H}_{N,g_{AB}}$
are the uncertainty Hamiltonians arising from $\beta_{\delta_{A}}$
and $\beta_{g_{AB}}$, respectively, which are given by

\begin{align}
\mathcal{H}_{N,\delta_{A}}/\hbar & =\beta_{\delta_{A}}\left(\frac{1}{2}I-\frac{1}{2}\sigma_{Z}\right),\label{eq:uty_H_detuning_2x2}\\
\mathcal{H}_{N,g_{AB}}/\hbar & =\beta_{g_{AB}}\sqrt{2}\sigma_{X}.\label{eq:uty_H_coupling_2x2}
\end{align}

We then employ the error contribution formula $J_{2,j}$ in Eq.\ (\ref{eq:J2_uty_general_format})
of the main text to evaluate the detuning and coupling errors. Since
nonzero Hamiltonian matrix elements exist only within the subspace
$\left\{ \left|1_{B}1_{A}\right\rangle ,\left|0_{B}2_{A}\right\rangle \right\} $,
the expression for $J_{2,j}$ can be rewritten as

\begin{align}
J_{2,j}= & \beta_{j}^{2}\int_{0}^{t_{f}}dt_{1}\int_{0}^{t_{1}}dt_{2}\nonumber \\
 & \cdot\left[\mathrm{Re}\left\{ \mathrm{Tr}\left[\left(R_{j}(t_{1})R_{j}(t_{2})\right)_{1\times1}\right]\right\} /2-\right.\nonumber \\
 & \left.\mathrm{Tr}\left(R_{j,1\times1}(t_{1})\right)\mathrm{Tr}\left(R_{j,1\times1}(t_{2})\right)/8\right],\label{eq:J2_uty_11_02_subspace}
\end{align}
where $[R_{j}(t_{1})R_{j}(t_{2})]_{1\times1}$ and $R_{j,1\times1}(t)$
denote the $1\times1$ matrices obtained by projecting $R_{j}(t_{1})R_{j}(t_{2})$
and $R_{j}(t)$, respectively, onto the subspace $\left\{ \left|1_{B}1_{A}\right\rangle \right\} $.
The matrix $R_{j}(t)=U_{I}^{\dagger}(t)H_{j}(t)U_{I}(t)$ is Hermitian
and can therefore be expressed in the basis $\left\{ \left|1_{B}1_{A}\right\rangle ,\left|0_{B}2_{A}\right\rangle \right\} $
as

\begin{equation}
R_{j}(t)=\left(\begin{array}{cc}
R_{j,00}(t) & R_{j,01}(t)\\
R_{j,01}^{\star}(t) & R_{j,11}(t)
\end{array}\right),\label{eq:R_2x2_matrix}
\end{equation}
where the diagonal elements $R_{j,00}(t)$ and $R_{j,11}(t)$ are
real, and the off-diagonal element $R_{j,01}(t)$ is complex, $R_{j,01}(t)\equiv R_{j,01}^{\mathrm{Re}}(t)+iR_{j,01}^{\mathrm{Im}}(t)$,
with $R_{j,01}^{\mathrm{Re}}(t)$ and $R_{j,01}^{\mathrm{Im}}(t)$
real. By substituting the matrix $R_{j}(t)$ from Eq.\ (\ref{eq:R_2x2_matrix})
into Eq.\ (\ref{eq:J2_uty_11_02_subspace}), the trace terms in $J_{2,j}$
become 
\begin{align}
 & \mathrm{Re}\left\{ \mathrm{Tr}\left[\left(R_{j}(t_{1})R_{j}(t_{2})\right)_{1\times1}\right]\right\} =R_{j,00}(t_{1})R_{j,00}(t_{2})\nonumber \\
 & +R_{j,01}^{\mathrm{Re}}(t_{1})R_{j,01}^{\mathrm{Re}}(t_{2})+R_{j,01}^{\mathrm{Im}}(t_{1})R_{j,01}^{\mathrm{Im}}(t_{2}),\label{eq:tr_R_R_matrix}
\end{align}
and 
\begin{equation}
\mathrm{Tr}\left[R_{j,1\times1}(t)\right]=R_{j,00}(t).\label{eq:tr_R_matrix}
\end{equation}
Therefore, $J_{2,j}$ can be decomposed as

\begin{equation}
J_{2,j}=J_{2,j}^{00}+J_{2,j}^{01,\mathrm{Re}}+J_{2,j}^{01,\mathrm{Im}},\label{eq:J2_uty_11_02_sum}
\end{equation}
where

\begin{align}
J_{2,j}^{00} & =\dfrac{3}{16}\beta_{j}^{2}\left[\int_{0}^{t_{f}}R_{j,00}(t)dt\right]^{2},\label{eq:J2_uty_11_02_sum_00}\\
J_{2,j}^{01,\mathrm{Re}} & =\dfrac{1}{4}\beta_{j}^{2}\left[\int_{0}^{t_{f}}R_{j,01}^{\mathrm{Re}}(t)dt\right]^{2},\label{eq:J2_uty_11_02_sum_01Re}\\
J_{2,j}^{01,\mathrm{Im}} & =\dfrac{1}{4}\beta_{j}^{2}\left[\int_{0}^{t_{f}}R_{j,01}^{\mathrm{Im}}(t)dt\right]^{2}.\label{eq:J2_uty_11_02_sum_01Im}
\end{align}

\subsection{Detuning error $J_{2,\delta_{A}}^{\mathrm{Est}}$}

For the detuning error, we first evaluate the matrix $R_{j}(t)=U_{I}^{\dagger}(t)H_{j}U_{I}(t)\equiv R_{\delta_{A}}(t)$
by substituting $H_{j}$$=\mathcal{H}_{N,\delta_{A}}/(\hbar\beta_{\delta_{A}})$,
where $\mathcal{H}_{N,\delta_{A}}$ is given by Eq.\ (\ref{eq:uty_H_detuning_2x2}),
together with $U_{I}(t)$ from Eq.\ (\ref{eq:ideal_U_2x2}). The
resulting $R_{\delta_{A}}(t)$ matrix elements are

\begin{align}
R_{\delta_{A},00}(t) & =\sin^{2}\theta\cdot\left(\dfrac{2g_{AB}^{2}}{2g_{AB}^{2}+\delta_{A}^{2}/4}\right),\label{eq:r_mat_detuning_00}\\
R_{\delta_{A},01}^{\mathrm{Re}}(t) & =\dfrac{1}{2}\sin^{2}\theta\cdot\left(\dfrac{\sqrt{2}\delta_{A}g_{AB}}{2g_{AB}^{2}+\delta_{A}^{2}/4}\right),\label{eq:r_mat_detuning_01_R}\\
R_{\delta_{A},01}^{\mathrm{Im}}(t) & =-\sin\theta\cos\theta\cdot\left(\dfrac{\sqrt{2}g_{AB}}{\sqrt{2g_{AB}^{2}+\delta_{A}^{2}/4}}\right).\label{eq:r_mat_detuning_01_I}
\end{align}
We then substitute $\beta_{j}=\beta_{\delta_{A}}$ and the matrix
elements of $R_{j}(t)=R_{\delta_{A}}(t)$ from Eqs.\ (\ref{eq:r_mat_detuning_00})-(\ref{eq:r_mat_detuning_01_I})
into the corresponding decomposed terms of $J_{2,j}$ in Eqs.\ (\ref{eq:J2_uty_11_02_sum_00})-(\ref{eq:J2_uty_11_02_sum_01Im})
to obtain the estimated detuning error, 
\begin{equation}
J_{2,\delta_{A}}^{\mathrm{Est}}=J_{2,\delta_{A}}^{00}+J_{2,\delta_{A}}^{01,\mathrm{Re}}+J_{2,\delta_{A}}^{01,\mathrm{Im}},\label{eq:J2_dA_est_all}
\end{equation}
where

\begin{align}
 & J_{2,\delta_{A}}^{00}=\nonumber \\
 & \dfrac{3}{16}\dfrac{\beta_{\delta_{A}}^{2}g_{AB}^{4}t_{f}^{2}}{\left(2g_{AB}^{2}+\delta_{A}^{2}/4\right)^{2}}\left[1-\dfrac{\sin\left(2\sqrt{2g_{AB}^{2}+\delta_{A}^{2}/4}\cdot t_{f}\right)}{2\sqrt{2g_{AB}^{2}+\delta_{A}^{2}/4}\cdot t_{f}}\right]^{2},\label{eq:J2_detuning_00}\\
 & J_{2,\delta_{A}}^{01,\mathrm{Re}}=\nonumber \\
 & \dfrac{1}{32}\dfrac{\beta_{\delta_{A}}^{2}g_{AB}^{2}\delta_{A}^{2}t_{f}^{2}}{\left(2g_{AB}^{2}+\delta_{A}^{2}/4\right)^{2}}\left[1-\dfrac{\sin\left(2\sqrt{2g_{AB}^{2}+\delta_{A}^{2}/4}\cdot t_{f}\right)}{2\sqrt{2g_{AB}^{2}+\delta_{A}^{2}/4}\cdot t_{f}}\right]^{2},\label{eq:J2_detuning_01_R}\\
 & J_{2,\delta_{A}}^{01,\mathrm{Im}}=\nonumber \\
 & \dfrac{1}{32}\dfrac{\beta_{\delta_{A}}^{2}g_{AB}^{2}}{\left(2g_{AB}^{2}+\delta_{A}^{2}/4\right)^{2}}\left[1-\cos(2\sqrt{2g_{AB}^{2}+\delta_{A}^{2}/4}\cdot t_{f})\right]^{2}.\label{eq:J2_detuning_01_I}
\end{align}

Under the strong $\left|1_{B}1_{A}\right\rangle $$\leftrightarrow$$\left|0_{B}2_{A}\right\rangle $
coupling condition, the detuning $\delta_{A}$ approaches zero, so
$J_{2,\delta_{A}}^{01,\mathrm{Re}}$ in Eq.\ (\ref{eq:J2_detuning_01_R}),
which is proportional to $\delta_{A}^{2}$, vanishes. Furthermore,
as $\delta_{A}\rightarrow0$, the CZ gate time $t_{f}$ approaches
its speed limit, satisfying $\sqrt{2}g_{AB}\cdot t_{f}\approx\pi$.
These limits imply that $2\sqrt{2g_{AB}^{2}+\delta_{A}^{2}/4}\cdot t_{f}\rightarrow2\pi$,
so both $[1-\cos(2\sqrt{2g_{AB}^{2}+\delta_{A}^{2}/4}\cdot t_{f})]$
in $J_{2,\delta_{A}}^{01,\mathrm{Im}}$ {[}Eq.\ (\ref{eq:J2_detuning_01_I}){]}
and $\sin(2\sqrt{2g_{AB}^{2}+\delta_{A}^{2}/4}\cdot t_{f})$ in $J_{2,\delta_{A}}^{00}$
{[}Eq.\ (\ref{eq:J2_detuning_00}){]} approach zero. Consequently,
we neglect $J_{2,\delta_{A}}^{01,\mathrm{Re}}$ and $J_{2,\delta_{A}}^{01,\mathrm{Im}}$,
and omit the sine term in $J_{2,\delta_{A}}^{00}$, yielding the approximate
expression for $J_{2,\delta_{A}}^{\mathrm{Est}}$ given in Eq.\ (\ref{eq:J2_dA_est})
of the main text.

\subsection{Coupling error $J_{2,g_{AB}}^{\mathrm{Est}}$}

For the coupling error, substituting $H_{j}=\mathcal{H}_{N,g_{AB}}/(\hbar\beta_{g_{AB}})$,
where $\mathcal{H}_{N,g_{AB}}$ is given by Eq.\ (\ref{eq:uty_H_coupling_2x2}),
together with $U_{I}(t)$ from Eq.\ (\ref{eq:ideal_U_2x2}), into
$R_{j}(t)=U_{I}^{\dagger}(t)H_{j}U_{I}(t)\equiv R_{g_{AB}}(t)$ yields
the matrix elements

\begin{align}
R_{g_{AB},00}(t) & =-\sin^{2}\theta\dfrac{2g_{AB}\delta_{A}}{2g_{AB}^{2}+\delta_{A}^{2}/4},\label{eq:r_mat_coupling_00}\\
R_{g_{AB},01}^{\mathrm{Re}}(t) & =\sqrt{2}\cos^{2}\theta+\sqrt{2}\sin^{2}\theta\dfrac{2g_{AB}^{2}-\delta_{A}^{2}/4}{2g_{AB}^{2}+\delta_{A}^{2}/4},\label{eq:r_mat_coupling_01_R}\\
R_{g_{AB},01}^{\mathrm{Im}}(t) & =-\sqrt{2}\sin\theta\cos\theta\dfrac{\delta_{A}}{\sqrt{2g_{AB}^{2}+\delta_{A}^{2}/4}}.\label{eq:r_mat_coupling_00_I}
\end{align}
Using $\beta_{j}=\beta_{g_{AB}}$ and the matrix elements of $R_{j}(t)=R_{g_{AB}}(t)$
from Eqs.\ (\ref{eq:r_mat_coupling_00})-(\ref{eq:r_mat_coupling_00_I})
in the corresponding decomposed terms of $J_{2,j}$ in Eqs.\ (\ref{eq:J2_uty_11_02_sum_00})-(\ref{eq:J2_uty_11_02_sum_01Im}),
the estimated coupling error is obtained as

\begin{equation}
J_{2,g_{AB}}^{\mathrm{Est}}=J_{2,g_{AB}}^{00}+J_{2,g_{AB}}^{01,\mathrm{Re}}+J_{2,g_{AB}}^{01,\mathrm{Im}},\label{eq:J2_gAB_est_all}
\end{equation}
where

\begin{align}
 & J_{2,g_{AB}}^{00}=\nonumber \\
 & \dfrac{3}{16}\dfrac{\beta_{g_{AB}}^{2}g_{AB}^{2}\delta_{A}^{2}t_{f}^{2}}{\left(2g_{AB}^{2}+\delta_{A}^{2}/4\right)^{2}}\left[1-\dfrac{\sin\left(2\sqrt{2g_{AB}^{2}+\delta_{A}^{2}/4}\cdot t_{f}\right)}{2\sqrt{2g_{AB}^{2}+\delta_{A}^{2}/4}\cdot t_{f}}\right]^{2},\label{eq:J2_coupling_00}\\
 & J_{2,g_{AB}}^{01,\mathrm{Re}}=\nonumber \\
 & 2\dfrac{\beta_{g_{AB}}^{2}g_{AB}^{4}t_{f}^{2}}{\left(2g_{AB}^{2}+\delta_{A}^{2}/4\right)^{2}}\left[1+\dfrac{\delta_{A}^{2}}{8g_{AB}^{2}}\dfrac{\sin\left(2\sqrt{2g_{AB}^{2}+\delta_{A}^{2}/4}\cdot t_{f}\right)}{2\sqrt{2g_{AB}^{2}+\delta_{A}^{2}/4}\cdot t_{f}}\right]^{2},\label{eq:J2_coupling_01_R}\\
 & J_{2,g_{AB}}^{01,\mathrm{Im}}=\nonumber \\
 & \dfrac{1}{32}\dfrac{\beta_{g_{AB}}^{2}\delta_{A}^{2}}{\left(2g_{AB}^{2}+\delta_{A}^{2}/4\right)^{2}}\left[1-\cos(2\sqrt{2g_{AB}^{2}+\delta_{A}^{2}/4}\cdot t_{f})\right]^{2}.\label{eq:J2_coupling_01_I}
\end{align}

The strong $\left|1_{B}1_{A}\right\rangle $$\leftrightarrow$$\left|0_{B}2_{A}\right\rangle $
coupling condition leads to the two limits, $\delta_{A}\rightarrow0$
and $2\sqrt{2g_{AB}^{2}+\delta_{A}^{2}/4}\cdot t_{f}\rightarrow2\pi$,
as discussed above for the detuning error. The former limit causes
$J_{2,g_{AB}}^{00}$ in Eq.\ (\ref{eq:J2_coupling_00}) and $J_{2,g_{AB}}^{01,\mathrm{Im}}$
in Eq.\ (\ref{eq:J2_coupling_01_I}) to vanish, since both are proportional
to $\delta_{A}^{2}$. The latter limit eliminates the sine term in
$J_{2,g_{AB}}^{00}$ in Eq.\ (\ref{eq:J2_coupling_00}), where $\sin\left(2\sqrt{2g_{AB}^{2}+\delta_{A}^{2}/4}\cdot t_{f}\right)\rightarrow0$.
Neglecting these terms yields the approximate expression for $J_{2,g_{AB}}^{\mathrm{Est}}$
given in Eq.\ (\ref{eq:J2_g_est}) of the main text.

\section{Ensemble-averaged detuning and coupling errors}

\label{appendix:precise_uty_cost_function}

In this appendix, we derive the ensemble-averaged detuning and coupling
errors, $\left\langle J_{2,\delta}\right\rangle $ and $\left\langle J_{2,g_{AB}}\right\rangle $,
respectively, which are employed as cost functions in the optimization
procedure discussed in Sec.\ \ref{sec:overall_infidelity_improvement}.

\subsection{Detuning error $\left\langle J_{2,\delta}\right\rangle $}

The diagonal uncertainties $\beta_{\delta}\in\left\{ \beta_{\alpha_{A}},\beta_{\alpha_{B}},\beta_{\omega_{0,A}},\beta_{\omega_{0,B}}\right\} $
are assumed to be mutually independent and normally distributed with
zero mean and a common standard deviation $\sigma_{\delta}$. Therefore,
the corresponding ensemble-averaged detuning errors can be evaluated
independently. Each diagonal uncertainty $\beta_{\delta}$ contributes
to the detuning uncertainties $\beta_{\delta_{A}}$, $\beta_{\delta_{B}}$,
and $\beta_{\delta_{\mathrm{01-10}}}$. The uncertainties $\beta_{\delta_{A}}$
and $\beta_{\delta_{B}}$ enter the Hamiltonian in the subspace $\left\{ \left|1_{B}1_{A}\right\rangle ,\left|0_{B}2_{A}\right\rangle ,\left|2_{B}0_{A}\right\rangle \right\} $,
whereas $\beta_{\delta_{\mathrm{01-10}}}$ enters the Hamiltonian
in the subspace $\left\{ \left|0_{B}1_{A}\right\rangle ,\left|1_{B}0_{A}\right\rangle \right\} $.
Since these Hamiltonian blocks commute, their contributions can be
evaluated separately. We first evaluate the contribution $\left\langle J_{2,\delta_{A}+\delta_{B}}\right\rangle $
arising from $\beta_{\delta_{A}}$ and $\beta_{\delta_{B}}$, and
then evaluate the contribution $\left\langle J_{2,\delta_{01-10}}\right\rangle $
associated with $\beta_{\delta_{\mathrm{01-10}}}$. The total ensemble-averaged
detuning error is then given by $\left\langle J_{2,\delta}\right\rangle =\left\langle J_{2,\delta_{A}+\delta_{B}}\right\rangle +\left\langle J_{2,\delta_{01-10}}\right\rangle $.

To evaluate the individual contributions to the ensemble-averaged
detuning error $\left\langle J_{2,\delta_{A}+\delta_{B}}\right\rangle $
associated with each diagonal uncertainty $\beta_{\delta}$, we first
define the uncertainty Hamiltonian $\mathcal{H}_{\delta}=\hbar\beta_{\delta}H_{\delta}$
from $\mathcal{H}_{I_{22}}$ in Eq.\ (7) of Ref.\ \citep{Huang2026}.
The operator $H_{\delta}$ is represented in the basis $\left\{ \left|1_{B}1_{A}\right\rangle ,\left|0_{B}2_{A}\right\rangle ,\left|2_{B}0_{A}\right\rangle \right\} $
as

\begin{equation}
H_{\delta}=\left(\begin{array}{ccc}
0 & 0 & 0\\
0 & \beta_{\delta_{A}}/\beta_{\delta} & 0\\
0 & 0 & \beta_{\delta_{B}}/\beta_{\delta}
\end{array}\right).\label{eq:uncertainty_Hamiltonian_j}
\end{equation}
For a given diagonal uncertainty $\beta_{\delta}$, the coefficients
$\beta_{\delta_{A}}/\beta_{\delta}$ and $\beta_{\delta_{B}}/\beta_{\delta}$
in $H_{\delta}$ are obtained from Eqs.\ (\ref{eq:uty_qbA_freq})-(\ref{eq:beta_dtB})
by retaining only the terms proportional to $\beta_{\delta}$. The
resulting detuning error is then evaluated by setting $\beta_{j}=\beta_{\delta}$
and $H_{j}=H_{\delta}$ in the general expression for $J_{2,j}$ given
in Eq.\ (\ref{eq:J2_uty_general_format}). Upon averaging $J_{2,j}$
over the uncertainty ensemble, the quadratic uncertainty term $\beta_{\delta}^{2}$
is replaced by its expectation value, $\left\langle \beta_{\delta}^{2}\right\rangle =\sigma_{\delta}^{2}$.
Therefore, the ensemble-averaged contribution can be obtained by substituting
$\beta_{j}=\sigma_{\delta}$ and $H_{j}=H_{\delta}$ into Eq.\ (\ref{eq:J2_uty_general_format}).
Applying this procedure to the four diagonal uncertainties yields
the contributions $\left\langle J_{2,\delta_{A}+\delta_{B},\alpha_{A}}\right\rangle $,
$\left\langle J_{2,\delta_{A}+\delta_{B},\alpha_{B}}\right\rangle $,
$\left\langle J_{2,\delta_{A}+\delta_{B},\omega_{0,A}}\right\rangle $,
and $\left\langle J_{2,\delta_{A}+\delta_{B},\omega_{0,B}}\right\rangle $,
corresponding to $\beta_{\alpha_{A}}$, $\beta_{\alpha_{B}}$, $\beta_{\omega_{0,A}}$,
and $\beta_{\omega_{0,B}}$, respectively. The total ensemble-averaged
detuning error associated with the detuning uncertainties $\beta_{\delta_{A}}$
and $\beta_{\delta_{B}}$ is therefore given by

\begin{align}
 & \left\langle J_{2,\delta_{A}+\delta_{B}}\right\rangle =\nonumber \\
 & \left\langle J_{2,\delta_{A}+\delta_{B},\alpha_{A}}\right\rangle +\left\langle J_{2,\delta_{A}+\delta_{B},\alpha_{B}}\right\rangle +\nonumber \\
 & \left\langle J_{2,\delta_{A}+\delta_{B},\omega_{0,A}}\right\rangle +\left\langle J_{2,\delta_{A}+\delta_{B},\omega_{0,B}}\right\rangle .\label{eq:ensemble_ave_J2_dt_dtAB}
\end{align}

For the ensemble-averaged detuning error $\left\langle J_{2,\delta_{01-10}}\right\rangle $,
we employ an alternative evaluation method. We begin with the realistic
Hamiltonian $\mathcal{H}_{1}$, expressed in the basis $\left\{ \left|0_{B}1_{A}\right\rangle ,\left|1_{B}0_{A}\right\rangle \right\} $,
given by 
\begin{equation}
\mathcal{H}_{1}=\hbar\left(\begin{array}{cc}
\Delta_{A}+\beta_{\Delta_{A}} & g_{AB}\\
g_{AB} & \Delta_{B}+\beta_{\Delta_{B}}
\end{array}\right),\label{eq:realistic_H_1}
\end{equation}
which is obtained from the ideal Hamiltonian $\mathcal{H}_{I_{1}}$
introduced in Sec.\ %
\mbox{%
II%
}\ A of Ref.\ \citep{Huang2026}. Here, the uncertainties $\beta_{\Delta_{A}}$
and $\beta_{\Delta_{B}}$, associated with the ideal qubit frequencies
$\Delta_{A}$ and $\Delta_{B}$, respectively, give rise to the detuning
uncertainty $\beta_{\delta_{\mathrm{01-10}}}=\beta_{\Delta_{A}}-\beta_{\Delta_{B}}$.
For each diagonal uncertainty $\beta_{\delta}$, the corresponding
terms $\beta_{\Delta_{A}}$ and $\beta_{\Delta_{B}}$ in $\mathcal{H}_{1}$
are obtained from Eqs.\ (\ref{eq:uty_qbA_freq})-(\ref{eq:uty_qbB_freq})
by retaining only the contributions proportional to $\beta_{\delta}$.
The realistic infidelity $\mathcal{I}$ is then simulated using the
resulting $\mathcal{H}_{1}$. During the simulation of $\mathcal{I}$,
the characterized single-qubit-Z gate $U_{Z}$ is incorporated to
compensate for the single-qubit phase error induced by $\beta_{\Delta_{A}}$
and $\beta_{\Delta_{B}}$. Consequently, the simulated infidelity
$\mathcal{I}$ is dominated by the intrinsic error $J_{1}$ and the
detuning error arising from $\beta_{\delta_{\mathrm{01-10}}}$. Upon
averaging $\mathcal{I}$ over the uncertainty ensemble, all odd-order
terms vanish because $\beta_{\delta}$ follows a zero-mean normal
distribution with standard deviation $\sigma_{\delta}$. Provided
that the uncertainty amplitude is sufficiently small, the leading
contribution to the detuning error is the second-order term proportional
to $\left\langle \beta_{\delta}^{2}\right\rangle =\sigma_{\delta}^{2}$.
This second-order term corresponds to the ensemble-averaged detuning
error represented by $\left\langle J_{2,j}\right\rangle $ according
to Eq.\ (\ref{eq:J2_uty_general_format}) and can be extracted from
$(\mathcal{I}_{+}+\mathcal{I}_{-})/2$, where $\mathcal{I}_{+}$ and
$\mathcal{I}_{-}$ denote the simulated infidelities obtained with
$\beta_{\delta}=+\sigma_{\delta}$ and $\beta_{\delta}=-\sigma_{\delta}$,
respectively. Because the odd-order terms have opposite signs for
$\mathcal{I}_{+}$ and $\mathcal{I}_{-}$, they cancel in the average,
such that $(\mathcal{I}_{+}+\mathcal{I}_{-})/2$ consists primarily
of $J_{1}$ and the second-order term proportional to $\sigma_{\delta}^{2}$.
The ensemble-averaged detuning error is therefore extracted as $(\mathcal{I}_{+}+\mathcal{I}_{-})/2-J_{1}$,
where subtracting the intrinsic error $J_{1}$ isolates the second-order
contribution. Applying this procedure to the four diagonal uncertainties
yields the individual ensemble-averaged detuning errors $\left\langle J_{2,\delta_{01-10},\alpha_{A}}\right\rangle $,
$\left\langle J_{2,\delta_{01-10},\alpha_{B}}\right\rangle $, $\left\langle J_{2,\delta_{01-10},\omega_{0,A}}\right\rangle $,
and $\left\langle J_{2,\delta_{01-10},\omega_{0,B}}\right\rangle $,
corresponding to $\beta_{\alpha_{A}}$, $\beta_{\alpha_{B}}$, $\beta_{\omega_{0,A}}$,
and $\beta_{\omega_{0,B}}$, respectively. The total ensemble-averaged
detuning error associated with the detuning uncertainty $\beta_{\delta_{\mathrm{01-10}}}$
is then given by 
\begin{align}
 & \left\langle J_{2,\delta_{01-10}}\right\rangle =\nonumber \\
 & \left\langle J_{2,\delta_{01-10},\alpha_{A}}\right\rangle +\left\langle J_{2,\delta_{01-10},\alpha_{B}}\right\rangle +\nonumber \\
 & \left\langle J_{2,\delta_{01-10},\omega_{0,A}}\right\rangle +\left\langle J_{2,\delta_{01-10},\omega_{0,B}}\right\rangle .\label{eq:ensemble_ave_J2_dt_dt01_10}
\end{align}

Finally, the contributions $\left\langle J_{2,\delta_{A}+\delta_{B}}\right\rangle $
and $\left\langle J_{2,\delta_{01-10}}\right\rangle $, obtained from
Eqs.\ (\ref{eq:ensemble_ave_J2_dt_dtAB}) and (\ref{eq:ensemble_ave_J2_dt_dt01_10}),
respectively, are combined to obtain the overall ensemble-averaged
detuning error,

\begin{equation}
\left\langle J_{2,\delta}\right\rangle =\left\langle J_{2,\delta_{A}+\delta_{B}}\right\rangle +\left\langle J_{2,\delta_{01-10}}\right\rangle .\label{eq:ensemble_ave_J2_dt_all}
\end{equation}

\subsection{Coupling error $\left\langle J_{2,g_{AB}}\right\rangle $}

A small residual error contribution remains in the coupling error
$J_{2,g_{AB}}$, leading to an inaccurate evaluation of the coupling
error. This is because the compensation of this residual contribution
by the characterized single-qubit-Z gate $U_{Z}$ in the realistic
CZ-gate implementation is not reflected in the evaluation based on
$J_{2,g_{AB}}$. To address this issue, we adopt the error-evaluation
method $(\mathcal{I}_{+}+\mathcal{I}_{-})/2-J_{1}$, introduced above
for evaluating $\left\langle J_{2,\delta_{01-10}}\right\rangle $,
to isolate the ensemble-averaged coupling error $\left\langle J_{2,g_{AB}}\right\rangle $.
This method removes the residual error contribution associated with
the evaluation of $J_{2,g_{AB}}$ and thereby improves the accuracy
of the extracted coupling error. Specifically, we employ the realistic
Hamiltonian $\mathcal{H}_{1+2}$, defined in the basis $\left\{ \left|0_{B}1_{A}\right\rangle ,\left|1_{B}0_{A}\right\rangle ,\left|1_{B}1_{A}\right\rangle ,\left|0_{B}2_{A}\right\rangle ,\left|2_{B}0_{A}\right\rangle \right\} $,
which is obtained by replacing the coupling constant $g_{AB}$ in
the ideal Hamiltonian $\mathcal{H}_{I_{1}}+\mathcal{H}_{I_{2}}$ with
$g_{AB}+\beta_{g_{AB}}$. The ideal Hamiltonians $\mathcal{H}_{I_{1}}$
and $\mathcal{H}_{I_{2}}$ are described in Sec.\ %
\mbox{%
II%
}\ A of Ref.\ \citep{Huang2026}. A normal distribution with zero
mean and standard deviation $\sigma_{g_{AB}}$ is assumed for the
uncertainty $\beta_{g_{AB}}$. Using the realistic Hamiltonian $\mathcal{H}_{1+2}$,
we simulate the infidelity $\mathcal{I}$ with $\beta_{g_{AB}}=+\sigma_{g_{AB}}$
and $\beta_{g_{AB}}=-\sigma_{g_{AB}}$ to obtain $\mathcal{I}_{+}$
and $\mathcal{I}_{-}$, respectively. Finally, by substituting the
simulated $\mathcal{I}_{\pm}$ and $J_{1}$ into $(\mathcal{I}_{+}+\mathcal{I}_{-})/2-J_{1}$,
we obtain the ensemble-averaged coupling error $\left\langle J_{2,g_{AB}}\right\rangle $.

\bibliographystyle{apsrev4-2}
\bibliography{citation_robust_gates}

\end{document}